\documentclass[aps,prb,twocolumn,reprint,showpacs,superscriptaddress,citeautoscript,longbibliography]{revtex4-1}
\usepackage{graphicx,amsmath,amssymb}
\usepackage{fixmath}

\begin{document}

\title{Angular momentum and topology in semiconducting single-wall carbon nanotubes}

\author{W. Izumida}
\email[]{izumida@cmpt.phys.tohoku.ac.jp}
\affiliation{Department of Physics, Tohoku University, Sendai 980-8578, Japan}
\author{R. Okuyama}
\email[]{okuyama@flex.phys.tohoku.ac.jp}
\affiliation{Department of Physics, Tohoku University, Sendai 980-8578, Japan}
\author{A. Yamakage}
\email[]{ai@rover.nuap.nagoya-u.ac.jp}
\affiliation{Department of Applied Physics, Nagoya University, Nagoya 464-8603, Japan}
\affiliation{Institute for Advanced Research, Nagoya University, Nagoya 464-8601, Japan}
\author{R. Saito}
\affiliation{Department of Physics, Tohoku University, Sendai 980-8578, Japan}

\date{
}

\begin{abstract}
  Semiconducting single-wall carbon nanotubes are classified into two
  types by means of orbital angular momentum of valley state, which is
  useful to study their low energy electronic properties in
  finite-length.
  The classification is given by an integer $d$, which is the greatest
  common divisor of two integers $n$ and $m$ specifying the chirality
  of nanotubes, by analyzing cutting lines.
  For the case that $d$ is equal to or greater than four, two lowest
  subbands from two valleys have different angular momenta with
  respect to the nanotube axis.
  Reflecting the decoupling of two valleys, discrete energy levels in
  finite-length nanotubes exhibit nearly fourfold degeneracy and its
  small lift by the spin--orbit interaction.
  For the case that $d$ is less than or equal to two, in which two
  lowest subbands from two valleys have the same angular momentum,
  discrete levels exhibit lift of fourfold degeneracy reflecting the
  coupling of two valleys.
  Especially, two valleys are strongly coupled when the chirality is
  close to the armchair chirality.
  An effective one-dimensional lattice model is derived by extracting
  states with relevant angular momentum, which reveals the valley
  coupling in the eigenstates.
  A bulk-edge correspondence, relationship between number of edge
  states and the winding number calculated in the corresponding bulk
  system, is analytically shown by using the argument principle, which
  enables us to estimate the number of edge states from the bulk
  property.
  The number of edge states depends not only on the chirality but also
  on the shape of boundary.
\end{abstract}

\pacs{73.63.Fg, 73.22.Dj}

\keywords{carbon nanotube, finite-length, degeneracy, cutting line, tight-binding calculation}

\maketitle

\section{Introduction}

Recent studies of quantum transport have revealed that single-wall
carbon nanotubes (SWNTs) contain richer phenomena,
especially in terms of the spin and valley degrees of
freedom.~\cite{RevModPhys.87.703}
Fourfold degeneracy of SWNTs, observed in the tunneling
spectroscopy,~\cite{Liang-2002-03,Cobden-2002-07,Jarillo-Herrero-2005-04,Moriyama-2005-05,Sapmaz-2005-04,Maki-2005-06,Cao-2005-09,Makarovski-2006-10,Moriyama-2007-04,Holm-2008-04}
has been widely considered as an intrinsic property of SWNTs,
reflecting the two degenerate $K$ and $K'$ valleys in the
two-dimensional (2D) Brillouin zone (BZ), together with two spin
states of up and down.
Measurements for ultraclean
SWNTs~\cite{Kuemmeth-2008-03,Jhang-2010-07,Jespersen-2011-04,Steele:2013fk}
show fine structures of lift of fourfold degeneracy caused by the
spin--orbit
interaction.~\cite{Ando-2000-06,Chico-2004-10,Huertas-Hernando-2006-10,Chico-2009-06,Izumida-2009-06,Jeong-2009-08}
Recent studies focused on the angular momenta of two valleys as a
conserved quantity in disorder-free finite-length metallic SWNTs
(m-SWNTs).~\cite{Izumida-2015-06,PhysRevB.92.075433}
It is shown that the two valleys have the same angular momentum for
the so-called metal-2 nanotubes,
whereas they have different angular momenta for the metal-1
nanotubes.~\cite{PhysRevB.71.125408,Izumida-2015-06}
(The definitions of metal-1 and metal-2 are given in
Sec. \ref{sec:AngularMomentum}.)
For finite-length m-SWNTs which has the same rotational symmetry with
the bulk, the two valleys are decoupled and the spin--orbit
interaction lifts the fourfold degeneracy for the metal-1 nanotubes,
whereas valley coupling due to the inter-valley scattering at the ends
lifts the fourfold degeneracy, regardless of spin--orbit interaction,
for the metal-2 nanotubes.~\cite{Izumida-2015-06,PhysRevB.92.075433}
Valley coupling could play a role for large lift of the fourfold
degeneracy in the
experiments.~\cite{Maki-2005-06,Makarovski-2006-10,Moriyama-2007-04,Holm-2008-04,Steele:2013fk,PhysRevB.91.155435,Ferrier:2016aa}

Many of the quantum transport measurements have been performed for the
m-SWNTs,~\cite{Liang-2002-03,Cobden-2002-07,Jarillo-Herrero-2005-04,Moriyama-2005-05,Sapmaz-2005-04,Maki-2005-06,Cao-2005-09,Makarovski-2006-10,Moriyama-2007-04,Holm-2008-04,Kuemmeth-2008-03,Jhang-2010-07,Jespersen-2011-04,Steele:2013fk,RevModPhys.87.703}
which could contain narrow energy gap induced by curvature of nanotube
surface.
Conducting of the semiconducting SWNTs (s-SWNTs) has also been
investigated by adjusting the gate
voltage.~\cite{Jarillo-Herrero-2004-05,Cao-2005-09,Deshpande-2008-04,RevModPhys.87.703}
In fact, the s-SWNTs are double abundant of the m-SWNTs.~\cite{Saito-1998}
However, the relationship between valley, angular momentum, and the
finite-length effect is not well analyzed for the s-SWNTs, even though
they play essential roles in the quantum transport measurements.
Here we will focus on the s-SWNTs.

In this paper, we will classify the s-SWNTs by means of the angular
momentum of the valley.
We will show that the classification is performed by an integer
$d=\mathrm{gcd}(n,m)$, which is the greatest common divisor (gcd) of
$n$ and $m$, where the integer set $(n,m)$ specifies the chiral
vector.~\cite{Saito-1998}
(See Sec. \ref{sec:AngularMomentum} for the definition of chiral
vector.)
For the s-SWNTs with $d \ge 4$, the two valleys have different angular
momenta and the two valleys are decoupled.
For this case the spin--orbit interaction lifts the fourfold
degeneracy in the finite-length nanotubes, as in the case of metal-1
nanotubes.~\cite{Izumida-2015-06}
On the other hand, for the s-SWNTs with $d =1$ and $d=2$, the two
valleys have the same angular momentum and valley coupling lifts the
fourfold degeneracy, as in the case of metal-2
nanotubes.~\cite{Izumida-2015-06}
Especially for the case of $|n - m| = 2$, which is close to the armchair
chirality, the two valleys are strongly coupled.
Note that $d=3$, more generally $\mathrm{mod} (d, 3) = 0$, holds only
for a part of m-SWNTs.
In order to understand the coupling of two valleys, an effective
one-dimensional (1D) lattice model is derived and analyzed.
Chirality and edge shape dependences of the number of edge states, as
another intrinsic feature in the finite-length nanotubes, are given by
expanding the theory for valley coupling.
Especially, a bulk-edge correspondence, relationship between the
number of edge states and the winding number given in the
corresponding bulk system, is analytically shown.

The present study reveals that the coupling of two valleys occurs in
the majority of SWNTs, {\it even for ultraclean tubes with clean
  edges}, which conserve the angular momentum of ideal bulk states of
electrons.
The valley coupling appears as the lift of fourfold degeneracy,
regardless of the spin--orbit interaction.
Moreover, the bulk-edge correspondence for edge states in the
semiconducting energy gap reveals another feature of SWNTs as
topological insulating materials.

This paper is organized as follows.
In Sec. \ref{sec:AngularMomentum}, classification of the s-SWNTs by
means of the angular momentum of the valley is given.
In Sec. \ref{sec:numCalc}, numerical calculation is given for
finite-length s-SWNTs, which shows the valley coupling for $d \le 2$
cases.
In Sec. \ref{sec:effctive_1D_model}, an effective 1D lattice model is
derived (Sec. \ref{subsec:effctive_1D_model}), eigenstates under
boundary is analyzed (Sec. \ref{subsec:Eigenstate}), and number of
edge states is given from both a mode analysis of boundary condition
and a topological viewpoint for the bulk 1D lattice model
(Sec. \ref{subsec:EdgeState}).
The conclusion is given in Sec. \ref{sec:Conclusion}.

\section{Angular momenta of two valleys}
\label{sec:AngularMomentum}

In order to discuss the angular momentum of SWNTs, we first redefine
the previous concept of cutting
line,~\cite{Samsonidze:2003-12-01T00:00:00:1533-4880:431} by using the
helical--angular construction.~\cite{Barros-2006-09}

Let us consider a SWNT defined by rolling up the graphene sheet in the
direction of the chiral vector $\mathbold{C}_h = n \mathbold{a}_1 + m
\mathbold{a}_2 \equiv (n,m)$, where $n$ and $m$ are integers
specifying the chirality of SWNT, $\mathbold{a}_1 = (\sqrt{3}/2,
1/2)a$ and $\mathbold{a}_2 = (\sqrt{3}/2, -1/2)a$ are the unit vectors
of graphene, $a=0.246$ nm is the lattice constant.~\cite{Saito-1998}
In this paper, we consider the rolling of the graphene from the front
to the back, and, unless otherwise indicated, the case of $n \ge m \ge
0$, which correspond to the ``zigzag-right handedness'', except for
the zigzag ($m=0$) and armchair ($n=m$) nanotubes, according to
Ref. \onlinecite{PhysRevB.69.205402}.
The nanotubes of opposite handedness, zigzag-left handedness, are
given for $(m,n)$, or rolling of the graphene sheet from the back to
the front.~\cite{PhysRevB.69.205402}

\subsection{Cutting line}
\label{subsec:cutting_line}

Here we introduce the cutting line defined under the helical--angular
construction,~\cite{Barros-2006-09} and show the angular momentum of
$K$ and $K'$ points for the m-SWNTs,~\cite{Izumida-2015-06} which are
known in the previous studies.

To discuss the angular momenta of two valleys, it is convenient to
introduce an alternative set of unit vectors, utilizing the rotational
and the helical symmetry of SWNT, $\mathbold{C}_h/d$ and $\mathbold{R}
= p \mathbold{a}_1 + q
\mathbold{a}_2$.~\cite{White-1993-03,PhysRevB.47.16671}
Here $d=\mathrm{gcd}(n,m)$ is the greatest common divisor (gcd) of $n$
and $m$, and the integers $p$ and $q$ are chosen to satisfy the
relation of 
\begin{equation}
  m p - n q = d.
  \label{eq:def_pq}
\end{equation}
Translation with $\mathbold{C}_h/d$ corresponds to the ${\cal C}_d$
rotation with respect to the nanotube axis.
The vector $\mathbold{R}$ has $a_z$ and $\Delta \theta d_t / 2$
components in the axis and circumference directions, respectively,
where 
\begin{equation}
  a_z = \frac{T d}{N}
  \label{eq:a_z}
\end{equation}
is the shortest distance between two A (B) atoms in the axis
direction, $T = a \sqrt{3(n^2 + m^2 + nm)} / d_R$ is the lattice
constant of the 1D nanotube, $N = 2 (n^2 + m^2 + nm) / d_R$ is the
number of A (B) atoms in the 1D nanotube unit cell,
$d_R=\mathrm{gcd}(2n+m,2m+n)$,
\begin{equation}
  \Delta \theta = 2 \pi \frac{t_1 q - t_2 p}{N}
  \label{eq:Delta_theta}
\end{equation}
is the azimuth component of $\mathbold{R}$ in the cylindrical
coordinates, $d_t = |\mathbold{C}_h|/\pi$ is the diameter of nanotube,
$t_1 = (2 m+ n) / d_R$, $t_2 = - (2 n + m) / d_R$.  
There is an arbitrarity to define $p$ and $q$ since $\mathbold{R} +
\alpha \mathbold{C}_h/d$ also satisfies the definition of
$\mathbold{R}$, with $\alpha = 0,1,\cdots,d-1$.
Two reciprocal lattice vectors, which are conjugate to
$\mathbold{C}_h/d$ and $\mathbold{R}$, are given by
\begin{equation}
  \mathbold{Q}_1 = -q \mathbold{b}_1 + p \mathbold{b}_2, ~~~~ 
  \mathbold{Q}_2 = \frac{m}{d} \mathbold{b}_1 - \frac{n}{d} \mathbold{b}_2,
  \label{eq:Q1Q2}
\end{equation}
where $\mathbold{b}_1 = (2 \pi/a) (1/\sqrt{3},1)$ and $\mathbold{b}_2 = (2 \pi/a)
(1/\sqrt{3},-1)$ are the reciprocal lattice vectors of graphene,
and the relations
$\mathbold{Q}_1 \cdot \mathbold{C}_h / d = \mathbold{Q}_2 \cdot \mathbold{R} = 2 \pi$,
$\mathbold{Q}_1 \cdot \mathbold{R} = \mathbold{Q}_2 \cdot \mathbold{C}_h /d = 0$ hold.
The wavevector in the first BZ is expressed by
\begin{equation}
  \mathbold{k} = \mu \frac{\mathbold{Q}_1}{d} + k \frac{\mathbold{Q}_2}{\left| \mathbold{Q}_2 \right|},
  \label{eq:kinBZ}
\end{equation}
with 
\begin{equation}
  \mu = 0, 1, \cdots, d-1,
  ~~\text{and,}~~
  0 \le k < \frac{2 \pi}{a_z}.
  \label{eq:cuttingLine}
\end{equation}
Note that $|\mathbold{Q}_2| = 2 \pi / a_z$.
The discretization of $\mu$ is due to the periodic boundary condition
in the circumference direction, which requires $\mathbold{k} \cdot
\mathbold{C}_h/2 \pi$ is an arbitrary integer.
The integer $\mu$ is the orbital angular momentum around the nanotube
axis, and each line in the 2D $k$-space specified by $\mu$ is called
cutting line.

In comparison to the conventional formulation,~\cite{Saito-1998} we
have the relations,
$\mathbold{K}_1 = \mathbold{Q}_1 / d + \mathbold{Q}_2 \Delta \theta / 2 \pi$ and 
$\mathbold{K}_2 = \mathbold{Q}_2 d/ N$, 
where $\mathbold{K}_1$ represents the separation of cutting lines and
$\mathbold{K}_2$ is the vector of short cutting lines in the
conventional formulation, and they are orthogonal each other.
The cutting lines in the conventional formulation are given by $\mu
\mathbold{K}_1 + k \mathbold{K}_2 / | \mathbold{K}_2 |$ with $\mu =
0,\cdots, N-1$ and $-\pi/T \le k < \pi/T$.
The present formulation of 2D BZ by Eq. \eqref{eq:kinBZ} is
essentially the same with that given in our previous
study~\cite{Izumida-2015-06}, in which the rectangular shape of 2D BZ
is given by the translation of the triangle at the corner of the
present parallelogram shape of 2D BZ by $\left( d \Delta \theta / 2
\pi \right) \mathbold{Q}_2$.
It is convenient to use the two reciprocal lattice vectors
$\mathbold{Q}_1$ and $\mathbold{Q}_2$ in the present formulation,
rather than $\mathbold{K}_1$ and $\mathbold{K}_2$, to specify not only
the angular momenta of two valleys,~\cite{Izumida-2015-06} but also
the closest positions of $K$ and $K'$ points on the cutting lines, as
shown below.

In general, a $K$ point, $\overrightarrow{\Gamma K} = \left( 2
\mathbold{b}_1 + \mathbold{b}_2 \right) / 3$, is expressed by,
\begin{equation}
  \overrightarrow{\Gamma K} = \frac{2n+m}{3} \frac{\mathbold{Q}_1}{d} + \frac{2p+q}{3} \mathbold{Q}_2.
  \label{eq:GammaK}
\end{equation}
A $K'$ point is specified by $\overrightarrow{\Gamma K'} =
-\overrightarrow{\Gamma K}$.
By comparing Eqs. (\ref{eq:kinBZ}) and (\ref{eq:GammaK}), it is
concluded that the SWNT is metallic when $\mathrm{mod} (2n + m, 3) =
0$,~\cite{Saito-1998} because the cutting lines with 
$(2n + m)/3$ and $-(2n + m)/3$, or equivalently, 
\begin{align}
  \mu_K & = \mathrm{mod} \left( \frac{2n + m}{3}, d \right), \\
  \mu_{K'} & = \mathrm{mod} \left( -\frac{2n + m}{3}, d \right),
\end{align}
where $0 \le \mu_K, \mu_{K'} \le d-1$, pass through the $K$ and $K'$
points, respectively, whereas it is semiconducting when $\mathrm{mod}
(2n + m, 3) \ne 0$ because there is no cutting line which passes the
$K$ or $K'$ points.
Here we used the property that the cutting line of $\mu$ is equivalent
to that of $\mathrm{mod}(\mu,d)$.
Further, the m-SWNTs are classified into metal-1 ($d_R = d$)
and metal-2 ($d_R = 3 d$).~\cite{Saito-1998}
Two different cutting lines $\mu_K \ne 0$ and $\mu_{K'} \ne 0$ pass
through $K$ and $K'$ points, respectively, for metal-1, while, the
cutting line $\mu_K = \mu_{K'} = 0$ passes through both $K$ and $K'$
points for metal-2, as already proven in
Ref. \onlinecite{Izumida-2015-06}.
The $K$ ($K'$) point on the line $\mu_K$ ($\mu_{K'}$) is given by
\begin{align}
 k_K & = \frac{2 \pi}{3 a_z} \mathrm{mod} (2p + q, 3), \\
 k_{K'} & = \frac{2 \pi}{3 a_z} \mathrm{mod} (- 2p - q, 3).
\end{align}
Especially for the metal-2 nanotubes, since the relation
$\mathrm{mod}(2p+q,3)={\rm mod}(2m/d,3)$ holds, $k_{K} = 4\pi/3a_z$
($2\pi/3a_z$) and $k_{K'} = 2\pi/3a_z$ ($4\pi/3a_z$) for
$\mathrm{mod}(m/d,3)=1$ ($2$), as shown in Ref.
\onlinecite{Izumida-2015-06}.

\subsection{Classification of s-SWNTs}
\label{subsec:classification_s-SWNTs}

\begin{figure}[htb]
  \includegraphics[width=8.5cm]{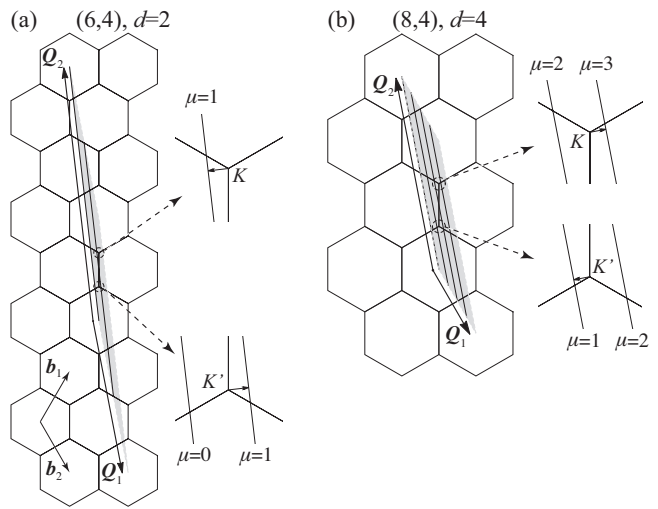}
  \caption{Cutting lines for (a) $(n,m)=(6,4)$ ($d=2$, type-1) and (b)
    $(n,m)=(8,4)$ ($d=4$, type-2) s-SWNT.
    The integers are chosen as (a) $(p,q)=(2,1)$ and (b)
    $(p,q)=(1,0)$.
    Shadow areas show 2D BZ given by two reciprocal lattice vectors
    $\mathbold{Q}_1$ and $\mathbold{Q}_2$.
    Dashed circles show $K$ and $K'$ points closest to the cutting
    lines.
    Right insets in each figure show enlarge near $K$ and $K'$ points.
    Arrows from $K$ and $K'$ points show $\pm \mathbold{K}_1 / 3$ to
    the nearest cutting lines.
  }
  \label{fig:cuttingLines}
\end{figure}

Using the new definition we classify the s-SWNTs.
For the s-SWNTs, the point closest to the $K$ point on the cutting
lines is given by,
\begin{equation}
  \overrightarrow{\Gamma K} \mp \frac{\mathbold{K}_1}{3} 
  = \frac{2n + m \mp 1}{3} \frac{\mathbold{Q}_1}{d} 
  + \frac{2p + q \mp \frac{\Delta \theta}{2 \pi} }{3} \mathbold{Q}_2,
  \label{eq:GammaK1}
\end{equation}
where $-$ and $+$ are applied for type-1 [$\mathrm{mod} (2n + m, 3) =
  1$] and type-2 [$\mathrm{mod} (2n + m, 3) = 2$]
s-SWNTs,~\cite{Jorio-2005-02} respectively (see
Fig. \ref{fig:cuttingLines}).
Here, we define $\mu$ specifying a cutting line which passes through a
point closest to a $K$ ($K'$) point the orbital angular momentum of
$K$ ($K'$) valley for the s-SWNTs.
From Eq. (\ref{eq:GammaK1}), the closest cutting line to the $K$ point
is given by
\begin{equation}
  \mu_1 = \mathrm{mod} \left( \frac{ 2n + m \mp 1 }{3}, d \right),
\end{equation}
and that to the $K'$ point is given by
\begin{equation}
  \mu_{1'} = \mathrm{mod} \left( -\frac{ 2n + m \mp 1 }{3}, d \right).
\end{equation}
Therefore, the condition that the two cutting lines $\mu_1$ and
$\mu_{1'}$ are identical is expressed by,
\begin{equation}
  \mathrm{mod} \left( 2 \times \frac{2n + m \mp 1}{3}, d \right) = 0.
  \label{eq:KK'identicalCondition}
\end{equation}
Since both $n$ and $m$ are multiple of $d$, Eq.
(\ref{eq:KK'identicalCondition}) holds only for $d = 1$ and $d = 2$.
For the s-SWNTs with $d \ge 4$, $\mu_1$ and $\mu_{1'}$ are
nonequivalent.
(See cutting lines in Fig. \ref{fig:cuttingLines}, as examples for
$d=2$ and $d=4$.)
Here we exclude the case of $d = 3$,
which holds only for a part of m-SWNTs.
For $d=1$, $\mu_1 = \mu_{1'} = 0$.
For $d=2$, we get $\mu_1 = \mu_{1'} = 1$ because they cannot be $0$ by
their definitions for $d \ge 2$.
It is also known from Eq. (\ref{eq:GammaK1}) that a point $k_1$
($k_{1'}$) on the line $\mu_1$ ($\mu_{1'}$), at which the bottom of
the conduction band and the top of the valence band appears from $K$
($K'$) valley, is given by
\begin{align}
  k_1 & = \frac{2 \pi}{3 a_z} \mathrm{mod} \left( 2p + q \mp \frac{\Delta \theta}{2 \pi}, 3 \right), \label{eq:k1}\\
  k_{1'} & = \frac{2 \pi}{3 a_z} \mathrm{mod} \left[ - \left( 2p + q \mp \frac{\Delta \theta}{2 \pi} \right), 3 \right]. \label{eq:k1d}
\end{align}
The cutting lines and the 1D $k$ points closest to $K$ and $K'$ points
are summarized in Table \ref{table:cuttingLine}.

\begin{table}[tbhp]
  \caption{
Cutting lines $\mu$ and 1D $k$ points closest to $K$ and $K'$ points.
The values of $\mu$ and $k$ are given in $\mu = 1,\cdots,d-1$ and $0
\le k < 2 \pi / a_z$, respectively.
Here $d=\mathrm{gcd}(n,m)$, $d_R=\mathrm{gcd}(2n+m,2m+n)$, $p$ and $q$
are given by Eq. \eqref{eq:def_pq}, $a_z$ and $\Delta \theta$ are
given in Eqs. \eqref{eq:a_z} and \eqref{eq:Delta_theta}, respectively.
The sign $-$ and $+$ in $\mu_1$, $\mu_{1'}$, $k_1$ and $k_{1'}$
applied for type-1 [$\mathrm{mod} (2n + m, 3) = 1$] and type-2
[$\mathrm{mod} (2n + m, 3) = 2$] s-SWNTs, respectively.
}
  \label{table:cuttingLine}
  \begin{tabular}{ll}
    \hline
    \hline
    Type & Cutting line and 1D $k$ point closest to $K$ and $K'$ \\
    \hline
    \multicolumn{2}{l}{Semiconducting SWNTs [$\mathrm{mod}(2m+n,3)=1,2$]} \\
    & $\mu_1 = \mathrm{mod} \left( \frac{ 2n + m \mp 1 }{3}, d \right)$, $\mu_{1'} = \mathrm{mod} \left( - \frac{ 2n + m \mp 1 }{3}, d \right)$ \\
    & $k_1 = \frac{2 \pi}{3 a_z} \mathrm{mod} \left( 2p + q \mp \frac{\Delta \theta}{2 \pi}, 3 \right)$ \\
    & $k_{1'} = \frac{2 \pi}{3 a_z} \mathrm{mod} \left[ - \left( 2p + q \mp \frac{\Delta \theta}{2 \pi} \right) , 3 \right]$ \\
    ~~$d = 1$ & $\mu_1 = \mu_{1'} = 0$ \\
    ~~$d = 2$ & $\mu_1 = \mu_{1'} = 1$ \\
    ~~$d \ge 4$ & $\mu_1 \neq \mu_{1'}$ \\
    \multicolumn{2}{l}{Metallic SWNTs [$\mathrm{mod}(2m+n,3)=0$]} \\
    & $\mu_K = \mathrm{mod} \left( \frac{2n + m}{3}, d \right)$, $\mu_{K'} = \mathrm{mod} \left( - \frac{2n + m}{3}, d \right)$ \\
    & $k_K = \frac{2 \pi}{3 a_z} \mathrm{mod} (2p + q, 3)$, $k_{K'} = \frac{2 \pi}{3 a_z} \mathrm{mod} (- 2p - q, 3)$ \\
    \multicolumn{2}{l}{~~metal-1 ($d_R = d$)} \\
    & $\mu_K \neq \mu_{K'}$ \\
    \multicolumn{2}{l}{~~metal-2 ($d_R = 3d$)} \\
    & $\mu_K = \mu_{K'} = 0$ \\
    & $k_{K} = \frac{4\pi}{3a_z}$, $k_{K'} = \frac{2\pi}{3a_z}$ $\left[ \text{if }\mathrm{mod} \left( \frac{m}{d},3 \right) = 1 \right]$ \\
    & $k_{K} = \frac{2\pi}{3a_z}$, $k_{K'} = \frac{4\pi}{3a_z}$ $\left[ \text{if }\mathrm{mod} \left( \frac{m}{d},3 \right) = 2 \right]$ \\
    \hline
    \hline
  \end{tabular}
\end{table}

\begin{figure}[htb]
  \includegraphics[width=8.5cm]{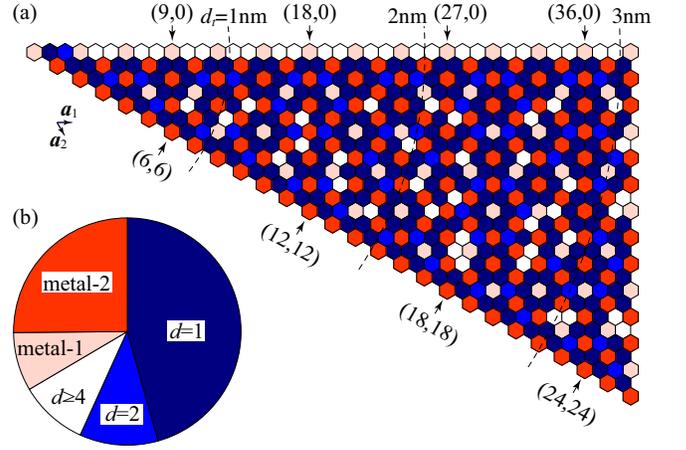}
  \caption{(Color online) Classification of single-wall carbon
    nanotubes by means of angular momenta of two valleys.  
    (a) A hexagon pointed by the chiral vector $n \mathbold{a}_1 + m
    \mathbold{a}_2$ from the leftmost hexagon represents $(n,m)$
    nanotube.
    Hexagons with pink (red) denotes metal-1 (metal-2) nanotubes.
    Hexagons with navy (blue) denotes $d=1$ ($d=2$) s-SWNTs, and 
    white hexagons denotes $d \ge 4$ s-SWNTs.
    Hexagons with dark colors (navy, blue and red) denote SWNTs, in
    which two valleys have the same angular momentum $\mu = 0$ (navy
    and red) or $\mu=1$ (blue).
    Hexagons with light colors (pink and white) denote SWNTs, in which
    two valleys have different angular momenta.
    The dashed curves show the borders indicating the corresponding
    diameters of nanotubes, $d_t=1$, $2$, and $3$ nm.
    (b) Pie chart of ratios of $d = 1$, $d = 2$, $d \ge 4$ of s-SWNTs,
    and metal-1 and metal-2 nanotubes for both left and right
    handedness in $0.5 \mathrm{nm} < d_t < 3 \mathrm{nm}$.
}
  \label{fig:nm_classification}
\end{figure}

Figure \ref{fig:nm_classification} shows the classification of SWNTs
up to $2n + m = 78$.
The ratio of the metal-2 nanotubes in the m-SWNTs, for both left and
right handedness,~\cite{PhysRevB.69.205402} in $0.5 \mathrm{nm} < d_t
< 3 \mathrm{nm}$ is about $75 \%$, and that of $d \le 2$ nanotubes in
the s-SWNTs is about $85 \%$.
This indicates that the two valleys have the same angular momentum in
the majority ($82 \%$) of SWNTs.
(See Fig. \ref{fig:nm_classification} (b) for the ratios.)

It is expected that finite-length s-SWNTs with $d \ge 4$, as well as
the metal-1 nanotubes with $d \ge 3$,~\cite{Izumida-2015-06} with
${\cal C}_d$ rotational symmetry exhibit fourfold degeneracy since
$\mu$ and $d-\mu$, which is equivalent to $-\mu$, states degenerate
under the time reversal symmetry,
and the cutting lines closest to $K$ and $K'$ are inequivalent for $d
\ge 3$.
It is important to note that the fourfold degeneracy is lifted by the
spin-orbit interaction that keeps the time-reversal
symmetry,~\cite{Izumida-2009-06} which will be confirmed in numerical
calculation presented in the Appendix \ref{sec:App:spectrum_0804}.
On the other hand, for finite-length s-SWNTs with $d \le 2$, as well
as the metal-2 nanotubes,~\cite{Izumida-2015-06} the strong valley
coupling can occur, which breaks the valley degeneracy regardless of
spin--orbit interaction, even the boundary is clean so as to keep the
rotational symmetry of bulk, which will be shown in the next section.

\section{Valley coupling in finite-length s-SWNTs}
\label{sec:numCalc}

In this section, we will show features of valley coupling in the
numerical calculation of the energy levels of finite-length s-SWNTs
with $d \le 2$.

\begin{figure}[htb]
  \includegraphics[width=8.5cm]{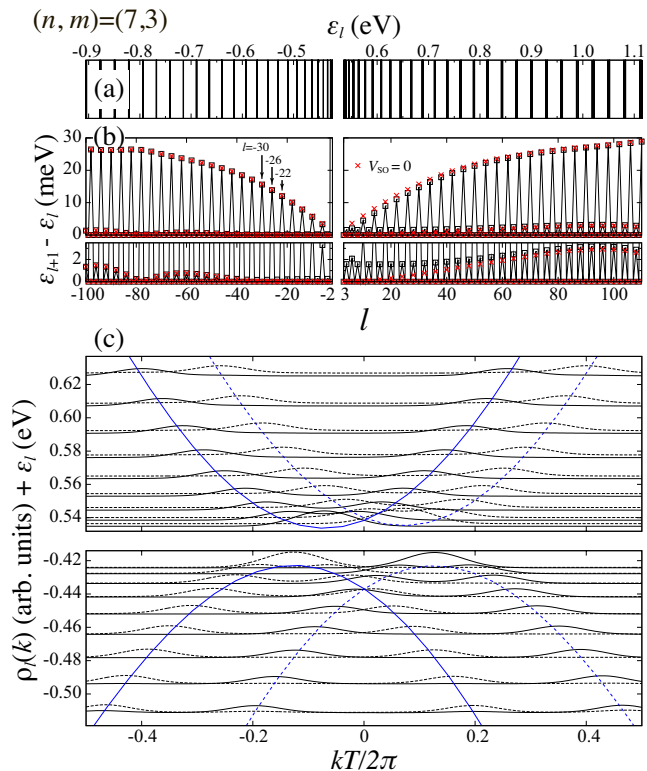}
  \caption{(Color online)
    Energy levels and eigenstates for $(7,3)$ nanotube ($d=1$, type-2) of
    $50.18$ nm length.
    (a) Energy levels $\varepsilon_l$ from valence ($\varepsilon_l \le
    -0.424$ eV, $l \le -2$) and conduction ($\varepsilon_l \ge 0.535$
    eV, $l \ge 3$) bands, where $l$ is the level index.
    (b) Level separation $\varepsilon_{l+1} - \varepsilon_l$ as a
    function of level index $l$.  
    The case of absence of spin-orbit interaction is shown by the red
    cross.
    To show the nearly fourfold degeneracy clearly, $l=-30, -26, -22$
    are indicated by arrows.
    (c) Intensity plot of Fourier transforms of wave functions at A
    sublattice for spin-up majority states.
    The energy for each level $\varepsilon_l$ is added for each
    intensity plot.
    They are presented by either solid or dashed lines in turn as
    increasing the energy to show them clearly.
    The blue solid and the blue dashed lines show the energy band
    calculation under the periodic boundary condition near the $K$ and
    the $K'$ valleys, respectively.
  }
  \label{fig:0703}
\end{figure}

In Fig. \ref{fig:0703} (a), we show the numerically calculated energy
levels $\varepsilon_l$ near the highest occupied molecular orbital
(HOMO), where $l$ is the index of the energy level numbered from HOMO
($l = 0$) in ascending order
of the energy for (7,3) nanotube of $50.18$ nm length.
Here the origin of the energy is set to be $(\varepsilon_{l=1} +
\varepsilon_{l=0})/2 = 0$.
The boundary is constructed by terminating at the plane orthogonal to
the nanotube axis and removing Klein-type terminations at which
terminated site neighbors two empty sites.
This is categorized as the minimal boundary, in which the edge
structure is selected to have minimum numbers of empty sites [dashed
  circles in Fig. \ref{fig:1D} (a)] and dangling bonds, and these
numbers are the same,~\cite{Akhmerov-2008-02} for both ends.
The diagrammatic picture of the left end is shown in Fig. \ref{fig:1D}
(a).
The calculation is done using the extended tight-binding method for
the finite-length nanotubes~\cite{Izumida-2015-06}, in which $\pi$ and
$\sigma$ orbitals of carbon atom are taken into account, and the
hopping and the overlap integrals between two atoms evaluated by the
{\it ab initio} calculation~\cite{Porezag-1995-05} for interatomic
distance up to 10 bohr ($\sim 5$\AA) in the three-dimensional
structure are included.
The optimized position of carbon atoms is utilized from the previous
calculation.~\cite{Izumida-2009-06}
To eliminate the dangling bonds in the numerical calculation, each
single dangling bond at the ends is terminated by a hydrogen atom.
The atomic spin--orbit interaction on each carbon atom are taken into
account.~\cite{Izumida-2009-06}
Unless otherwise indicated, the value of $V_\mathrm{SO} = 6$ meV for
the atomic spin--orbit interaction is used in the calculation.

The energy levels of $\varepsilon_l \le -0.424$ eV ($l \le -2$), which
are originated from the valence band, and $\varepsilon_l \ge 0.535$ eV
($l \ge 3$), originated from the conduction band, are shown in
Fig. \ref{fig:0703} (a).
In the energy gap, there are fourfold degenerate localized states near
the ends ($l = -1, 0, 1, 2$), 
two spin degenerate states polarized at
B sublattices at left end and two spin degenerate states polarized at
A sublattices at right end,
at $\varepsilon = 0$, not shown in the figure), known as the edge
states,~\cite{Fujita-1996,Klusek2000508,PhysRevB.71.193406,PhysRevB.73.085421}
which will be discussed in Sec. \ref{subsec:EdgeState}.
Due to the hopping to next nearest neighbor and farther sites, the
localized states appear below the center of the energy
gap.~\cite{Sasaki-2006-11}
As increasing (decreasing) the energy from the bottom of the
conduction band (the top of the valence band), the level separation,
$\varepsilon_{l+1} - \varepsilon_l$, gradually increases to the
constant value reflecting the changing from the quadratic to linear
energy dispersion of the energy bands, as shown in Fig. \ref{fig:0703}
(b).
The energy levels are nearly degenerate in fourfold.
In Fig.  \ref{fig:0703} (c), we show intensity plot of Fourier
transform of wavefunction on A sublattice for each level as a function
of $k$.
[A tiny magnetic field ($B = 10^{-6}$ T, corresponding spin splitting
  is $\sim 10^{-8}$ meV) parallel to the nanotube axis is applied to
  extract the spin-up majority states in the numerical calculation.
  The intensity plot is zone-folded in the 1D BZ of nanotube and is
  smoothed by a Gaussian function.]
The fourfold degeneracy reflects the decoupling of the two valleys,
which is confirmed that each eigenstate is formed from leftgoing and
rightgoing states in a single valley, as shown in Fig. \ref{fig:0703}
(c).
The small lift of the fourfold degeneracy, shown in lower panel of
Fig. \ref{fig:0703} (b), is due to the spin--orbit interaction and the
small valley coupling in the finite-length effect, as seen in the
finite-length metal-2 nanotubes.~\cite{Izumida-2015-06}

\begin{figure}[htb]
  \includegraphics[width=8.5cm]{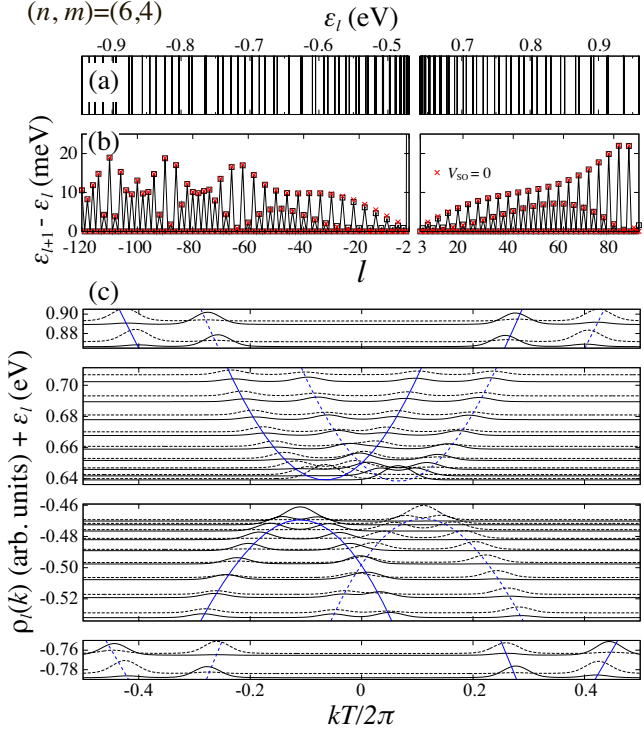}
  \caption{(Color online)
    Energy levels and eigenstates for $(6,4)$ nanotube ($d=2$) of
    $50.14$ nm length.
    (a) Energy levels $\varepsilon_l$ from valence ($\varepsilon_l \le
    -0.469$ eV, $l \le -2$) and conduction ($\varepsilon_l \ge 0.639$
    eV, $l \ge 3$) bands, where $l$ is the level index.
    (b) Level separation $\varepsilon_{l+1} - \varepsilon_l$ as a
    function of level index $l$.
    (c) Intensity plot of Fourier transforms of wave functions of A
    sublattice for spin-up majority states.
    The blue solid and dashed lines show the energy band calculation
    under the periodic boundary condition near the $K$ and $K'$
    valleys, respectively.
  }
  \label{fig:0604}
\end{figure}

For comparison in Fig. \ref{fig:0703}, the energy levels
$\varepsilon_l$ and level separation $\varepsilon_{l+1} -
\varepsilon_l$ for (6,4) nanotube of $50.14$ nm length with a minimal
boundary are shown in Figs. \ref{fig:0604} (a) and \ref{fig:0604} (b),
respectively.
As in the case of Fig. \ref{fig:0703}, the energy levels for the four
edge states exist within the energy gap at $\varepsilon_l = 0$ for $l
= -1, 0, 1, 2$,
two states polarized at B sublattices at left
end and two states polarized at A sublattices at right end
(not shown in the figure).
In contrast to the case of Fig. \ref{fig:0703}, except near the bottom
or the top of the energy gap, the energy levels show twofold
degeneracy reflecting the strongly couple of two valleys, which is
confirmed in the intensity plot of Fig. \ref{fig:0604} (c).
The oscillatory behavior of two- and fourfold degeneracies reflects
the asymmetric group velocity in the same valley, as seen in the
m-SWNTs.~\cite{Izumida-2012-04,Izumida-2015-06}
The behaviors of decoupling or strong coupling of two valleys
demonstrated by the numerical calculation can be understand by the
effective 1D lattice model, which is discussed in the next section.

\section{Effective 1D lattice model}
\label{sec:effctive_1D_model}

An effective 1D lattice model, which includes only the nearest
neighbor hopping term for the spinless case, exhibits not only the
microscopic mechanism of valley coupling but also the number of edge
states appearing in the energy gap.

\subsection{Derivation of effective 1D lattice model}
\label{subsec:effctive_1D_model}

First, we derive an effective 1D lattice model for each angular
momentum (e.g., $\mu=0$ for $d=1$, $\mu=0$ and $\mu=1$ for $d=2$
s-SWNTs) from the nearest neighbor tight-binding Hamiltonian.
To derive the effective model, we restrict ourselves in the nearest
neighbor tight-binding model for $\pi$ electrons,
$H = \sum_{\mathbold{r}} \sum_{j=1}^3 \gamma_j c_{\mathrm{A}, \mathbold{r}}^\dagger
c_{\mathrm{B}, \mathbold{r} + \mathbold{\Delta}_j} + \mathrm{H.c.}$,
where $c_{\sigma, \mathbold{r}}^\dagger$ ($c_{\sigma, \mathbold{r}}$)
is the creation (annihilation) operator of $\pi$ electron on $\sigma$
atom ($\sigma=$A, B) at site $\mathbold{r}$, $\mathbold{\Delta}_j$ is
the vector from A to nearest $j$-th B atom ($j=1,2,3$), and $\gamma_j$
is the hopping integral between A and $j$-th B atom.
The atom position is expressed by the two unit vectors as
$\mathbold{r} = \nu \mathbold{C}_h / d + \ell \mathbold{R}$, where
$\nu = 0,1,\cdots,d-1$, and the integer $\ell$ points the lattice
position in the axis direction in units of $a_z$.

By employing the Fourier transform in the $\mathbold{Q}_1$ direction
to the operator,
\begin{equation}
  c_{\sigma (\mu, \ell)} 
  = \frac{1}{\sqrt{d}} \sum_{\nu=0}^{d-1}
  \exp \left( -\mathrm{i} \frac{\mu}{d} \mathbold{Q}_1 \cdot \nu \frac{\mathbold{C}_h}{d} \right)
  c_{\sigma, \mathbold{r}},
\end{equation}
the Hamiltonian is decomposed into projected Hamiltonian for $\mu$-th
angular momentum, 
$H = \sum_{\mu=0}^{d-1} H_{\mathrm{o}, \mu}$, 
$H_{\mathrm{o}, \mu} = 
\sum_\ell \sum_{j=1}^3 \gamma_j \mathrm{e}^{\mathrm{i} \frac{2 \pi}{d} \Delta \nu_j \mu }
c_{\mathrm{A} (\mu, \ell)}^\dagger c_{\mathrm{B} (\mu, \ell + \Delta \ell_j)} + \mathrm{H.c.}$
Here $\Delta \nu_j$ and $\Delta \ell_j$ for the three nearest neighbor
carbon atoms are the $\mathbold{C}_h/d$ and $\mathbold{R}$ components
of $\mathbold{\Delta}_j$, respectively, and their explicit forms are
$\Delta \nu_1 = \left( p-q \right) /3$, $\Delta \nu_2 = - \left( 2p+q
\right)/3$, $\Delta \nu_3 = 2q+p/3$, and $\Delta \ell_j = -(n-m)/3d$,
$(2n+m)/3d$, $-(2m+n)/3d$ for $j=1, 2, 3$, respectively.
By extracting the relevant $H_{\mathrm{o}, \mu}$, one gets an
effective 1D Hamiltonian.

\begin{figure}[htb]
  \includegraphics[width=7cm]{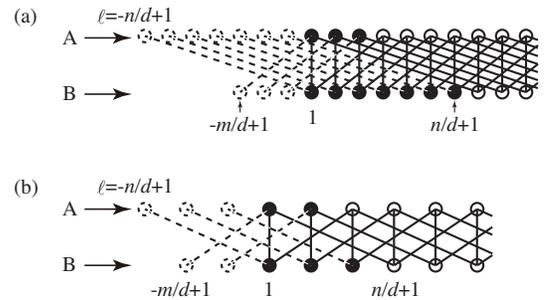}
  \caption{Bond connection diagram of effective 1D lattice model
    $H_\mu$ and left end with minimal boundary for (a) $(n/d,m/d) =
    (7,3)$, and (b) $(n/d,m/d) = (3,2)$.
  }
  \label{fig:1D}
\end{figure}

For the s-SWNTs, as well as the metal-1 nanotubes, since $\Delta
\ell_j$ and $\Delta \nu_j$ are fractional numbers with three as the
denominator, translation conversion of the coordinate for B sublattice
such as $\mathbold{\Delta}_1 \rightarrow 0$ would be convenient.
This conversion makes A and B sublattices terminated at the same
$\ell=1$ site, which is convenient for topological discussion which
will be given in Sec. \ref{subsec:EdgeState}.
Further, this conversion makes the hopping integrals in the
Hamiltonian real numbers for $d=1$ and $d=2$ cases, as shown below.
The Hamiltonian is modified under this conversion, $H_{\mathrm{o},
  \mu} \rightarrow H_\mu$, where
\begin{align}
  H_\mu = 
  \sum_\ell \sum_{j=1}^3 \gamma_j \mathrm{e}^{\mathrm{i} \frac{2 \pi}{d} \Delta \nu_j' \mu }
  c_{\mathrm{A} (\mu, \ell)}^\dagger c_{\mathrm{B} (\mu, \ell + \Delta \ell_j')}' + \mathrm{H.c.} 
  \label{eq:H_mu}
\end{align}
where
\begin{equation}
  \Delta \ell_1' = 0, ~~~~ \Delta \ell_2' = \frac{n}{d}, ~~~~ \Delta \ell_3' = -\frac{m}{d},
  \label{eq:Delta_elld}
\end{equation}
and
\begin{equation}
  \Delta \nu_1' = 0,
  ~~~~ \Delta \nu_2' = -p,
  ~~~~ \Delta \nu_3' = q.
  \label{eq:Delta_nud}
\end{equation}
The connection diagram of the effective 1D lattice model $H_\mu$
is depicted in Figs. \ref{fig:1D} (a) and \ref{fig:1D} (b) for
$(n/d,m/d)=(7,3)$ and $(3,2)$, respectively.
The solid lines show the hopping between atoms.
The solid circles represent the carbon atoms at the boundary, the
dashed circles represent the empty atomic sites and the dashed lines
represent the missing bonds.

\subsection{Eigenstate}
\label{subsec:Eigenstate}

To understand the coupling of two valleys in the eigenstates, we will
show how the eigenstates are constructed from the leftgoing and
rightgoing traveling waves in the two valleys using the effective 1D
model.
Hereafter, 1D lattice site of the effective 1D model is measured in
units of $a_z$, for simplicity, and 1D wavenumber is therefore
measured in units of $1/a_z$.

For an eigenstate with energy $\varepsilon$ of $H_\mu$, which has the
amplitude $\phi_{\sigma \ell}$ at $\sigma$-atom on $\ell$-site
satisfying the solution of $\phi_{\sigma \ell + t} = \lambda^t
\phi_{\sigma \ell}$, $\phi_{\mathrm{B} \ell} = \eta \phi_{\mathrm{A}
  \ell}$,~\cite{Izumida-2015-06} we have the following simultaneous
equations for $\lambda$ and $\eta$,
\begin{align}
  & 1
  + \mathrm{e}^{\mathrm{i} \frac{2 \pi}{d} p \mu} \lambda^{-\frac{n}{d}} 
  + \mathrm{e}^{-\mathrm{i} \frac{2 \pi}{d} q \mu} \lambda^{\frac{m}{d}} 
  = \frac{\varepsilon}{\gamma} \eta, 
  \label{eq:lambdaEq_A} \\
  & 1
  + \mathrm{e}^{-\mathrm{i} \frac{2 \pi}{d} p \mu} \lambda^{\frac{n}{d}} 
  + \mathrm{e}^{\mathrm{i} \frac{2 \pi}{d} q \mu} \lambda^{-\frac{m}{d}} 
  = \frac{\varepsilon}{\gamma} \frac{1}{\eta}.
  \label{eq:lambdaEq_B}
\end{align}
Assumption of using the same hopping integrals $\gamma = \gamma_j$ is
safely acceptable for the s-SWNTs, unlike the
m-SWNTs~\cite{Izumida-2015-06} in which narrow gap is induced by the
small difference among $\gamma_j$ reflecting the curvature of nanotube
surface.
The simultaneous equations have $2(n/d + m/d)$ sets of solutions
$(\lambda, \eta)$ because each equation is the $(n/d + m/d)$th degree
equation.
We call A (B) mode for a mode with $|\eta| < 1$ ($|\eta| > 1$) because
the wavefunction is polarized at the A (B) sublattice.
A mode with $| \lambda | < 1$ $(| \lambda | > 1)$ is a evanescent mode
at the left (right) side, and a mode with $|\lambda| = |\eta| = 1$ is
a traveling mode which extends in whole of the system.

In the low energy limit $|\varepsilon / \gamma | \ll 1$, the roots of
left hand side of Eq. \eqref{eq:lambdaEq_A}
[Eq. \eqref{eq:lambdaEq_B}] gives $\lambda$ for the A [B]
modes.
As shown in the Appendix \ref{sec:App:numEva}, the number of evanescent
modes at the left end for $\sigma$ sublattice, $N_\sigma$, are given
by counting the number of roots in between two outermost roots, which
correspond to the modes with the longest decay length, in the unit
circle centered at the origin in the complex plane, as follows;
\begin{equation}
  N_\mathrm{A} = \left\lfloor \frac{2n + m}{3d} \right\rfloor + n_\mathrm{A}, ~~
  N_\mathrm{B} = \left\lfloor \frac{n + 2m}{3d} \right\rfloor + n_\mathrm{B},
  \label{eq:NA_NB}
\end{equation}
where $\lfloor x \rfloor$ is the floor function giving the greatest
integer that is less than or equal to $x$, and $(n_\mathrm{A},
n_\mathrm{B}) = (0,1)$ [$(1,0)$] for the cases of
$\mathrm{mod}(2n/d+m/d,3)=1$ and $| \mu/d - 1/2 | > 1/6$ [$< 1/6$],
or, $\mathrm{mod}(2n/d+m/d,3)=2$ and $| \mu/d - 1/2 | < 1/6$ [$>
  1/6$].
The integers $n_{\sigma}$ are summarized in Table \ref{table:na_nb}.

\begin{table}[tbhp]
  \caption{Integers $n_{\sigma}$ in Eq. \eqref{eq:NA_NB} for given
    type of s-SWNTs, $\mathrm{mod}(2n/d+m/d,3)$ and for given angular
    momentum $\mu$, at $\varepsilon = 0$.}
  \label{table:na_nb}
  \begin{tabular}{ccccc}
    \hline
    \hline
    \multicolumn{1}{c}{} 
    & \multicolumn{2}{c}{$\mathrm{mod}(2\frac{n}{d}+\frac{m}{d},3)=1$} 
    & \multicolumn{2}{c}{$\mathrm{mod}(2\frac{n}{d}+\frac{m}{d},3)=2$} \\
    & $0 \le \frac{\mu}{d} < \frac{1}{3}$ &
    $\frac{1}{3} < \frac{\mu}{d} < \frac{2}{3}$ ~~~~&
    $0 \le \frac{\mu}{d} < \frac{1}{3}$ &
    $\frac{1}{3} \le \frac{\mu}{d} < \frac{2}{3}$ \\
    & or $\frac{2}{3} < \frac{\mu}{d} < 1$ & & or $\frac{2}{3} < \frac{\mu}{d} < 1$ & \\
    \hline
    $n_\mathrm{A}$ & 0 & 1 & 1 & 0 \\
    $n_\mathrm{B}$ & 1 & 0 & 0 & 1 \\
    \hline
    \hline
  \end{tabular}
\end{table}

\begin{figure}[htb]
  \includegraphics[width=7cm]{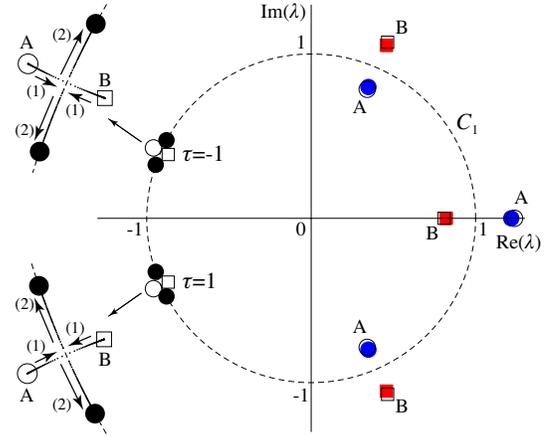}
  \caption{(Color online) Solutions $\lambda$ for
    Eqs. \eqref{eq:lambdaEq_A} and \eqref{eq:lambdaEq_B} in the
    complex plane for $(n,m)=(6,4)$, ($d=2$, type-1), $\mu=1$, $p=2$,
    $q=1$.
    Modes near $\lambda = \lambda_\tau = \mathrm{e}^{\mathrm{i} \tau k_1}$
    are labeled by $\tau = \pm 1$, where $k_1 = 1.13 \pi$.
    Dashed circle $C_1$ shows the unit circle centered at the origin.
    Blue and red filled marks are the solutions for A and B modes,
    respectively, and black filled circles on $C_1$ are the solutions
    for the traveling modes, at $|\varepsilon / \gamma| = 0.4$.
    Open circles (squares) show the solutions for A (B) modes at
    $\varepsilon = 0$.
    Insets show the enlargements near $\lambda_{\tau=\pm 1}$, and the
    dots between the open and filled marks show the solutions for the
    energies in $0 \le |\varepsilon/\gamma| \le 0.4$ with energy
    interval of $0.4 |\gamma| /100$.
    Transition from evanescent modes to traveling modes occurs at
    $|\varepsilon/\gamma| = \varepsilon_\mathrm{gap}/2 = 0.21$.
    Arrows and numbers show the changing of the modes when
    $|\varepsilon|$ increases in (1) $0 \le |\varepsilon| \le
    \varepsilon_\mathrm{gap}/2$, and (2) $|\varepsilon| \ge
    \varepsilon_\mathrm{gap}/2$.
  }
  \label{fig:lambdaSol}
\end{figure}

When the energy increases (decreases) from $\varepsilon = 0$, the
evanescent modes with the longest decay length change to the traveling
modes, as shown in Fig. \ref{fig:lambdaSol}.
Let us explicitly consider the cases of $d=1$ where $\mu=0$ is only
the independent cutting line, and $\mu=1$ for $d=2$, for which
$H_\mu$ contains the states closest to both $K$ and $K'$ points, as
discussed in Sec. \ref{sec:AngularMomentum}.
Since the modes we focus on are closest to
\begin{equation}
  \lambda_\tau = \exp \left( \mathrm{i} \tau k_1 \right),
  \label{eq:lambda_tau}
\end{equation}
where $\tau = 1$ ($-1$) denotes the $K$ ($K'$) valley and $k_1$ is the
wavenumber for the state closest to the $K$ point given in
Eq. \eqref{eq:k1}, by employing the similar analysis for the m-SWNTs
with narrow gap given in Appendix D of
Ref. \onlinecite{Izumida-2015-06}, these modes are given by expanding
Eqs. (\ref{eq:lambdaEq_A}) and (\ref{eq:lambdaEq_B}) near
$\lambda_{\tau}$.
(The detailed calculation for the modes is given in Appendix
\ref{sec:App:eva-travel}.)

At $\varepsilon = 0$, the B (A) modes with 
\begin{equation}
  \lambda_{\tau, 0} = \exp \left( \mathrm{i} \tau k_1 - \frac{2a_z}{3d_t} \right)
  \label{eq:lambda_0}
\end{equation}
are the modes with the longest decay length for the type-1 (type-2)
s-SWNTs.
When the energy increases (decreases) to $\varepsilon = b
\varepsilon_\mathrm{gap} / 2$, which corresponds to the energy bottom
(top) of the conduction (valence) band for $b=1$ ($b=-1$), the mode of
the longest decay length together with the conjugated mode [$(\lambda,
  \eta) \leftrightarrow (1/\lambda^*, 1/\eta^*)$] are continuously
changing to the doubly degenerate modes with $\lambda =
\lambda_{\tau}$ at each $\tau$ valley, where $\varepsilon_\mathrm{gap} =
2 |\gamma| a / \sqrt{3} d_t$.
When the energy increases or decreases further to
\begin{equation}
  \varepsilon = b |\gamma|
  \frac{\sqrt{3}a}{2 a_z} \sqrt{k^2 + \left( \frac{2a_z}{3 d_t} \right)^2},
  \label{eq:epsilon_k}
\end{equation}
the doubly degenerate modes split into the leftgoing ($r=-1$) and
rightgoing ($r=1$) traveling modes with 
\begin{equation}
  \lambda_{\tau r k} = \exp \left[ \mathrm{i} \left( \tau k_1 + r k \right) \right]
  \label{eq:lambda_k}
\end{equation}
at each $\tau$ valley, where $k > 0$.
The above changing of the modes is seen in Fig. \ref{fig:lambdaSol} in
the $\lambda$ complex plane for $(n,m)=(6,4)$, $d=2$ and $\mu=1$ [see
  the modes labeled by $\tau = 1$ ($\tau = -1$), which correspond to
  the modes of $K$ ($K'$) valley].
During the energy changing, the other modes remain almost unchanged,
as shown that the filled and open marks are overlapped in
Fig. \ref{fig:lambdaSol}.
Since the two evanescent modes of B [A] sublattice at the zero energy
change to the traveling modes above and below the energy gap, the
integers $N_\sigma$ is now given by Eq. (\ref{eq:NA_NB}) with
$(n_\mathrm{A}, n_\mathrm{B}) = (0,-1)$ [$(-1,0)$] for the type-1
[type-2] s-SWNTs.
The integers $n_{\sigma}$ for the energy inside and outside the energy
gap are summarized in Table \ref{table:na_nb_energy}.

\begin{table}[tbhp]
  \caption{Integers $n_{\sigma}$ in Eq. \eqref{eq:NA_NB} for the
    energy inside and outside the energy gap, for $d=1$ ($\mu=0$) and
    $\mu=1$ of $d=2$ cases.}
  \label{table:na_nb_energy}
  \begin{tabular}{ccccc}
    \hline
    \hline
    \multicolumn{1}{c}{} & \multicolumn{2}{c}{Type-1 s-SWNTs} & \multicolumn{2}{c}{Type-2 s-SWNTs} \\
    & $|\varepsilon| < \frac{\varepsilon_\mathrm{gap}}{2}$ ~~&
    $|\varepsilon| > \frac{\varepsilon_\mathrm{gap}}{2}$ ~~~~&
    $|\varepsilon| < \frac{\varepsilon_\mathrm{gap}}{2}$ ~~&
    $|\varepsilon| > \frac{\varepsilon_\mathrm{gap}}{2}$ \\
    \hline
    $n_\mathrm{A}$ & 0 & 0 & 1 & $-1$ \\
    $n_\mathrm{B}$ & 1 & $-1$ & 0 & 0 \\
    \hline
    \hline
  \end{tabular}
\end{table}

The wavefunction near the left end is written as the linear
combination of the relevant modes ($|\lambda| \le 1$).
Above and below the energy gap, the wavefunction is written as
\begin{equation}
  \phi_{\sigma k}^b \left( \ell \right)
  = \sum_{\tau} g_{\sigma \tau k}^b(\ell) \mathrm{e}^{\mathrm{i} \tau k_1 \ell}
  + \sum_{m_\sigma=1}^{N_\sigma} c_{\sigma m_\sigma} \left( \lambda_{\sigma m_\sigma}^{<} \right)^\ell, 
  \label{eq:phi}
\end{equation}
where $\lambda_{\sigma m_\sigma}^{<}$ denotes the evanescent modes of
$\sigma$ sublattice at the left end which satisfy $\left
|\lambda_{\sigma m_\sigma}^{<} \right| < 1$, and
\begin{equation}
  g_{\sigma \tau k}^b(\ell) 
  = 
  \sum_r c_{\sigma \tau r k}^b \mathrm{e}^{\mathrm{i} r k \ell}
  \label{eq:envelopeF}
\end{equation}
is the envelope function for the fast oscillating mode,
$\mathrm{e}^{\mathrm{i} \tau k_1 \ell}$.
Here we have the relation 
\begin{equation}
  c_{\mathrm{B} \tau r k}^b 
  = \mathrm{e}^{\mathrm{i} \Phi_{\tau r k}^b} c_{\mathrm{A} \tau r k}^b,
  \label{eq:cAcB}
\end{equation}
where $\Phi_{\tau r k}^b$ is the phase difference between A and B
sublattices for the traveling mode.
[The explicit expression of $\Phi_{\tau r k}^b$ is given in
  Eq. \eqref{eq:app:Phi_trkb} in Appendix \ref{sec:App:eva-travel}.]

Under the boundary conditions, constriction on the coefficients of the
traveling modes is determined.
As shown in Figs. \ref{fig:1D} (a) and (b), the following conditions
hold for the minimal boundary,
\begin{align}
  & \phi_{\mathrm{A} k}^b (\ell) = 0 \text{ at }\ell=-n/d+1,\cdots,0, \label{eq:BC_A} \\ 
  & \phi_{\mathrm{B} k}^b (\ell) = 0 \text{ at }\ell=-m/d+1,\cdots,0. \label{eq:BC_B}
\end{align}
For the case of $n > m \ge 0$, except for $m=n-2$, the number of the
boundary conditions for A sublattice, $n/d$, is larger or equal to
that of the relevant modes, the two fast oscillating modes and
$N_\mathrm{A}$ evanescent modes, $N_\mathrm{A} + 2$.
For $(n,m)=(7,3)$ as an example, the number of relevant modes for A
sublattice is $N_\mathrm{A} + 2 = 6$ and that of boundary conditions
for A sublattice is $n/d = 7$ in $|\varepsilon| > \varepsilon_{\rm
  gap}/2$.
For this case, the boundary conditions Eq. \eqref{eq:BC_A} act as the
``fixed boundary condition'' for the envelope function of A
sublattice, which force the envelope function zero near the left end,
$g_{\mathrm{A} \tau k}^b(\ell \rightarrow 0) = 0$.
To satisfy this condition, the wavefunctions are formed from the
leftgoing and rightgoing traveling modes within the same valley with
the same amplitude, in order to have the following functional form,
\begin{equation}
  g_{\mathrm{A} \tau k}^b(\ell)
  \propto 
  \sin (k \ell).
  \label{eq:envelope_sin}
\end{equation}
Since the boundary conditions are satisfied by constructing the
eigenfunctions in each valley separately, two valleys are decoupled
and the valley degeneracy occurs.
When the spin--orbit interaction turns on, Eq. \eqref{eq:envelope_sin}
is satisfied for one of the two valleys while the component of another
valley is vanished, as seen in Fig. \ref{fig:0703} (c).

For the case of $m=n-2$, the number of boundary conditions is one less
than that of the relevant modes for both A and B sublattice, $n/d =
N_\mathrm{A} + 1$ and $m/d = N_\mathrm{B} + 1$.
For the case of $\mu=1$ of $(n,m)=(6,4)$, as an example, the numbers
of boundary conditions for the A and B sublattices are $n/d = 3$ and
$m/d = 2$, respectively, and that of relevant modes in A and B
sublattices are $N_\mathrm{A} + 2 = 4$ and $N_\mathrm{B} + 2 = 3$,
respectively, in $|\varepsilon| > \varepsilon_\mathrm{gap}/2$.
Therefore, the wavefunctions for A and B sublattices can be
determined with an arbitrarity of the amplitude.
In addition, by employing the similar analysis of our previous study
for the m-SWNTs,~\cite{Izumida-2015-06} which is given in Appendix
\ref{sec:App:valleyCoupling}, it is shown that the coefficients
satisfy $|c_{\sigma, \tau=1, r, k}^b| = |c_{\sigma, \tau=-1, -r,
  k}^b|$ and $c_{\sigma, \tau=1, -r, k}^b = c_{\sigma, \tau=-1, r,
  k}^b = 0$, or equivalently, the first term in Eq. \eqref{eq:phi} is
written as
\begin{equation}
  {\phi'}_{\sigma k}^b \left( \ell \right)
  \propto
  \mathrm{e}^{\mathrm{i} \left( k_1 + r k \right) \ell}
  + \mathrm{e}^{-\mathrm{i} \left( k_1 + r k \right) \ell}
  \mathrm{e}^{-\mathrm{i} \varphi} 
\end{equation}
for either $r = 1$ or $r = -1$ in the linear dispersion region in
which the relation $|\Phi_{\tau r k}^b - \Phi_{\tau -r k}^b| = \pi$
holds, where $\varphi$ is the phase difference between the two
traveling modes, $\varphi = \arg c_{\sigma, \tau=1, r, k}^b - \arg
c_{\sigma, \tau=-1, -r, k}^b$.
Therefore, we have the strong coupling of two valleys, as in the case
of the armchair nanotubes or the chiral metal-2 SWNTs with so-called
the orthogonal boundary.~\cite{Izumida-2015-06}
Near the top of the valence band and bottom of the conduction band, on
the other hand, two valleys are decoupled since the eigenstates are
constructed from the leftgoing and rightgoing states in the same
valley to satisfy the boundary conditions, as also discussed in
details in Appendix \ref{sec:App:valleyCoupling}.
It should be noted that the strong valley coupling could occur for the
other chiralities than $m = n - 2$ for the s-SWNTs with other
boundaries, as in the case of the metal-2
nanotubes.~\cite{Izumida-2015-06}

For the case of $m > n \ge 0$, the case of zigzag-left handedness, the
similar discussion with above is employed: we have the fixed boundary
condition for the envelop function of B sublattice at the left end,
except for the case of $m = n + 2$ in which the strong valley coupling
occurs.

\subsection{Edge state}
\label{subsec:EdgeState}

The number of edge states appeared at $\varepsilon = 0$ is evaluated
by the difference between the number of independent evanescent modes
and the number of boundary conditions, as in the case of previous
discussions for graphene with certain
boundaries~\cite{Akhmerov-2008-02} and for the
m-SWNTs.~\cite{Izumida-2015-06}
For the case of $n > m \ge 0$, the number of edge states which are
localized at A and B sublattices at the left end are given by,
\begin{equation}
  N_\mathrm{edge}^{(\mathrm{L},\mathrm{A})} = 0, ~~~~
  N_\mathrm{edge}^{(\mathrm{L},\mathrm{B})} = N_\mathrm{B} - \frac{m}{d},
  \label{eq:N_edgeStates}
\end{equation}
respectively, for each $\mu$ and each spin state.
Since the number of boundary conditions is larger or equal to that of
the relevant modes for A sublattice, the number of edge states for A
sublattice is zero.
As shown in Appendix \ref{sec:App:EdgeState}, these edge states appear
only at $\varepsilon = 0$.
The numbers of edge states are estimated using Eq. \eqref{eq:NA_NB}
with Table \ref{table:na_nb}.
For instance, $N_\mathrm{edge}^{(\mathrm{L},\mathrm{B})} = 1$ in
$(n,m)=(7,3)$ nanotube.
For $(6,4)$ nanotube, $N_\mathrm{edge}^{(\mathrm{L},\mathrm{B})} = 0$
for $\mu=0$ and $N_\mathrm{edge}^{(\mathrm{L},\mathrm{B})} = 1$ for
$\mu=1$.
Since the numbers estimated by Eq. \eqref{eq:N_edgeStates} give the
edge states for each spin and each end, therefore total number of edge
states is $4$ for both $(7,3)$ and $(6,4)$ nanotubes, which is
consistent with the numerical calculation in Figs. \ref{fig:0703} and
\ref{fig:0604}.

As an alternative way to evaluate the number of edge states, we can
employ the theory of topological categorization for the bulk
systems.~\cite{Wen1989641,PhysRevLett.89.077002}
The winding number is frequently discussed in topological insulating
materials.
For the SWNTs, the winding number for the angular momentum $\mu$,
$w_\mu$, is given by
\begin{equation}
  w_\mu 
  = 
  \frac{1}{2 \pi} \int_0^{2 \pi} dk \frac{\partial \arg f_\mu \left( k \right)}{\partial k},
  \label{eq:windingNumber}
\end{equation}
where $f_\mu(k)$ is the off-diagonal term of the Hamiltonian matrix of
$H_\mu$,
\begin{equation}
  f_\mu(k) = \langle \mathrm{A} k \mu | H_\mu | \mathrm{B} k \mu \rangle 
  = \sum_{j=1}^3 \gamma_j 
  \mathrm{e}^{\mathrm{i} \frac{2 \pi}{d} \Delta \nu_j' \mu } \mathrm{e}^{\mathrm{i} k \Delta \ell_j'},
  \label{eq:f_mu_k}
\end{equation}
$| \sigma k \mu \rangle$ is the Bloch state of $\sigma$ sublattice
with wavenumber $k$ and the angular momentum $\mu$.
Since the integral of Eq. \eqref{eq:f_mu_k} gives phase accumulation
from $k=0$ to $k=2 \pi$ for the function with $2 \pi$ periodicity, the
winding number of Eq. \eqref{eq:windingNumber} is an integer.
Further, under continuous changing of the system parameters such as
the hopping integrals $\gamma_j$, the winding number is invariant,
staying at an integer value, as long as it is well-defined value.
In fact, the winding number is well-defined value for a system with a
finite energy gap, in which $f_\mu (k)$ is finite value for any $k$
and then $\arg f_\mu (k)$ is well-defined value for any $0 \le k < 2
\pi$.
Note that energy gap closes at $k$, at which $f_\mu (k) = 0$ and thus
$\arg f_\mu (k)$ cannot be defined.
When a metallic phase appears during the changing of system
parameters, we may have a topological phase transition, which could be
induced by applying the magnetic field for m-SWNTs with narrow energy
gap.~\cite{Izumida-2015-06,PhysRevB.71.195401}

The winding number has been widely used as the number of edge states
for the topological systems, as well as other topological invariants,
in the context of categorization of materials by their topological
properties.~\cite{PhysRevB.78.195125,Kitaev-AIPConfProc2009,Ryu-NJP-2010}
For the integer quantum Hall systems, it is well-known that the
filling number of the Landau levels as the bulk quantity, which is
known to be a topological invariant called Chern
number,~\cite{PhysRevLett.49.405,KOHMOTO1985343} and the number of
edge channels have one-to-one
correspondence.~\cite{PhysRevB.23.5632,PhysRevB.25.2185,PhysRevB.48.11851,PhysRevLett.71.3697}
For 1D systems, on the other hand, only a few
examples~\cite{PhysRevLett.89.077002,PhysRevB.83.085426,PhysRevB.83.224511,PhysRevB.84.125132}
are known to show exact one-to-one correspondence of the winding
number and the number of edge states, so-called bulk-edge
correspondence, which would be important for better understanding of
the topological materials.
Here we will show the bulk-edge correspondence: the winding number is
the number of edge states for the s-SWNTs by using the results from
the analysis in Sec. \ref{subsec:Eigenstate} with a theorem in the
complex analysis.

\begin{figure}[htb]
  \includegraphics[width=8cm]{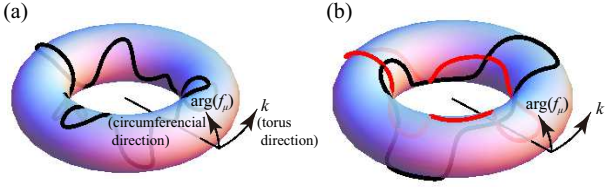}
  \caption{(Color online) Phase of matrix element $f_\mu$ on torus of
    $k$--$\arg(f_\mu)$ face for (a) $(n,m)=(7,3)$ and for (b) $(6,4)$,
    $\mu=0$ (black) and $\mu=1$, $p=2$, $q=1$ (red).}
  \label{fig:0703_0604_windingN}
\end{figure}

First, we show the numerically evaluated $w_\mu$ values for s-SWNTs of
$(n,m)=(7,3)$ and $(6,4)$.
Figure \ref{fig:0703_0604_windingN} shows the phases of $f_\mu(k)$ in
for (a) $(n,m)=(7,3)$ and (b) $(6,4)$, which is plotted on torus in
which circumferential direction correspond to argument of $f_\mu(k)$
as a function of $k$ in the direction around the torus, to show the
``winding'' property clearly, in $0 \le k < 2 \pi$ and $0 \le \arg
f_\mu < 2 \pi$.
For both cases, the total winding number, how many times the curves
winds around the circumference direction, is one.
Therefore, the total edge states including both ends and two spin
states is estimated to be four, which is consistent with the numerical
calculation showing the edge states at $l = -1, 0, 1, 2$.

Let us show that the number of edge states estimated from the mode
analysis shown in Eq. (\ref{eq:N_edgeStates}) and the winding number
in Eq. (\ref{eq:windingNumber}) is identical for the s-SWNTs.
This ``bulk-edge correspondence'' is proven by using the following
theorem known as Cauchy's argument principle in the complex analysis,
\begin{equation}
  \oint_C d \lambda \frac{\partial \log F(\lambda)}{\partial \lambda}
  = 2 \pi \mathrm{i} (n_\mathrm{r} - n_\mathrm{p}), 
  \label{eq:ArgPrinciple}
\end{equation}
where $n_\mathrm{r}$ and $n_\mathrm{p}$ are the number of complex roots and
poles, respectively, of a 
function $F(\lambda)$ inside a closed contour $C$ in $\lambda$ plane.
Using the same hopping integrals $\gamma = \gamma_j$ for the s-SWNTs,
Eq. (\ref{eq:windingNumber}) is written as
\begin{equation}
  w_\mu 
  = 
  \frac{1}{2 \pi \mathrm{i}} \oint_{C_1} d \lambda \frac{\partial \log F_\mu(\lambda)}{\partial \lambda}, 
  \label{eq:windingNumber_02}
\end{equation}
where $C_1$ is the unit circle centered at the origin as shown in
Fig. \ref{fig:lambdaSol}, $F_\mu(\lambda)$ is given by $\lambda =
\mathrm{e}^{\mathrm{i} k}$ in Eq. (\ref{eq:f_mu_k}) divided by the
constant $\gamma$,
\begin{equation}
  F_\mu(\lambda)
  = 
  1
  + \mathrm{e}^{-\mathrm{i} \frac{2 \pi}{d} p \mu} \lambda^{\frac{n}{d}} 
  + \mathrm{e}^{\mathrm{i} \frac{2 \pi}{d} q \mu} \lambda^{-\frac{m}{d}}.
  \label{eq:f_mu_lambda}
\end{equation}
The right hand side of Eq. (\ref{eq:f_mu_lambda}) is exactly the left
hand side of Eq. (\ref{eq:lambdaEq_B}), which has $m/d$ poles at
$\lambda = 0$ and, from the analysis in Sec. \ref{subsec:Eigenstate},
$N_\mathrm{B}$ roots inside $C_1$.
Using the argument principle of Eq. (\ref{eq:ArgPrinciple}) to
Eq. (\ref{eq:windingNumber_02}), we have
\begin{equation}
  w_\mu = N_\mathrm{B} - \frac{m}{d},
\end{equation}
which gives the same result with
$N_\mathrm{edge}^{(\mathrm{L},\mathrm{B})}$ in
Eq. (\ref{eq:N_edgeStates}), showing the bulk-edge correspondence,
\begin{equation}
  w_\mu = N_\mathrm{edge}^{(\mathrm{L},\mathrm{B})}.
  \label{eq:w_N_B}
\end{equation}
Equation \eqref{eq:w_N_B} shows that the winding number is the number of
edge states for B sublattice at left end, which gives more precise
relation than other cases which show the relation of winding number
and the difference of number of edge states between two
sublattices.~\cite{PhysRevB.83.085426,PhysRevB.83.224511,PhysRevB.84.125132}

It would be worthful to mention the case of zigzag-left handedness, $m
> n \ge 0$.
For this case, the number of edge states are given by,
\begin{equation}
  N_\mathrm{edge}^{(\mathrm{L},\mathrm{A})} = N_\mathrm{A} - \frac{n}{d}, ~~~~
  N_\mathrm{edge}^{(\mathrm{L},\mathrm{B})} = 0,
  \label{eq:N_edgeStates_leftHanded}
\end{equation}
since the number of boundary conditions is larger or equal to that of
the relevant modes for B sublattice.
Using the relation, $N_\mathrm{B} - m/d = - \left( N_\mathrm{A} - n/d
\right)$, which holds at $\varepsilon=0$ since $n_\mathrm{A} = 1 -
n_\mathrm{B}$ (see Table \ref{table:na_nb}), regardless of left and
right handedness, we get the following bulk-edge correspondence for
zigzag-left handedness,
\begin{equation}
  w_\mu = - N_\mathrm{edge}^{(\mathrm{L},\mathrm{A})}.
  \label{eq:w_N_A}
\end{equation}

We now get the bulk-edge correspondence for both zigzag-right and left
handedness cases.
When sign of the winding number is positive (negative), the number of
edge states for B (A) sublattice is given by the absolute value of the
winding number, while that of A (B) sublattice is zero.
The bulk-edge correspondence and the number of edge states for A and B
sublattices are summarized in Table \ref{table:bulk-edge}.

\begin{table}[tbhp]
  \caption{Bulk-edge correspondence and number of edge states for A
    and B sublattices for the minimal boundary condition.}
  \label{table:bulk-edge}
  \begin{tabular}{ccc}
    \hline
    \hline
    & Zigzag-right handedness
    & Zigzag-left handedness \\
    & $n > m \ge 0$
    & $m > n \ge 0$ \\
    \hline
    $w_\mu$ & $N_\mathrm{edge}^{(\mathrm{L},\mathrm{B})}$ & $-N_\mathrm{edge}^{(\mathrm{L},\mathrm{A})}$ \\
    $N_\mathrm{edge}^{(\mathrm{L},\mathrm{A})}$ & 0 & $N_\mathrm{A} - \frac{n}{d}$ \\
    $N_\mathrm{edge}^{(\mathrm{L},\mathrm{B})}$ & $N_\mathrm{B} - \frac{m}{d}$ & 0 \\
    \hline
    \hline
  \end{tabular}
\end{table}

\begin{figure}[htb]
  \includegraphics[width=7cm]{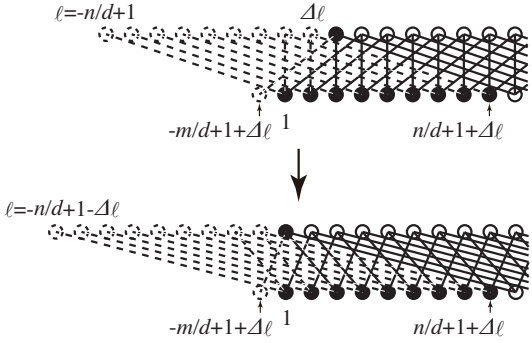}
  \caption{Transformation of effective 1D lattice model $H_\mu
    \rightarrow H_\mu'$ to calculate winding number for number of edge
    states for $(n/d,m/d) = (7,3)$ and $\Delta \ell = 2$.}
  \label{fig:1D_mod}
\end{figure}

In general, the number of edge states depends on the boundary shape.
For the case that the termination of A sublattice is at $\ell = \Delta
\ell$ and that at B sublattice is at $\ell = 1$, for instance, the
winding number which gives the number of edge states should be
calculated for the different bulk Hamiltonian as follows (the
transformation of the Hamiltonian from $H_\mu$ to $H_\mu'$ is depicted
in Fig. \ref{fig:1D_mod});
\begin{align}
  H_\mu' = 
  \sum_\ell \sum_{j=1}^3 \gamma_j \mathrm{e}^{\mathrm{i} \frac{2 \pi}{d} \Delta \nu_j' \mu }
  c_{\mathrm{A} (\mu, \ell)}^\dagger c_{\mathrm{B} (\mu, \ell + \Delta \ell_j' + \Delta \ell)}' + \mathrm{H.c.}
  \label{eq:H_mu_d}
\end{align}
The winding number for Eq. \eqref{eq:H_mu_d} is then calculated as,
\begin{align}
  w_\mu'
  & = 
  \frac{1}{2 \pi} \int_0^{2 \pi} dk \frac{\partial \arg f_\mu' \left( k \right)}{\partial k} \nonumber \\
  & = 
  w_\mu + \Delta \ell,
  \label{eq:windingNumber_d}
\end{align}
where
\begin{equation}
  f_\mu(k)' = \langle \mathrm{A} k \mu | H_\mu' | \mathrm{B} k \mu \rangle 
  = 
  \mathrm{e}^{\mathrm{i} k \Delta \ell} f_\mu(k).
  \label{eq:f_mu_k_d}
\end{equation}
Note that both Eqs. \eqref{eq:H_mu} and \eqref{eq:H_mu_d} gives the
same energy band and the same number of evanescent modes for the same
chirality $(n,m)$, since their difference appears only on the phase of
off-diagonal term of the Hamiltonian matrix.
The number of boundary conditions for the A sublattice increases by
$\Delta \ell$ while that for the B sublattice decreases by $\Delta
\ell$.
Therefore, it can be checked that the number of edge states is given
by Eq. \eqref{eq:windingNumber_d}.
Equation \eqref{eq:windingNumber_d} explains the numerically observed
difference of the numbers of edge states between different boundaries
shown in Figs. 2 and 3 in our previous study.~\cite{Izumida-2015-06}
Note that Eq. (\ref{eq:windingNumber}) itself can be applied to
evaluate the number of edge states for m-SWNTs with appropriate values
of $\gamma_j$ to reproduce the narrow gap induced by the curvature and
the spin--orbit interaction.~\cite{Izumida-2009-06}
It is shown in the Appendix \ref{sec:App:symmetry} that $w_\mu$ can be
non-zero values, except for the armchair nanotubes, by employing an
analysis with viewpoint of the symmetry of SWNTs.

Finally, we comment on the applicability of the simplified 1D model.
Perturbations could modified the edge states quantitatively.
For the metallic SWNTs with narrow energy gap, some edge states could
exist in the energy band region because of the effect of the hopping
to next nearest neighbor and farther sites.~\cite{Sasaki-2006-11}
Breaking of the ${\cal C}_d$ rotational symmetry also affects as a
perturbation to the edge states, which mixes the edge states with
different angular momenta.
Even though such perturbations changes the energies of edge states,
the number of edge states could keeps from the unperturbed case as
long as keeping finite energy gap, as shown in Fig. 8 in
Ref. \onlinecite{Izumida-2015-06}.

\section{Conclusion}
\label{sec:Conclusion}

In summary, we studied the angular momentum of two valleys in the
s-SWNTs.
The classification of the s-SWNTs in the sense of the angular momentum
of valley was given by the integer $d$.
For the case of $d \ge 4$, the two valleys are decoupled in the
finite-length nanotubes which keep ${\cal C}_d$ rotational symmetry.
Lift of the fourfold degeneracy is caused by the spin--orbit
interaction.
On the other hand, for the cases of $d =1$ and $d = 2$, the coupling
of two valleys lifts the fourfold degeneracy.
Especially, when $|n - m| = 2$, near the armchair chirality, they are
strongly coupled and the effect of the spin--orbit interaction is
hidden by the large lift of the degeneracy by valley coupling.
The effective 1D lattice model was introduced by extracting relevant
angular momentum states, which explained the valley coupling in the
eigenfunctions.
The analysis on the winding number provided the bulk-edge
correspondence for the edge states in the s-SWNTs.

The presented study showed that the valley coupling in the eigenstates
and the edge states in the semiconducting energy gap strongly depends
on the chirality and boundary shape.
The valley coupling occurs in the majority of both metallic and
semiconducting SWNTs, even they are defect free and they have clean
edges which conserve the angular momentum of bulk states.
The recent progress on the separation~\cite{Liu:2011aa} and
synthesis~\cite{Sanchez-Valencia:2014aa} of single-chirality SWNTs, or
simultaneous measurement of quantum transport, chirality and the
boundary shape by high resolution measurements in atomic scale such as
the scanning tunneling spectroscopy~\cite{Wilder-1998-01,odom-1998-01}
or high-resolution Raman spectroscopy~\cite{PhysRevLett.90.095503}
would enable to observe the chirality and boundary dependences of the
valley coupling and the edge
states.~\cite{Klusek2000508,PhysRevB.71.193406,PhysRevB.73.085421}

\begin{acknowledgments}
  We acknowledge JSPS KAKENHI Grants (No. 15K05118 for W. I.,
  No. 25286005 for R. S.), and MEXT KAKENHI Grants (No. 25107001 and
  No. 25107005 for R. O. and R. S.), Japan.
  A. Y. is grateful to S. Kobayashi for fruitful discussion.
\end{acknowledgments}

\appendix

\section{Discrete energy levels for finite-length valley decoupled s-SWNTs}
\label{sec:App:spectrum_0804}

\begin{figure}[htb]
  \includegraphics[width=8.5cm]{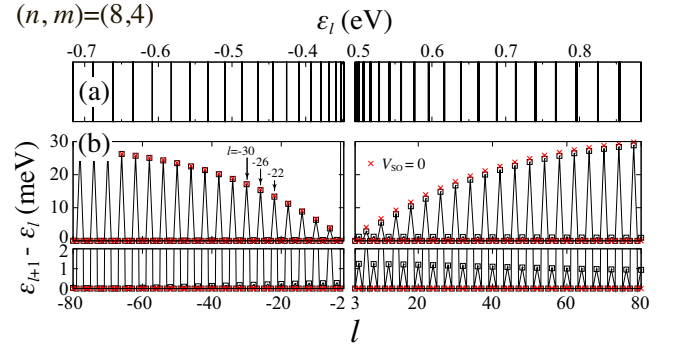}
  \caption{(Color online) 
    Energy levels for $(8,4)$ nanotube of $50.07$ nm length.
    (a) Energy levels $\varepsilon_l$ from valence ($\varepsilon_l \le
    -0.348$ eV, $l \le -2$) and conduction ($\varepsilon_l \ge 0.496$
    eV, $l \le 3$) bands, where $l$ is the level index.
    There are also energy levels at $\varepsilon_l = 0$ at $l = -1, 0,
    1, 2$, which are localized at the ends (not shown).
    (b) Level separation $\varepsilon_{l+1} - \varepsilon_l$ as a
    function of level index $l$.
  }
  \label{fig:app:0804}
\end{figure}

In Fig. \ref{fig:app:0804}, we show numerical calculation of the
discrete energy levels for $(8,4)$ nanotube, which has $d=4$, as an
example for the case of $d \ge 4$.
The calculation is done by the same method with Figs. \ref{fig:0604}
and \ref{fig:0703}, the extended tight-binding method, for the
finite-length of $50.07$ nm, with a minimal boundary for both ends
which keep the bulk ${\cal C}_{d=4}$ rotational symmetry.
The calculated energy levels show fourfold degeneracy.
This clearly shows the decoupling of two valley.
In the actual situation, the spin--orbit interaction lifts the
fourfold degeneracy.

\section{Analysis of effective 1D lattice model}
\label{sec:App:eff1D}

Here we give the detailed calculation for
Secs. \ref{subsec:Eigenstate} and \ref{subsec:EdgeState}.

\subsection{Number of evanescent modes}
\label{sec:App:numEva}

\begin{figure}[htb]
  \includegraphics[width=7cm]{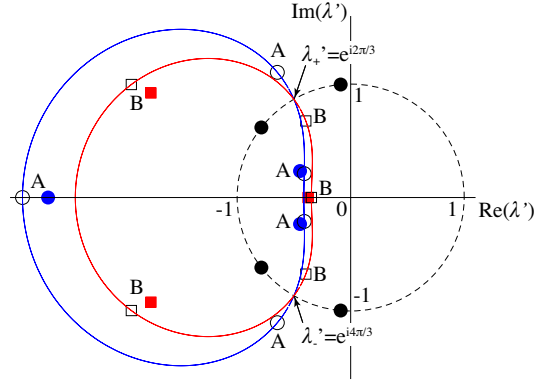}
  \caption{(Color online) Solutions $\lambda'$ for
    Eqs. \eqref{eq:lambdaEq_A} and \eqref{eq:lambdaEq_B} in the
    complex plane for $(n,m)=(6,4)$, ($d=2$, type-1) $\mu=1$, $p=2$,
    $q=1$.
    The marks show the solutions at $\varepsilon = 0$ and
    $|\varepsilon / \gamma| = 0.4$.
    The same symbols with Fig. \ref{fig:lambdaSol} are used.
    Closed blue and red curves give the condition of Eq.
    (\ref{eq:app:lambdadEq_A_e=0_Abs}) for A mode and the
    corresponding one for B mode, respectively, and open marks on the
    curves are the solutions at $\varepsilon = 0$.
  }
  \label{fig:lambdadSol}
\end{figure}

First, we give the detailed derivation of Eq. \eqref{eq:NA_NB}.
In the low energy limit $|\varepsilon / \gamma | \ll 1$ and by using
the same hopping integrals $\gamma = \gamma_j$, we get the following
equations for the A and B modes, respectively, from
Eqs. (\ref{eq:lambdaEq_A}) and (\ref{eq:lambdaEq_B}),
\begin{align}
  &  \mathrm{e}^{-\mathrm{i} \frac{2 \pi}{d} \mu} \lambda'^{\frac{n}{d}} 
  = \left( -1 - \lambda' \right)^{\frac{n}{d} + \frac{m}{d}},
  \label{eq:app:lambdadEq_A_e=0} \\
  &  \mathrm{e}^{\mathrm{i} \frac{2 \pi}{d} \mu} \lambda'^{\frac{m}{d}} 
  = \left( -1 - \lambda' \right)^{\frac{n}{d} + \frac{m}{d}}.
  \label{eq:app:lambdadEq_B_e=0}
\end{align}
where 
\begin{equation}
  \lambda' 
  = 
  \mathrm{e}^{-\mathrm{i} \frac{2 \pi (p + q) \mu}{d} }
  \lambda^{\frac{n}{d} + \frac{m}{d}}
  \label{eq:app:lambdad_def}
\end{equation}
is introduced since this conversion aligns the evanescent modes of
$|\lambda| < 1$, which also satisfy $|\lambda'| < 1$, on a curve
between $\lambda_+' = \mathrm{e}^{\mathrm{i} 2\pi/3}$ and $\lambda_-'
= \mathrm{e}^{\mathrm{i} 4\pi/3}$, as shown in
Fig. \ref{fig:lambdadSol}.
Hereafter we explicitly show the analysis for A mode.
Eq. \eqref{eq:app:lambdadEq_A_e=0} can be separated into two equations
for absolute and phase values as follows;
\begin{align}
  & \left| \lambda' \right|^{\frac{n}{d}}
  = \left| 1 + \lambda' \right|^{\frac{n}{d} + \frac{m}{d}},
  \label{eq:app:lambdadEq_A_e=0_Abs} \\
  & \left( \frac{n}{d} + \frac{m}{d} \right) \arg \left( -1 - \lambda' \right)
  - \frac{n}{d} \arg \lambda' + \frac{2 \pi}{d} \mu
  = 2 \pi l',
  \label{eq:app:lambdadEq_A_e=0_Phase}
\end{align}
where $l'$ is an arbitrary integer.
Since Eqs. (\ref{eq:app:lambdadEq_A_e=0_Abs}) and
(\ref{eq:app:lambdadEq_A_e=0_Phase}) have the same forms with the
equations discussed for the metallic
condition,~\cite{Akhmerov-2008-02} the similar discussion can be
employed for counting the evanescent modes, as shown below.

The condition (\ref{eq:app:lambdadEq_A_e=0_Abs}) gives a closed curve
passing $\lambda_+'$ and $\lambda_-'$ in the complex plane, as shown
by the blue curve in Fig. \ref{fig:lambdadSol}.
Then, $n/d+m/d$ positions satisfying
Eq. (\ref{eq:app:lambdadEq_A_e=0_Phase}) on the curve give the
solutions of $\lambda'$ of Eq. (\ref{eq:app:lambdadEq_A_e=0}), as
shown by the open marks on the blue curve in
Fig. \ref{fig:lambdadSol}.
The number of evanescent modes at the left end, which satisfy
$|\lambda| < 1$, is given by counting the number of modes on the curve
in between $\lambda_+'$ and $\lambda_-'$.
The left hand side of Eq. (\ref{eq:app:lambdadEq_A_e=0_Phase})
monotonically decreases when $\lambda'$ moves from $\lambda_+'$ and
$\lambda_-'$ on the curve, since $\arg (- 1 - \lambda')$ decreases
when $\arg (\lambda')$ increases.
Note that $\arg (\lambda_-')$ and $\arg (- 1 - \lambda')$ are defined
in $0 \le \arg (\lambda_-'), \arg (- 1 - \lambda') < 2 \pi$ for this
case.
The values of $l'$ satisfying Eq. \eqref{eq:app:lambdadEq_A_e=0_Phase}
at $\lambda_+'$ and $\lambda_-'$ are given by
\begin{equation}
  l'_+ = \frac{n + 2m}{3d} + \frac{\mu}{d},
  ~~~~
  l'_- = - \frac{n - m}{3d} + \frac{\mu}{d},
  \label{eq:app:l'+l'-}
\end{equation}
respectively.
Note that $l'_\pm$ are fractional numbers, unlike the metallic
case.~\cite{Akhmerov-2008-02}
By counting the integers $l'$ in $l'_- < l' < l'_+$ for the cases of
$\mathrm{mod}(n/d+m/d,3)=1,2$ for $0 \le \mu/d < 1/3$, $1/3 < \mu/d <
2/3$, $2/3 < \mu/d < 1$, we get $N_\mathrm{A}$.
By employing the similar discussion with above, we also get
$N_\mathrm{B}$.
The results are summarized in Eq. \eqref{eq:NA_NB} and in Table
\ref{table:na_nb}.

\subsection{Evanescent modes and traveling modes}
\label{sec:App:eva-travel}

Here we will give the evanescent modes with the longest decay length
at $\varepsilon = 0$, Eq. \eqref{eq:lambda_0}, and their changing to
traveling modes in Eq. \eqref{eq:lambda_k} when energy increases or
decreases to $|\varepsilon| > \varepsilon_\mathrm{gap}/2$.
We restrict $d=1$ and $d=2$ cases, in which $(\lambda^*, \eta^*)$ is
also a set of solution for a solution $(\lambda, \eta)$ since
$\gamma_j$, $\varepsilon$ and $\mathrm{e}^{-\mathrm{i} \frac{2 \pi}{d}
  \Delta \nu_j' \mu}$ ($j=1,2,3$) are real numbers.

First, we mention the relation between $\lambda$ and $\lambda'$, which
is defined by Eq.  \eqref{eq:app:lambdad_def}, for the mode of
$\lambda = \lambda_\tau = \mathrm{e}^{\mathrm{i} \tau k_1}$.
After some algebraic calculation, we get the corresponding phase of
$\lambda = \lambda_\tau$ in $\lambda'$ as follows;
\begin{align}
  \left. \arg \lambda' \right|_{\lambda=\lambda_\tau} & = 
  \tau k_1 \left( \frac{n}{d} + \frac{m}{d} \right) - \frac{2 \pi \left( p + q \right) \mu}{d} \nonumber \\
  & = \tau \left( \frac{2 \pi}{3} + \delta \theta \right) + 2 \pi \alpha,
  \label{eq:app:k1_2pi3_relation}
\end{align}
for both $d=1$ and $\mu=1$ of $d=2$ cases, where $\alpha$ is an
integer, and
\begin{equation}
  \delta \theta = -t_y \frac{2 \pi}{3} \frac{n - m}{2 \left( n^2 + m^2 + nm \right)},
  \label{eq:app:deltatheta}
\end{equation}
where 
\begin{equation}
  t_y = \left\{
  \begin{array}{cl}
    +1 & \text{~~for type-1}, \\
    -1 & \text{~~for type-2}, \\
  \end{array}
  \right.
  \label{eq:app:ty_definition}
\end{equation}
is introduced for type-1 [$\mathrm{mod} (2n + m, 3) = 1$] and type-2
[$\mathrm{mod} (2n + m, 3) = 2$] s-SWNTs.
Since Eq. \eqref{eq:app:deltatheta} is the small correction, the
solution near $\lambda'_+$ ($\lambda'_-$) corresponds that at $K$
($K'$) valley.

The evanescent modes with the longest decay length at $\varepsilon =
0$ have $\lambda'$ closest to $\lambda'_\tau$, and they are expressed
by
\begin{equation}
  \lambda' 
  = 
  (1 - \delta r) \exp \left[ \mathrm{i} \tau \left( \frac{2 \pi}{3} + \delta \varphi \right) \right],
  \label{eq:app:lambdad_e=0}
\end{equation}
with small values of $| \delta r | \ll 1$ and $| \delta \varphi | \ll
1$.
For small $\delta r$ and $\delta \varphi$, we have the relation,
$-1-\lambda'=(1 - \delta r') \mathrm{e}^{-\mathrm{i} \tau ( 2 \pi/3 -
  \delta \varphi' )}$, where $\delta r' = \delta r / 2 + \sqrt{3}
\delta \varphi / 2$ and $\delta \varphi' = - \sqrt{3} \delta r / 2 +
\delta \varphi / 2$.
Eq. \eqref{eq:app:lambdadEq_A_e=0_Abs} gives the following relation,
\begin{equation}
  \left( n - m \right) \delta r 
  = \sqrt{3} \left( n + m \right) \delta \varphi,
  \label{eq:app:deltar_deltatheta_01}
\end{equation}
and Eq. \eqref{eq:app:lambdadEq_A_e=0_Phase} gives the following
relation,
\begin{equation}
  \frac{\sqrt{3}}{2} \left( n + m \right) \delta r 
  + \frac{1}{2} \left( n - m \right) \delta \varphi
  = 2 \pi \tau \left( l'_\tau - l'_{0, \tau} \right) d,
  \label{eq:app:deltar_deltatheta_02}
\end{equation}
where $l'_{0, \tau}$ is the integer closest to $l'_\tau$.
By doing the similar analysis with that after
Eq. \eqref{eq:app:l'+l'-} in Appendix \ref{sec:App:numEva}, we get
\begin{equation}
\tau \left( l'_\tau - l'_{0, \tau} \right) d = -t_y \frac{1}{3}.
\end{equation}
From Eqs. \eqref{eq:app:deltar_deltatheta_01} and
\eqref{eq:app:deltar_deltatheta_02}, we get
\begin{equation}
  \delta r = -t_y \frac{2 a_z}{3 d_t} \left( \frac{n}{d} + \frac{m}{d} \right), 
  ~~
  \delta \varphi = -t_y \frac{2 \pi}{3} \frac{n - m}{2 \left( n^2 + m^2 + nm \right)}.
  \label{eq:app:deltar_deltatheta}
\end{equation}
In fact, $\delta \varphi = \delta \theta$.
The similar calculation for the B mode gives the results which has
opposite sign for $\delta r$, $-t_y \rightarrow t_y$ in
Eq. \eqref{eq:app:deltar_deltatheta} while the same equation holds for
$\delta \varphi$.
The conversion from $\lambda'$ to $\lambda$ for
Eq. \eqref{eq:app:lambdad_e=0} is performed by using
Eqs. \eqref{eq:app:lambdad_def} and \eqref{eq:app:k1_2pi3_relation},
then we get
\begin{equation}
  \lambda_{\tau, 0}^\sigma = \exp \left( \mathrm{i} \tau k_1 + t_y \sigma \frac{2a_z}{3d_t} \right),
  \label{eq:app:lambda_0}
\end{equation}
where $\sigma = 1$ ($\sigma = -1$) for A (B) mode.
For the evanescent modes at the left end, which satisfy $|\lambda| <
1$, we get Eq. \eqref{eq:lambda_0}.
Substituting $\lambda_{\tau, 0}^{\sigma=1}$ of
Eq. \eqref{eq:app:lambda_0} for Eq. \eqref{eq:lambdaEq_A} with
$\varepsilon = 0$ and expanding in the first order of $a_z/d_t$, we
get the following relation,
\begin{align}
  & 1
  + \mathrm{e}^{\mathrm{i} \frac{2 \pi}{d} p \mu} \mathrm{e}^{-\mathrm{i} \tau k_1 \frac{n}{d}}
  + \mathrm{e}^{-\mathrm{i} \frac{2 \pi}{d} q \mu} \mathrm{e}^{\mathrm{i} \tau k_1 \frac{m}{d}} \nonumber \\
  = & t_y \mathrm{e}
  ^{-\mathrm{i} \left[ - \frac{2 \pi}{d} p \mu + \tau \left( k_1 \frac{n}{d} + \theta \right) \right]}
  \frac{a}{\sqrt{3} d_t},
  \label{eq:app:lambda_relation}
\end{align}
which will be used in the following analysis, where $\theta = \arccos
\frac{2n + m}{2 \sqrt{n^2 + m^2 + nm}}$ is the chiral angle.

The evanescent modes in the energy gap are expressed by
\begin{equation}
  \lambda_{\tau \kappa} = \exp \left( \mathrm{i} \tau k_1 - \kappa \right),
  \label{eq:app:lambda_kappa}
\end{equation}
where $|\kappa| \le 2 a_z / 3 d_t$.
Substituting Eq. \eqref{eq:app:lambda_kappa} for
Eqs. \eqref{eq:lambdaEq_A} and \eqref{eq:lambdaEq_B}, expanding in the
first order of $\kappa$, and using Eq. \eqref{eq:app:lambda_relation},
we get the energy
\begin{equation}
  \varepsilon_\kappa^b = b |\gamma|
  \frac{\sqrt{3}a}{2 a_z} \sqrt{ \left( \frac{2a_z}{3 d_t} \right)^2 - \kappa^2},
  \label{eq:app:epsilon_kappa}
\end{equation}
and $\eta$,
\begin{align}
  \eta_{\tau \kappa}^b 
  = &
  b \exp \left\{
    - \mathrm{i} \left[ - \frac{2 \pi}{d} p \mu + \tau \left( k_1 \frac{n}{d} + \theta \right) 
    \right] \right\}
 \mathrm{sgn} \left( \gamma \right) \nonumber \\
 & \times
 t_y 
 \sqrt{
   \frac{ \frac{2a_z}{3d_t} + t_y \kappa }{ \frac{2a_z}{3d_t} - t_y \kappa }
   }
  \label{eq:app:eta_tkappab}
\end{align}

For a state in $| \varepsilon | \ge \varepsilon_\mathrm{gap}/2$, the
$\lambda$ solution of Eqs. \eqref{eq:lambdaEq_A} and
\eqref{eq:lambdaEq_B} has the following form,
\begin{equation}
  \lambda_{\tau r k} = \exp \left[ \mathrm{i} \left( \tau k_1 + r k \right) \right].
  \label{eq:app:lambda_k}
\end{equation}
By the similar calculation with that for
Eq. \eqref{eq:app:epsilon_kappa},
we get the energy
\begin{equation}
  \varepsilon_k^b = b |\gamma|
  \frac{\sqrt{3}a}{2 a_z} \sqrt{k^2 + \left( \frac{2a_z}{3 d_t} \right)^2},
  \label{eq:app:epsilon_k}
\end{equation}
and $\eta$,
\begin{equation}
  \eta_{\tau r k}^b = \exp \left( \mathrm{i} \Phi_{\tau r k}^b \right),
  \label{eq:app:eta_trkb}
\end{equation}
where
\begin{equation}
  \Phi_{\tau r k}^b 
  = 
  \frac{2 \pi}{d} p \mu 
  - \tau \left( k_1 \frac{n}{d} + \theta \right)
  + \arg \left[ b \gamma \left( t_y \frac{2 a_z}{3 d_t} - \mathrm{i} r k \right) \right].
  \label{eq:app:Phi_trkb}
\end{equation}

\subsection{Strong coupling of two valleys}
\label{sec:App:valleyCoupling}

The wavefunctions of A and B sublattices above and below the energy
gap near the left end are written as
\begin{align}
  \phi_{\mathrm{A} k}^b \left( \ell \right)
  & = \sum_{\tau} g_{\mathrm{A} \tau k}^b \lambda_\tau^\ell
  + \sum_{m_\mathrm{A}=1}^{N_\mathrm{A}} c_{\mathrm{A} m_\mathrm{A}} \left( \lambda_{\mathrm{A} m_\mathrm{A}}^{<} \right)^\ell, 
  \label{eq:app:phi_A} \\
  \phi_{\mathrm{B} k}^b \left( \ell \right)
  & = \sum_{\tau} g_{\mathrm{B} \tau k}^b \lambda_\tau^\ell
  + \sum_{m_\mathrm{B}=1}^{N_\mathrm{B}} c_{\mathrm{B} m_\mathrm{B}} \left( \lambda_{\mathrm{B} m_\mathrm{A}}^{<} \right)^\ell.
  \label{eq:app:phi_B}
\end{align}
The envelope functions $g_{\sigma \tau k}^b$ are regarded as constant
values near the left end.
The boundary conditions of Eqs. \eqref{eq:BC_A} and \eqref{eq:BC_B}
for the wavefunctions of Eqs. \eqref{eq:app:phi_A} and
\eqref{eq:app:phi_B} are expressed by the following matrix form,
\begin{equation}
  D_\mathrm{o} \mathbold{c}_\mathrm{o} = 0,
  \label{eq:app:D_oC_o=0}
\end{equation}
where 
\begin{equation}
  D_\mathrm{o} =
  \left(
  \begin{array}{cccccc}
    A_+ & A_- & 0   & 0   & D_\mathrm{A} & 0 \\
    0   & 0   & B_+ & B_+ & 0 & D_\mathrm{B} \\
  \end{array}
  \right),
  \label{eq:app:D_o}
\end{equation}
with the submatrices,
\begin{align}
  & A_\tau =
  \left(
  \begin{array}{c}
    1 \\
    \left( \lambda_\tau \right)^{-1} \\
    \vdots \\
    \left( \lambda_\tau \right)^{-\frac{n}{d}+1} \\
  \end{array}
  \right),
  ~~
  B_\tau =
  \left(
  \begin{array}{c}
    1 \\
    \left( \lambda_\tau \right)^{-1} \\
    \vdots \\
    \left( \lambda_\tau \right)^{-\frac{m}{d}+1} \\
  \end{array}
  \right),
  \label{eq:app:AB_tau} \\
  & D_\mathrm{A} =
  \left(
  \begin{array}{ccc}
    1 & \cdots & 1 \\
    \left( \lambda_{\mathrm{A} 1}^{<} \right)^{-1} & \cdots & \left( \lambda_{\mathrm{A} N_\mathrm{A}}^{<} \right)^{-1} \\
    \vdots & \ddots & \vdots \\
    \left( \lambda_{\mathrm{A} 1}^{<} \right)^{-\frac{n}{d}+1} & \cdots & \left( \lambda_{\mathrm{A} N_\mathrm{A}}^{<} \right)^{-\frac{n}{d}+1} \\
  \end{array}
  \right),
  \label{eq:app:D_A} \\
  & D_\mathrm{B} =
  \left(
  \begin{array}{ccc}
    1 & \cdots & 1 \\
    \left( \lambda_{\mathrm{B} 1}^{<} \right)^{-1} & \cdots & \left( \lambda_{\mathrm{B} N_\mathrm{B}}^{<} \right)^{-1} \\
    \vdots & \ddots & \vdots \\
    \left( \lambda_{\mathrm{B} 1}^{<} \right)^{-\frac{m}{d}+1} & \cdots & \left( \lambda_{\mathrm{B} N_\mathrm{B}}^{<} \right)^{-\frac{m}{d}+1} \\
  \end{array}
  \right),
  \label{eq:app:D_B}
\end{align}
and 
\begin{align}
  \mathbold{c}_\mathrm{o}^\mathrm{T} =
  & 
  \left(
  g_{\mathrm{A}, +, k}^b, 
  g_{\mathrm{A}, -, k}^b, 
  g_{\mathrm{B}, +, k}^b, 
  g_{\mathrm{B}, -, k}^b, \right. 
  \nonumber \\
  & 
  \left.
  c_{\mathrm{A} 1}, 
  \cdots,
  c_{\mathrm{A} N_\mathrm{A}},
  c_{\mathrm{B} 1},
  \cdots,
  c_{\mathrm{B} N_\mathrm{B}}
  \right).
  \label{eq:app:C_o_vec}
\end{align}
In general, we have the following relation between two complex numbers,
\begin{equation}
  g_{\mathrm{B} \tau k}^b = r_\tau g_{\mathrm{A} \tau k}^b \mathrm{e}^{\mathrm{i} \Phi_{\tau}},
  \label{eq:app:coef_B-A}
\end{equation}
where $r_\tau \ge 0$, $0 \le \Phi_{\tau} < 2 \pi$.
Using Eq. \eqref{eq:app:coef_B-A}, Eq. \eqref{eq:app:D_oC_o=0} is
rewritten as
\begin{equation}
  D \mathbold{c} = 0,
  \label{eq:app:DC=0}
\end{equation}
where
\begin{equation}
  D =
  \left(
  \begin{array}{cccc}
    A_+ & A_- & D_\mathrm{A} & 0 \\
    r_+ \mathrm{e}^{\mathrm{i} \Phi_+} B_+ & r_- \mathrm{e}^{\mathrm{i} \Phi_-} B_- & 0 & D_\mathrm{B} \\
  \end{array}
  \right),
  \label{eq:app:D}
\end{equation}
and
\begin{equation}
  \mathbold{c}^\mathrm{T} =
  \left(
  g_{\mathrm{A}, +, k}^b, 
  g_{\mathrm{A}, -, k}^b, 
  c_{\mathrm{A} 1}, 
  \cdots,
  c_{\mathrm{A} \frac{n}{d}-1},
  c_{\mathrm{B} 1},
  \cdots,
  c_{\mathrm{B} \frac{m}{d}-1}
  \right).
  \label{eq:app:C_vec}
\end{equation}
The matrix $D$ is the $\left( n/d+m/d \right) \times \left( n/d+m/d
\right)$ square matrix.

Hereafter let us focus on the case of $m = n - 2$, in which
$N_\mathrm{A} = n/d - 1$ and $N_\mathrm{B} = m/d - 1$.
Since the number of boundary conditions is one less than that of the
relevant modes for both A and B sublattice, both A and B sublattices
have non-trivial solutions, $g_{\sigma \tau k}^b \neq 0$ for both
$\sigma=\mathrm{A}$ and B.
For this case, $r_+$ and $r_-$ are finite positive values ($r_\tau >
0$).
To satisfy Eq. \eqref{eq:app:DC=0} for non-trivial coefficients,
$\mathbold{c} \neq 0$, the determinant of the matrix $D$ should be
zero, $\left| D \right| = 0$.
We will show the condition for $\left| D \right| = 0$.

\newpage

Before the evaluation of $\left| D \right|$, we will show some relations, which
will be used later.
By using the roots of the left hand side of Eq. \eqref{eq:lambdaEq_A},
it is rewritten as,
\begin{align}
  & 1 
  + \mathrm{e}^{\mathrm{i} \frac{2 \pi}{d} p \mu} \lambda^{-\frac{n}{d}} 
  + \mathrm{e}^{-\mathrm{i} \frac{2 \pi}{d} q \mu} \lambda^{\frac{m}{d}}
  \nonumber \\
  = & 
  \frac{\mathrm{e}^{-\mathrm{i} \frac{2 \pi}{d} q \mu}}{\lambda^{\frac{n}{d}}}
  \left[ \lambda^{\frac{n}{d}+\frac{m}{d}} 
  + \mathrm{e}^{\mathrm{i} \frac{2 \pi}{d} q \mu} \lambda^{\frac{n}{d}} 
  + \mathrm{e}^{\mathrm{i} \frac{2 \pi}{d} \left( p + q \right) \mu} \right]
  \nonumber \\
  = & 
  \frac{\mathrm{e}^{-\mathrm{i} \frac{2 \pi}{d} q \mu}}{\lambda^{\frac{n}{d}}}
  \prod_{m_1 = 1}^{\frac{n}{d} - 1} \left( \lambda - \lambda_{\mathrm{A} m_1}^< \right)
  \nonumber \\
  & \times 
  \prod_{m_2 = 1}^{\frac{m}{d} - 1} \left( \lambda - \lambda_{\mathrm{A} m_2}^> \right)
  \prod_{\tau=\pm} \left( \lambda - \lambda_{\tau,0} \right)
  \nonumber \\
  = & 
  \frac{\mathrm{e}^{-\mathrm{i} \frac{2 \pi}{d} q \mu}}{\lambda} 
  \prod_{m_1 = 1}^{\frac{n}{d} - 1} \left( -\lambda_{\mathrm{A} m_1}^< \right)
  \prod_{m_1 = 1}^{\frac{n}{d} - 1} \left( \frac{1}{\lambda} - \frac{1}{\lambda_{\mathrm{A} m_1}^<} \right) 
  \nonumber \\
  & \times 
  \prod_{m_2 = 1}^{\frac{m}{d} - 1} \left( \lambda - \lambda_{\mathrm{A} m_2}^> \right)
  \prod_{\tau=\pm} \left( \lambda - \lambda_{\tau,0} \right).
  \label{eq:app:lambdaEq_A_LHS}
\end{align}
The value $\prod_{m_1 = 1}^{\frac{n}{d} - 1} \left(
-\lambda_{\mathrm{A} m_1}^< \right)$ in the right hand side of
Eq. \eqref{eq:app:lambdaEq_A_LHS} is a real number.
This is because there is another root, $\lambda_{\mathrm{A}
  m_1}^{<*}$, for a complex root $\lambda_{\mathrm{A} m_1}^<$, for
$d=1$ and $d=2$ since $\mathrm{e}^{\mathrm{i} \frac{2 \pi}{d} q \mu}$
and $\mathrm{e}^{\mathrm{i} \frac{2 \pi}{d} \left( p + q \right)
  \mu}$, which appear in the left hand side of
Eq. \eqref{eq:app:lambdaEq_A_LHS}, are real numbers.
For the traveling mode $(\lambda,\eta) = (\lambda_{\tau r k},
\eta_{\tau r k}^b)$, a part of right hand side of
\eqref{eq:app:lambdaEq_A_LHS} is calculated as,
\begin{equation}
  \frac{1}{\lambda_{\tau r k}} \prod_{\tau=\pm} \left( \lambda_{\tau r k} - \lambda_{\tau,0} \right)
  =
  - \left( 
  t_y \frac{2 a_z}{3 d_t} - \mathrm{i} r k \right) 2 \mathrm{i} \tau \sin k_1.
\end{equation}
Therefore, we have the following relation,
\begin{align}
  & \prod_{m_1 = 1}^{\frac{n}{d} - 1} \left( \frac{1}{\lambda_{\tau r k}} - \frac{1}{\lambda_{\mathrm{A} m_1}^<} \right) 
  \prod_{m_2 = 1}^{\frac{m}{d} - 1} \left( \lambda_{\tau r k} - \lambda_{\mathrm{A} m_2}^> \right)
  \nonumber \\
  = & R e^{{\rm i} \frac{2 \pi}{d} \Delta \nu_3 \mu} e^{{\rm i} \left( \frac{\pi}{2} + \Phi_{\tau r k=0}^b \right) }
  \label{eq:app:relation_01}
\end{align}
where $R$ is a finite real number, which does not depend on $\tau$,
$r$, $k$ and $b$.
Further, the evanescent modes of B sublattice, $\lambda_{\mathrm{B}
  m_\mathrm{B}}^{<}$ in \eqref{eq:app:D_B}, are replaced by the
evanescent modes of A sublattice at the right end,
$\lambda_{\mathrm{A} m_\mathrm{A}}^{>}$, which satisfy $\left|
\lambda_{\mathrm{A} m_\mathrm{A}}^{>} \right| > 1$, that is,
\begin{equation}
  \lambda_{\mathrm{B} m'}^{<} = \frac{1}{\lambda_{\mathrm{A} m'}^{>}},
  \label{eq:app:lambda_A-B}
\end{equation}
for $m'=1,\cdots,N_\mathrm{B}$, because there is one-to-one
correspondence between the evanescent mode of B sublattice at the left
end and that of A sublattice at the right end, since $(1/\lambda^*,
1/\eta^*)$ is the conjugated mode of $(\lambda, \eta)$ for
Eqs. \eqref{eq:lambdaEq_A} and \eqref{eq:lambdaEq_B}, and, there is
another root, $\lambda_{\mathrm{A} m_1}^{<*}$, for a complex root
$\lambda_{\mathrm{A} m_1}^<$ for $d=1$ and $d=2$, as mentioned before.

The determinant of $D$ is expanded as,
\begin{equation}
  \left| D \right| 
  = r_+ \mathrm{e}^{\mathrm{i} \Phi_+} \left| A_- D_\mathrm{A} \right| \left| B_+ D_\mathrm{B} \right|
  - 
  r_- \mathrm{e}^{\mathrm{i} \Phi_-} \left| A_+ D_\mathrm{A} \right| \left| B_- D_\mathrm{B} \right|.
  \label{eq:app:detD_expand}
\end{equation}
By using Eq. \eqref{eq:app:lambda_A-B}, and the relation on the
Vandermonde matrix, the determinant is calculated to be,
\begin{widetext}
  \begin{align}
    \left| D \right| = & 
    \left( -1 \right)^{ \frac{\frac{n}{d} (\frac{n}{d} - 1)}{2} + \frac{\frac{m}{d} (\frac{m}{d} - 1)}{2} }
    \prod_{1 \le m_1 < m_2 \le \frac{n}{d} - 1} \left( \frac{1}{\lambda_{\mathrm{A} m_1}^<} - \frac{1}{\lambda_{\mathrm{A} m_2}^<} \right)
    \prod_{1 \le m_1 < m_2 \le \frac{m}{d} - 1} \left( \lambda_{\mathrm{A} m_1}^> - \lambda_{\mathrm{A} m_2}^> \right)
    \nonumber \\
    & \times 
    \left\{
    r_+ e^{{\rm i} \Phi_+}
    \prod_{m_1 = 1}^{\frac{n}{d} - 1} \left( \frac{1}{\lambda_-} - \frac{1}{\lambda_{\mathrm{A} m_1}^<} \right) 
    \prod_{m_2 = 1}^{\frac{m}{d} - 1} \left( \lambda_- - \lambda_{\mathrm{A} m_2}^> \right)
    -
    r_- e^{-{\rm i} \Phi_-}
    \prod_{m_1 = 1}^{\frac{n}{d} - 1} \left( \frac{1}{\lambda_+} - \frac{1}{\lambda_{\mathrm{A} m_1}^<} \right) 
    \prod_{m_2 = 1}^{\frac{m}{d} - 1} \left( \lambda_+ - \lambda_{\mathrm{A} m_2}^> \right)
    \right\}. 
    \label{eq:app:detD}
  \end{align}
\end{widetext}
Using Eq. \eqref{eq:app:relation_01}, we have
\begin{equation}
  \left| D \right|
  \propto 
  e^{{\rm i} \left( \Phi_+ + \Phi_{\tau=-1, r, k=0}^b \right)}
  + r_- e^{{\rm i} \left( \Phi_- + \Phi_{\tau=1, r, k=0}^b \right)}.
\end{equation}
Therefore, the condition of $\left| D \right| = 0$ is expressed by the
following relations,
\begin{equation}
  r_+ = r_-,
  \label{eq:app:r+r-}
\end{equation}
and
\begin{equation}
  \Phi_+ - \Phi_- 
  =
  \Phi_{\tau=1, r, k=0}^b - \Phi_{\tau=-1, r, k=0}^b + \pi + 2 \pi \alpha,
  \label{eq:app:Phi+_Phi-}
\end{equation}
where $\alpha$ is an arbitrary integer.

In fact, the envelope functions are constructed from the leftgoing and
rightgoing modes as expressed in Eq. \eqref{eq:envelopeF}, that is,
\begin{align}
  g_{\mathrm{A} \tau k}^b
  & = 
  c_{\mathrm{A} \tau r k} + c_{\mathrm{A} \tau -r k},
  \label{eq:app:coef_A} \\
  g_{\mathrm{B} \tau k}^b
  & = 
  \mathrm{e}^{\mathrm{i} \Phi_{\tau r k}^b} \left[ c_{\mathrm{A} \tau r k} + \mathrm{e}^{\mathrm{i} \left( \Phi_{\tau -r k}^b - \Phi_{\tau r k}^b \right)} c_{\mathrm{A} \tau -r k} \right],
  \label{eq:app:coef_B}
\end{align}
By converting Eqs. \eqref{eq:app:coef_A} and \eqref{eq:app:coef_B}
into the form of Eq \eqref{eq:app:coef_B-A}, we can check the explicit
form of the envelope functions which satisfy Eqs. \eqref{eq:app:r+r-}
and \eqref{eq:app:Phi+_Phi-}.

Let us consider the following two cases, (i) near the top of the
valence band and bottom of the conduction band, $k \ll a_z/d_t$, and,
(ii) the linear dispersion region, $k \gg a_z/d_t$.
For the case of (i), since $\Phi_{\tau -r k}^b \simeq \Phi_{\tau r
  k}^b$, from Eqs. \eqref{eq:app:coef_A} and \eqref{eq:app:coef_B} we
have
\begin{equation}
  g_{\mathrm{B} \tau k}^b
  = 
  \mathrm{e}^{\mathrm{i} \Phi_{\tau r k=0}^b} 
  g_{\mathrm{A} \tau k}^b.
  \label{eq:app:coef_B-A_k=0}
\end{equation}
By comparing Eqs. \eqref{eq:app:coef_B-A} and
\eqref{eq:app:coef_B-A_k=0}, we have $r_\tau = 1$ and $\Phi_\tau =
\Phi_{\tau r k=0}^b$.
For this case Eq. \eqref{eq:app:Phi+_Phi-} cannot be satisfied, that
is, $\left| D \right| \neq 0$.
Therefore, $\mathbold{c} = 0$, to satisfy the boundary conditions.
This means that the following relation holds,
\begin{equation}
  g_{\sigma \tau k}^b \propto \sin \left( k \ell \right)
  \label{eq:app:envelope_k=0}
\end{equation}
for the envelope functions of both $\sigma = \mathrm{A}$ and $\sigma =
\mathrm{B}$.
Similar to the case of $m \neq n - 2$, the boundary conditions are
satisfied by constructing the eigenstate in each valley separately.
Therefore, two valleys are decoupled and the valley degeneracy
occurs.

For the case of (ii) the linear dispersion region, in which the
relation $|\Phi_{\tau r k}^b - \Phi_{\tau -r k}^b| = \pi$ holds,
Eq. \eqref{eq:app:coef_B} is rewritten as
\begin{equation}
  g_{\mathrm{B} \tau k}^b
  = 
  \mathrm{e}^{\mathrm{i} \Phi_{\tau r k}^b} \left( c_{\mathrm{A} \tau r k} - c_{\mathrm{A} \tau -r k} \right).
  \label{eq:app:coef_B_k_linear}
\end{equation}
To satisfy Eq. \eqref{eq:app:Phi+_Phi-}, we have
\begin{equation}
  c_{\mathrm{A} \tau r k}^b \neq 0, 
  ~~\text{and}~~c_{\mathrm{A}, \tau, -r, k}^b = 0, 
  \label{eq:app:coef_klinear}
\end{equation}
for either $r=1$ or $r=-1$, or, 
\begin{equation}
  \arg \left( c_{\mathrm{A} \tau r k}^b \right) - \arg \left( c_{\mathrm{A} \tau -r k}^b \right) = \pi \alpha,
  \label{eq:app:argcoef_klinear}
\end{equation}
where $\alpha$ is an arbitrary integer.

When we apply the boundary conditions for the right end, the second
condition of Eq. \eqref{eq:app:argcoef_klinear} is too strict to
determine quantized wavenumber.
For $d=1$ and $d=2$ cases, the following relation on the wavefunctions
reflecting the parity symmetry of the effective 1D Hamiltonian, which
corresponds to the ${\cal C}_2'$ rotational symmetry around the axis
perpendicular to the nanotube axis, is utilized for the boundary
conditions of the right end;~\cite{Izumida-2015-06}
\begin{equation}
  \phi_{{\rm A} k}^{b} (\ell) = p_a \phi_{{\rm B} k}^b (N_s + 1 - \ell) ~~\text{for any}~\ell, 
  \label{eq:app:parityRelation}
\end{equation}
where $p_a = \pm 1$ are the parity eigenvalues, $\ell = N_s$ is the
right end.
Equation \eqref{eq:app:parityRelation} is calculated as,
\begin{equation}
  c_{\sigma \tau r k}^b
  = 
  p_a \mathrm{e}^{\mathrm{i} \Phi_{-\tau -r k}^b} \mathrm{e}^{-\mathrm{i} \left( k_1 + r k \right) \left( N_s + 1 \right)}
  c_{\sigma -\tau -r k}^b,
  \label{eq:parityRelation_02}
\end{equation}
For the case of Eq. \eqref{eq:app:argcoef_klinear},
Eq. \eqref{eq:parityRelation_02} is applied for both $r=1$ and $r=-1$.
This is overcomplete to determine the quantized wavenumber.
Therefore, the condition of Eq. \eqref{eq:app:coef_klinear}, which
means the strong valley coupling, is realized for the finite-length
systems.

\subsection{Edge states at zero energy}
\label{sec:App:EdgeState}

By employing the similar discussion with Appendix
\ref{sec:App:valleyCoupling}, we will show the condition of emerging
of edge states.

For long nanotubes, the eigenfunctions within the energy gap, $\left|
\varepsilon \right| < \varepsilon_\mathrm{gap}/2$, are written as,
\begin{align}
  \phi_\mathrm{A} \left( \ell \right)
  & 
  = \sum_{m_\mathrm{A} = 1}^{N_\mathrm{A}} c_{\mathrm{A} m_\mathrm{A}} \left( \lambda_{\mathrm{A} m_\mathrm{A}}^{<} \right)^\ell
  + \sum_{m_\mathrm{B} = 1}^{N_\mathrm{B}} c_{\mathrm{B} m_\mathrm{B}} \frac{1}{\eta_{\mathrm{B} m_\mathrm{B}}} \left( \lambda_{\mathrm{B} m_\mathrm{B}}^{<} \right)^\ell, 
  \label{eq:app:phi_A_eva} \\
  \phi_\mathrm{B} \left( \ell \right)
  & 
  = \sum_{m_\mathrm{A} = 1}^{N_\mathrm{A}} c_{\mathrm{A} m_\mathrm{A}} \eta_{\mathrm{A} m_\mathrm{A}} \left( \lambda_{\mathrm{A} m_\mathrm{A}}^{<} \right)^\ell
  + \sum_{m_\mathrm{B} = 1}^{N_\mathrm{B}} c_{\mathrm{B} m_\mathrm{B}} \left( \lambda_{\mathrm{B} m_\mathrm{B}}^{<} \right)^\ell, 
  \label{eq:app:phi_B_eva}
\end{align}
where $\left| \eta_{\mathrm{A} m_\mathrm{A}} \right| < 1$ and $\left|
1 / \eta_{\mathrm{B} m_\mathrm{B}} \right| < 1$, and especially,
$\eta_{\mathrm{A} m_\mathrm{A}} = 1 / \eta_{\mathrm{B} m_\mathrm{B}} =
0$ at $\varepsilon = 0$.
By employing the boundary conditions of Eqs. \eqref{eq:BC_A} and
\eqref{eq:BC_B} to the eigenfunctions \eqref{eq:app:phi_A_eva} and
\eqref{eq:app:phi_B_eva}, we get the following relation,
\begin{equation}
  D_\mathrm{e} \mathbold{c}_\mathrm{e} = 0,
  \label{eq:app:DC_e=0}
\end{equation}
where
\begin{equation}
  D_\mathrm{e} =
  \left(
  \begin{array}{cccc}
    D_\mathrm{A} & E_\mathrm{B} \\
    E_\mathrm{A} & D_\mathrm{B} \\
  \end{array}
  \right),
  \label{eq:app:D_e=0}
\end{equation}
and
\begin{equation}
  \mathbold{c}_\mathrm{e}^\mathrm{T} =
  \left(
  c_{\mathrm{A} 1}, 
  \cdots,
  c_{\mathrm{A} N_\mathrm{A}},
  c_{\mathrm{B} 1},
  \cdots,
  c_{\mathrm{B} N_\mathrm{B}}
  \right),
  \label{eq:app:C_e_vec}
\end{equation}
where $D_\mathrm{A}$ and $D_\mathrm{B}$ are given in
Eqs. \eqref{eq:app:D_A} and \eqref{eq:app:D_B}, and
\begin{align}
  & E_\mathrm{A} =
  \left(
  \begin{array}{ccc}
    \eta_{\mathrm{A} 1} & \cdots & \eta_{\mathrm{A} N_\mathrm{A}} \\
    \eta_{\mathrm{A} 1} \left( \lambda_{\mathrm{A} 1}^{<} \right)^{-1} & \cdots & \eta_{\mathrm{A} N_\mathrm{A}} \left( \lambda_{\mathrm{A} N_\mathrm{A}}^{<} \right)^{-1} \\
    \vdots & \ddots & \vdots \\
    \eta_{\mathrm{A} 1} \left( \lambda_{\mathrm{A} 1}^{<} \right)^{-\frac{n}{d}+1} & \cdots & \eta_{\mathrm{A} N_\mathrm{A}} \left( \lambda_{\mathrm{A} N_\mathrm{A}}^{<} \right)^{-\frac{n}{d}+1} \\
  \end{array}
  \right),
  \label{eq:app:E_A} \\
  & E_\mathrm{B} =
  \left(
  \begin{array}{ccc}
    \frac{1}{\eta_{\mathrm{B} 1}} & \cdots & \frac{1}{\eta_{\mathrm{B} N_\mathrm{B}}} \\
    \frac{1}{\eta_{\mathrm{B} 1}} \left( \lambda_{\mathrm{B} 1}^{<} \right)^{-1} & \cdots & \frac{1}{\eta_{\mathrm{B} N_\mathrm{B}}} \left( \lambda_{\mathrm{B} N_\mathrm{B}}^{<} \right)^{-1} \\
    \vdots & \ddots & \vdots \\
    \frac{1}{\eta_{\mathrm{B} 1}} \left( \lambda_{\mathrm{B} 1}^{<} \right)^{-\frac{m}{d}+1} & \cdots & \frac{1}{\eta_{\mathrm{B} N_\mathrm{B}}} \left( \lambda_{\mathrm{B} N_\mathrm{B}}^{<} \right)^{-\frac{m}{d}+1} \\
  \end{array}
  \right),
  \label{eq:app:E_B}
\end{align}
The condition that Eq. \eqref{eq:app:DC_e=0} has non-trivial
solutions, $\mathbold{c}_\mathrm{e} \neq 0$, is the determinant of the
matrix $D_\mathrm{e}$ is zero, $\left| D_\mathrm{e} \right| = 0$.
This condition is satisfied when $\eta_{\mathrm{A} m_\mathrm{A}} =
1/\eta_{\mathrm{B} m_\mathrm{B}} = 0$, which is the case of
$\varepsilon = 0$, since $E_\mathrm{A} = 0$ and $E_\mathrm{B} = 0$,
and both $D_\mathrm{A}$ and $D_\mathrm{B}$ are not square matrices.
At finite energy $\varepsilon \neq 0$, $\eta_{\mathrm{A}
  m_\mathrm{A}}$ and $\eta_{\mathrm{B} m_\mathrm{B}}$ have finite
values, then the determinant cannot be zero, $\left| D_\mathrm{e}
\right| \neq 0$, except for accidental cases.

\section{Winding number under symmetry of SWNTs}
\label{sec:App:symmetry}

It is convenient to consider topological invariants in each subspace
of the Hilbert space under a symmetry.~\cite{PhysRevB.90.115207}
The topological property of the subsystem $H_\mu$ is characterized by
the winding number $w_\mu$ defined
by,~\cite{Wen1989641,PhysRevLett.89.077002}
\begin{equation}
  w_\mu 
  = 
  \frac{\mathrm{i}}{4 \pi} \int_0^{2 \pi} dk 
  \mathrm{Tr} 
  \left\{
    \Gamma_\mu(k) \left[ H_\mu(k) \right]^{-1} \frac{\partial H_\mu(k)}{\partial k}
    \right\},
    \label{eq:app:windingNumber}
\end{equation}
where 
\begin{equation}
  H_\mu(k)
  = 
  \left( 
  \begin{array}{cc}
    0         & f_\mu(k) \\
    f_\mu(k)^* & 0        \\
  \end{array}
  \right)
  \label{eq:app:H_mu_k_22}
\end{equation}
is the $2 \times 2$ Hamiltonian matrix formed by the base of Bloch
functions $| \sigma k \mu \rangle$ with $\sigma=\mathrm{A}$ and ${\rm
  B}$, and
\begin{equation}
  \Gamma_\mu(k)
  = 
  \left( 
  \begin{array}{cc}
    1 & 0  \\
    0 & -1 \\
  \end{array}
  \right)
  \label{eq:app:Gamma_mu_22}
\end{equation}
is known as so-called the chiral operator which multiple $-1$ on $|
\mathrm{B} k \mu \rangle$ in the operated state.
Since $w_\mu$ can be either positive or negative integers, the number
of total edge states at each end and each spin is given by $\sum_\mu
\left| w_\mu \right|$.
We put $\mu$ and $k$ on the operator $\Gamma_\mu(k)$, to emphasize
that the operation acts in $(k,\mu)$ subspace of the Hilbert space.
The above defined winding number is a topological invariant since the
bipartite system $H_\mu$, as well as the superconductors within the
mean field theory, satisfies so-called sublattice (or chiral)
symmetry, $\left\{ \Gamma_\mu(k), H_\mu(k) \right\} = 0$.

The winding number corresponds to the number of edge states on an end
in which A and B sublattices are terminated at the same position. 
[As examples, see left ends in Figs. \ref{fig:1D} (a) and \ref{fig:1D}
  (b)].
In other words, for a case that A and B sublattices are terminated at
different positions in the original 1D model, a modified model, in
which translation of B sublattice is performed for the original model
so as to have an end terminating both sublattices at the same
position, should be employed for calculating the winding number.
In general, the numbers of edge states for the different terminations
are different each other, such as the minimal and orthogonal
boundaries discussed in Ref. \onlinecite{Izumida-2015-06} for the same
bulk systems.
Substituting Eqs. (\ref{eq:app:H_mu_k_22}) and
(\ref{eq:app:Gamma_mu_22}) for Eq. (\ref{eq:app:windingNumber}), one
gets Eq. (\ref{eq:windingNumber}) which is convenient to evaluate the
winding number numerically.
Since $f_\mu(k)$ is a periodic function of $k$ of period $2 \pi$, the
winding number is an integer.
Hereafter we will show the winding number is a nontrivial value, that
is, $w_\mu$ can be non-zero, under the symmetry of SWNTs.
In the following, we employ a generalized discussion for application
to not only the SWNTs but also other materials.
The following discussion is restricted for the spinless case.

For the time reversal operation, ${\cal T}$, which changes $(k,\mu)$
to $(-k,-\mu)$, we have
\begin{equation}
  {\cal T} H_\mu(k) {\cal T}^{-1} = H_{-\mu}(-k)
  \label{eq:app:T_H}
\end{equation}
and 
\begin{equation}
  {\cal T} \Gamma_\mu(k) {\cal T}^{-1} = p_t \Gamma_{-\mu}(-k).
  \label{eq:app:T_Gamma}
\end{equation}
where $p_t = 1$ for the SWNTs.
When the operation also changes the``pseudo-spin'' $\sigma$ to
$-\sigma$, Eq. (\ref{eq:app:T_Gamma}) shows anticommutation relation
of $p_t = -1$, which occurs for so-called class DIII and CI
superconductors in which the Nambu pseudo-spin for the BCS
particle-hole index changes the sign under the ${\cal T}$
operation.~\cite{PhysRevB.78.195125,Kitaev-AIPConfProc2009,Ryu-NJP-2010}
Using Eqs. (\ref{eq:app:T_H}) and (\ref{eq:app:T_Gamma}) to
Eq. (\ref{eq:app:windingNumber}), it is shown that the winding number
of $-\mu$ subspace, $w_{-\mu}$, has the following relation,
\begin{equation}
  w_{-\mu} = p_t w_\mu,
  \label{eq:app:w_T_1}
\end{equation}
where we used the relation of $\langle \beta | O | \alpha \rangle =
\langle \tilde{\alpha} | {\cal T} O^{\dagger} {\cal T}^{-1} |
\tilde{\beta} \rangle$ for the antilinear operator ${\cal T}$ and the
anticommutation relation of $\Gamma_\mu(k)$ and $H_\mu(k)$,
where $O$ is a linear operator and tilde on the states indicates the
antilinear operated states, $| \tilde{\alpha} \rangle = {\cal T} |
\alpha \rangle$ and $| \tilde{\beta} \rangle = {\cal T} | \beta
\rangle$.~\cite{JJSakurai-MQP-2010}

In the sense of symmetry of point group, chiral SWNTs contains ${\cal
  C}_2'$ rotational symmetry around the axis perpendicular to the
nanotube axis, in addition to ${\cal C}_d$ rotational symmetry.
The ${\cal C}_2'$ operation changes $(k,\mu)$ to $(-k,-\mu)$, and the
sublattice index $\sigma$ to $-\sigma$, where $-\sigma = \mathrm{B}$
(A) for $\sigma = \mathrm{A}$ (B).
We have the relations
\begin{equation}
  C_2' H_\mu(k) C_2'^{-1} = H_{-\mu}(-k)
  \label{eq:app:C_2_H}
\end{equation}
and 
\begin{equation}
  C_2' \Gamma_\mu(k) C_2'^{-1} = p_2 \Gamma_{-\mu}(-k),
  \label{eq:app:C_2_Gamma}
\end{equation}
where $p_2 = -1$.
The anticommutation relation of Eq. (\ref{eq:app:C_2_Gamma}) reflects
that ${\cal C}_2'$ exchanges the A and B sublattices. 
Then, we have the following relation for the winding number,
\begin{equation}
  w_{-\mu} = - p_2 w_\mu.
  \label{eq:app:w_C_2}
\end{equation}
Since Eqs. (\ref{eq:app:w_T_1}) and (\ref{eq:app:w_C_2}) are identical
for the present case, which do not give extra restriction for the
winding number, it is concluded that $w_\mu$ can be a non-zero value.
For a case of $p_2 = 1$ reflecting the ``pseudo-spin'' keeping ${\cal
  C}_2'$ operation, for instance, gives the opposite result that
$w_\mu = 0$, even though this is not the present case.

For the achiral SWNTs, the armchair and the zigzag nanotubes, there
exist extra symmetries.
Let us consider mirror reflection $\sigma_v$ with a vertical mirror
plane including the nanotube axis, which changes $\mu$ to $-\mu$ while
the direction of $k$ is kept unchanged.
Strictly speaking, $\sigma_v$ changes $(k,\mu)$ to $(k',-\mu)$ where
$k' = k - 2 \Delta \theta \mu$ has the additional constant to $k$
reflecting that $\mathbold{Q}_1 / d$, which separates the two neighbor
cutting lines, has $-\Delta \theta / 2 \pi$ components in $\mathbold{Q}_2$
direction in the oblique coordinates for $(k,\mu)$
(see Sec. \ref{subsec:cutting_line}.)
Further, $\sigma_v$ exchanges the sublattice index $\mathrm{A}
\leftrightarrow \mathrm{B}$ for the armchair nanotubes while it is
unchanged for the zigzag nanotubes.
We have the relations
\begin{equation}
  \sigma_v H_\mu(k) \sigma_v^{-1} = H_{-\mu}(k')
  \label{eq:app:sigma_v_H}
\end{equation}
and 
\begin{equation}
  \sigma_v \Gamma_\mu(k) \sigma_v^{-1} = p_v \Gamma_{-\mu}(k'),
  \label{eq:app:sigma_v_Gamma}
\end{equation}
where $p_v = -1$ ($1$) for the armchair (zigzag) nanotubes.
Using Eqs. (\ref{eq:app:sigma_v_H}) and (\ref{eq:app:sigma_v_Gamma})
to (\ref{eq:app:windingNumber}), we have the following relation,
\begin{equation}
  w_{-\mu} = p_v w_\mu,
  \label{eq:app:w_sigma_v}
\end{equation}
The relation (\ref{eq:app:w_sigma_v}), combined with
Eq. (\ref{eq:app:w_T_1}), gives $w_\mu = 0$ for the armchair nanotubes
while it does not give extra restriction for the zigzag nanotubes.
For the other symmetry, mirror symmetry with a plane perpendicular to
the nanotube axis, $\sigma_h$, we have $\sigma_h \Gamma_\mu(k)
\sigma_h^{-1} = p_h \Gamma_\mu(-k')$, where $p_h=1$ ($-1$) for the
armchair (zigzag) nanotubes.
The similar calculation with above gives
\begin{equation}
  w_\mu = -p_h w_\mu,
\end{equation}
which accidentally
gives the same results from Eqs. (\ref{eq:app:w_C_2}) and
(\ref{eq:app:w_sigma_v}).
The inversion symmetry does not give any further extra restrictions
from above since the operation is represented by the combination of
$C_2'$ and $\sigma_v$ operations.

To complete the symmetric property of the winding number, we also add
the property from the $d$-fold rotational symmetry, which requires
$[C_d, H_\mu(k)] = 0$.
From the relation $C_d \Gamma_\mu C_d^{-1} = p_d \Gamma_\mu$, where
$p_d = 1$ for the SWNTs, and Eq. (\ref{eq:app:windingNumber}), we have
\begin{equation}
  w_\mu = p_d w_\mu.
\end{equation}
This is nontrivial, however for the other case of $p_d=-1$ where the
$d$-fold rotation exchanges the A and B sublattices, we have the
winding number $w_\mu = 0$.

The obtained necessary conditions for the nontrivial winding numbers
are summarized as,
\begin{equation}
  p_d = -p_2 = p_v  = -p_h = 1, ~~~~(p_t = 1),
  \label{eq:app:w_p_condition01}
\end{equation}
for the SWNTs and other bipartite insulators, 
\begin{equation}
   p_d = p_2 = -p_v = -p_h = 1, ~~~~(p_t = -1),
  \label{eq:app:w_p_condition02}
\end{equation}
for the topological superconductors.~\cite{Yamakage-inPrep}

As mentioned above, the SWNTs, except for the armchair nanotubes, have
nontrivial winding number.
Here the indices of the SWNTs are summarized as follows;
\begin{equation}
  p_t = 1, \text{ and, } p_d = -p_2 = 1,
  \label{eq:app:p_allSWNTs}
\end{equation}
for the all SWNTs.
In addition to Eq. (\ref{eq:app:p_allSWNTs}),
\begin{equation}
  p_v = -p_h = 1,
  \label{eq:app:p_zigzag}
\end{equation}
for the zigzag nanotubes, and 
\begin{equation}
  p_v = -p_h = -1,
  \label{eq:app:p_armchair}
\end{equation}
for the armchair nanotubes.
As another example of the application of
Eq. (\ref{eq:app:w_p_condition01}), we employ it to so-called the
Su-Schrieffer-Heeger model for the
polyacetylene.~\cite{PhysRevLett.42.1698}
The system has no rotational symmetry ($d=1$), $p_2 = -1$, $p_v = 1$
and $p_t = 1$ are all the relevant indices for the system, then the
system has nontrivial winding number, which is consistent with the
well-known properties of zero-energy edge states for the
Su-Schrieffer-Heeger
model.~\cite{RevModPhys.82.3045,RevModPhys.83.1057}

The result that there is no edge states for the armchair nanotubes
from the above discussion is consistent with the numerical calculation
in Ref. \onlinecite{Izumida-2015-06} which does not show any
evanescent modes for the armchair nanotube with the minimal boundary.
Note that the small energy gap of the order of sub-milli-electron-volt
is induced by the spin--orbit interaction.~\cite{Izumida-2009-06}
This property for the armchair nanotubes is contrast to the other
topological
materials,~\cite{Kane-2005-11,PhysRevLett.115.036806,Yamakage-JPSJ-2016-01}
in which edge states appears in the energy gap induced by the
spin--orbit interaction.
We confirmed by the extended tight-binding calculation with large
spin--orbit interaction ($V_\mathrm{SO}=1$ eV) for $(n,m)=(6,6)$, $50$ nm
length armchair nanotube that there is no edge state in the energy gap
of $\sim 0.13$ eV.
The effective 1D model with additional imaginary hopping
terms,~\cite{Ando-2000-06} to reproduces the spin--orbit effect, also
showed winding number being zero for the armchair nanotubes.


\begin{thebibliography}{72}%
\makeatletter
\providecommand \@ifxundefined [1]{%
 \@ifx{#1\undefined}
}%
\providecommand \@ifnum [1]{%
 \ifnum #1\expandafter \@firstoftwo
 \else \expandafter \@secondoftwo
 \fi
}%
\providecommand \@ifx [1]{%
 \ifx #1\expandafter \@firstoftwo
 \else \expandafter \@secondoftwo
 \fi
}%
\providecommand \natexlab [1]{#1}%
\providecommand \enquote  [1]{``#1''}%
\providecommand \bibnamefont  [1]{#1}%
\providecommand \bibfnamefont [1]{#1}%
\providecommand \citenamefont [1]{#1}%
\providecommand \href@noop [0]{\@secondoftwo}%
\providecommand \href [0]{\begingroup \@sanitize@url \@href}%
\providecommand \@href[1]{\@@startlink{#1}\@@href}%
\providecommand \@@href[1]{\endgroup#1\@@endlink}%
\providecommand \@sanitize@url [0]{\catcode `\\12\catcode `\$12\catcode
  `\&12\catcode `\#12\catcode `\^12\catcode `\_12\catcode `\%12\relax}%
\providecommand \@@startlink[1]{}%
\providecommand \@@endlink[0]{}%
\providecommand \url  [0]{\begingroup\@sanitize@url \@url }%
\providecommand \@url [1]{\endgroup\@href {#1}{\urlprefix }}%
\providecommand \urlprefix  [0]{URL }%
\providecommand \Eprint [0]{\href }%
\providecommand \doibase [0]{http://dx.doi.org/}%
\providecommand \selectlanguage [0]{\@gobble}%
\providecommand \bibinfo  [0]{\@secondoftwo}%
\providecommand \bibfield  [0]{\@secondoftwo}%
\providecommand \translation [1]{[#1]}%
\providecommand \BibitemOpen [0]{}%
\providecommand \bibitemStop [0]{}%
\providecommand \bibitemNoStop [0]{.\EOS\space}%
\providecommand \EOS [0]{\spacefactor3000\relax}%
\providecommand \BibitemShut  [1]{\csname bibitem#1\endcsname}%
\let\auto@bib@innerbib\@empty
\bibitem [{\citenamefont {Laird}\ \emph {et~al.}(2015)\citenamefont {Laird},
  \citenamefont {Kuemmeth}, \citenamefont {Steele}, \citenamefont
  {Grove-Rasmussen}, \citenamefont {Nyg\aa{}rd}, \citenamefont {Flensberg},\
  and\ \citenamefont {Kouwenhoven}}]{RevModPhys.87.703}%
  \BibitemOpen
  \bibfield  {author} {\bibinfo {author} {\bibfnamefont {E.~A.}\ \bibnamefont
  {Laird}}, \bibinfo {author} {\bibfnamefont {F.}~\bibnamefont {Kuemmeth}},
  \bibinfo {author} {\bibfnamefont {G.~A.}\ \bibnamefont {Steele}}, \bibinfo
  {author} {\bibfnamefont {K.}~\bibnamefont {Grove-Rasmussen}}, \bibinfo
  {author} {\bibfnamefont {J.}~\bibnamefont {Nyg\aa{}rd}}, \bibinfo {author}
  {\bibfnamefont {K.}~\bibnamefont {Flensberg}}, \ and\ \bibinfo {author}
  {\bibfnamefont {L.~P.}\ \bibnamefont {Kouwenhoven}},\ }\bibfield  {title}
  {\enquote {\bibinfo {title} {Quantum transport in carbon nanotubes},}\
  }\href@noop {} {\bibfield  {journal} {\bibinfo  {journal} {Rev. Mod. Phys.}\
  }\textbf {\bibinfo {volume} {87}},\ \bibinfo {pages} {703} (\bibinfo {year}
  {2015})}\BibitemShut {NoStop}%
\bibitem [{\citenamefont {Liang}\ \emph {et~al.}(2002)\citenamefont {Liang},
  \citenamefont {Bockrath},\ and\ \citenamefont {Park}}]{Liang-2002-03}%
  \BibitemOpen
  \bibfield  {author} {\bibinfo {author} {\bibfnamefont {W.}~\bibnamefont
  {Liang}}, \bibinfo {author} {\bibfnamefont {M.}~\bibnamefont {Bockrath}}, \
  and\ \bibinfo {author} {\bibfnamefont {H.}~\bibnamefont {Park}},\ }\bibfield
  {title} {\enquote {\bibinfo {title} {Shell filling and exchange coupling in
  metallic single-walled carbon nanotubes},}\ }\href@noop {} {\bibfield
  {journal} {\bibinfo  {journal} {Phys. Rev. Lett.}\ }\textbf {\bibinfo
  {volume} {88}},\ \bibinfo {pages} {126801} (\bibinfo {year}
  {2002})}\BibitemShut {NoStop}%
\bibitem [{\citenamefont {Cobden}\ and\ \citenamefont
  {Nyg{\aa}rd}(2002)}]{Cobden-2002-07}%
  \BibitemOpen
  \bibfield  {author} {\bibinfo {author} {\bibfnamefont {D.~H.}\ \bibnamefont
  {Cobden}}\ and\ \bibinfo {author} {\bibfnamefont {J.}~\bibnamefont
  {Nyg{\aa}rd}},\ }\bibfield  {title} {\enquote {\bibinfo {title} {Shell
  filling in closed single-wall carbon nanotube quantum dots},}\ }\href@noop {}
  {\bibfield  {journal} {\bibinfo  {journal} {Phys. Rev. Lett.}\ }\textbf
  {\bibinfo {volume} {89}},\ \bibinfo {pages} {046803} (\bibinfo {year}
  {2002})}\BibitemShut {NoStop}%
\bibitem [{\citenamefont {Jarillo-Herrero}\ \emph {et~al.}(2005)\citenamefont
  {Jarillo-Herrero}, \citenamefont {Kong}, \citenamefont {van~der Zant},
  \citenamefont {Dekker}, \citenamefont {Kouwenhoven},\ and\ \citenamefont
  {De~Franceschi}}]{Jarillo-Herrero-2005-04}%
  \BibitemOpen
  \bibfield  {author} {\bibinfo {author} {\bibfnamefont {P.}~\bibnamefont
  {Jarillo-Herrero}}, \bibinfo {author} {\bibfnamefont {J.}~\bibnamefont
  {Kong}}, \bibinfo {author} {\bibfnamefont {H.~S.~J.}\ \bibnamefont {van~der
  Zant}}, \bibinfo {author} {\bibfnamefont {C.}~\bibnamefont {Dekker}},
  \bibinfo {author} {\bibfnamefont {L.~P.}\ \bibnamefont {Kouwenhoven}}, \ and\
  \bibinfo {author} {\bibfnamefont {S.}~\bibnamefont {De~Franceschi}},\
  }\bibfield  {title} {\enquote {\bibinfo {title} {Electronic transport
  spectroscopy of carbon nanotubes in a magnetic field},}\ }\href@noop {}
  {\bibfield  {journal} {\bibinfo  {journal} {Phys. Rev. Lett.}\ }\textbf
  {\bibinfo {volume} {94}},\ \bibinfo {pages} {156802} (\bibinfo {year}
  {2005})}\BibitemShut {NoStop}%
\bibitem [{\citenamefont {Moriyama}\ \emph {et~al.}(2005)\citenamefont
  {Moriyama}, \citenamefont {Fuse}, \citenamefont {Suzuki}, \citenamefont
  {Aoyagi},\ and\ \citenamefont {Ishibashi}}]{Moriyama-2005-05}%
  \BibitemOpen
  \bibfield  {author} {\bibinfo {author} {\bibfnamefont {S.}~\bibnamefont
  {Moriyama}}, \bibinfo {author} {\bibfnamefont {T.}~\bibnamefont {Fuse}},
  \bibinfo {author} {\bibfnamefont {M.}~\bibnamefont {Suzuki}}, \bibinfo
  {author} {\bibfnamefont {Y.}~\bibnamefont {Aoyagi}}, \ and\ \bibinfo {author}
  {\bibfnamefont {K.}~\bibnamefont {Ishibashi}},\ }\bibfield  {title} {\enquote
  {\bibinfo {title} {Four-electron shell structures and an interacting
  two-electron system in carbon-nanotube quantum dots},}\ }\href@noop {}
  {\bibfield  {journal} {\bibinfo  {journal} {Phys. Rev. Lett.}\ }\textbf
  {\bibinfo {volume} {94}},\ \bibinfo {pages} {186806} (\bibinfo {year}
  {2005})}\BibitemShut {NoStop}%
\bibitem [{\citenamefont {Sapmaz}\ \emph {et~al.}(2005)\citenamefont {Sapmaz},
  \citenamefont {Jarillo-Herrero}, \citenamefont {Kong}, \citenamefont
  {Dekker}, \citenamefont {Kouwenhoven},\ and\ \citenamefont {van~der
  Zant}}]{Sapmaz-2005-04}%
  \BibitemOpen
  \bibfield  {author} {\bibinfo {author} {\bibfnamefont {S.}~\bibnamefont
  {Sapmaz}}, \bibinfo {author} {\bibfnamefont {P.}~\bibnamefont
  {Jarillo-Herrero}}, \bibinfo {author} {\bibfnamefont {J.}~\bibnamefont
  {Kong}}, \bibinfo {author} {\bibfnamefont {C.}~\bibnamefont {Dekker}},
  \bibinfo {author} {\bibfnamefont {L.~P.}\ \bibnamefont {Kouwenhoven}}, \ and\
  \bibinfo {author} {\bibfnamefont {H.~S.~J.}\ \bibnamefont {van~der Zant}},\
  }\bibfield  {title} {\enquote {\bibinfo {title} {Electronic excitation
  spectrum of metallic carbon nanotubes},}\ }\href@noop {} {\bibfield
  {journal} {\bibinfo  {journal} {Phys. Rev. B}\ }\textbf {\bibinfo {volume}
  {71}},\ \bibinfo {pages} {153402} (\bibinfo {year} {2005})}\BibitemShut
  {NoStop}%
\bibitem [{\citenamefont {Maki}\ \emph {et~al.}(2005)\citenamefont {Maki},
  \citenamefont {Ishiwata}, \citenamefont {Suzuki},\ and\ \citenamefont
  {Ishibashi}}]{Maki-2005-06}%
  \BibitemOpen
  \bibfield  {author} {\bibinfo {author} {\bibfnamefont {H.}~\bibnamefont
  {Maki}}, \bibinfo {author} {\bibfnamefont {Y.}~\bibnamefont {Ishiwata}},
  \bibinfo {author} {\bibfnamefont {M.}~\bibnamefont {Suzuki}}, \ and\ \bibinfo
  {author} {\bibfnamefont {K.}~\bibnamefont {Ishibashi}},\ }\bibfield  {title}
  {\enquote {\bibinfo {title} {Electronic transport of a carbon nanotube
  quantum dot in different coupling regimes},}\ }\href@noop {} {\bibfield
  {journal} {\bibinfo  {journal} {Jpn. J. Appl. Phys.}\ }\textbf {\bibinfo
  {volume} {44}},\ \bibinfo {pages} {4269} (\bibinfo {year}
  {2005})}\BibitemShut {NoStop}%
\bibitem [{\citenamefont {Cao}\ \emph {et~al.}(2005)\citenamefont {Cao},
  \citenamefont {Wang},\ and\ \citenamefont {Dai}}]{Cao-2005-09}%
  \BibitemOpen
  \bibfield  {author} {\bibinfo {author} {\bibfnamefont {J.}~\bibnamefont
  {Cao}}, \bibinfo {author} {\bibfnamefont {Q.}~\bibnamefont {Wang}}, \ and\
  \bibinfo {author} {\bibfnamefont {H.}~\bibnamefont {Dai}},\ }\bibfield
  {title} {\enquote {\bibinfo {title} {Electron transport in very clean,
  as-grown suspended carbon nanotubes},}\ }\href@noop {} {\bibfield  {journal}
  {\bibinfo  {journal} {Nat. Mater.}\ }\textbf {\bibinfo {volume} {4}},\
  \bibinfo {pages} {745} (\bibinfo {year} {2005})}\BibitemShut {NoStop}%
\bibitem [{\citenamefont {Makarovski}\ \emph {et~al.}(2006)\citenamefont
  {Makarovski}, \citenamefont {An}, \citenamefont {Liu},\ and\ \citenamefont
  {Finkelstein}}]{Makarovski-2006-10}%
  \BibitemOpen
  \bibfield  {author} {\bibinfo {author} {\bibfnamefont {A.}~\bibnamefont
  {Makarovski}}, \bibinfo {author} {\bibfnamefont {L.}~\bibnamefont {An}},
  \bibinfo {author} {\bibfnamefont {J.}~\bibnamefont {Liu}}, \ and\ \bibinfo
  {author} {\bibfnamefont {G.}~\bibnamefont {Finkelstein}},\ }\bibfield
  {title} {\enquote {\bibinfo {title} {Persistent orbital degeneracy in carbon
  nanotubes},}\ }\href@noop {} {\bibfield  {journal} {\bibinfo  {journal}
  {Phys. Rev. B}\ }\textbf {\bibinfo {volume} {74}},\ \bibinfo {pages} {155431}
  (\bibinfo {year} {2006})}\BibitemShut {NoStop}%
\bibitem [{\citenamefont {Moriyama}\ \emph {et~al.}(2007)\citenamefont
  {Moriyama}, \citenamefont {Fuse},\ and\ \citenamefont
  {Ishibashi}}]{Moriyama-2007-04}%
  \BibitemOpen
  \bibfield  {author} {\bibinfo {author} {\bibfnamefont {S.}~\bibnamefont
  {Moriyama}}, \bibinfo {author} {\bibfnamefont {T.}~\bibnamefont {Fuse}}, \
  and\ \bibinfo {author} {\bibfnamefont {K.}~\bibnamefont {Ishibashi}},\
  }\bibfield  {title} {\enquote {\bibinfo {title} {Shell structures and
  electron-spin configurations in single-walled carbon nanotube quantum
  dots},}\ }\href@noop {} {\bibfield  {journal} {\bibinfo  {journal} {Phys.
  Stat. Sol. B}\ }\textbf {\bibinfo {volume} {244}},\ \bibinfo {pages} {2371}
  (\bibinfo {year} {2007})}\BibitemShut {NoStop}%
\bibitem [{\citenamefont {Holm}\ \emph {et~al.}(2008)\citenamefont {Holm},
  \citenamefont {J{\o}rgensen}, \citenamefont {Grove-Rasmussen}, \citenamefont
  {Paaske}, \citenamefont {Flensberg},\ and\ \citenamefont
  {Lindelof}}]{Holm-2008-04}%
  \BibitemOpen
  \bibfield  {author} {\bibinfo {author} {\bibfnamefont {J.~V.}\ \bibnamefont
  {Holm}}, \bibinfo {author} {\bibfnamefont {H.~I.}\ \bibnamefont
  {J{\o}rgensen}}, \bibinfo {author} {\bibfnamefont {K.}~\bibnamefont
  {Grove-Rasmussen}}, \bibinfo {author} {\bibfnamefont {J.}~\bibnamefont
  {Paaske}}, \bibinfo {author} {\bibfnamefont {K.}~\bibnamefont {Flensberg}}, \
  and\ \bibinfo {author} {\bibfnamefont {P.~E.}\ \bibnamefont {Lindelof}},\
  }\bibfield  {title} {\enquote {\bibinfo {title} {Gate-dependent
  tunneling-induced level shifts observed in carbon nanotube quantum dots},}\
  }\href@noop {} {\bibfield  {journal} {\bibinfo  {journal} {Phys. Rev. B}\
  }\textbf {\bibinfo {volume} {77}},\ \bibinfo {pages} {161406(R)} (\bibinfo
  {year} {2008})}\BibitemShut {NoStop}%
\bibitem [{\citenamefont {Kuemmeth}\ \emph {et~al.}(2008)\citenamefont
  {Kuemmeth}, \citenamefont {Ilani}, \citenamefont {Ralph},\ and\ \citenamefont
  {McEuen}}]{Kuemmeth-2008-03}%
  \BibitemOpen
  \bibfield  {author} {\bibinfo {author} {\bibfnamefont {F.}~\bibnamefont
  {Kuemmeth}}, \bibinfo {author} {\bibfnamefont {S.}~\bibnamefont {Ilani}},
  \bibinfo {author} {\bibfnamefont {D.~C.}\ \bibnamefont {Ralph}}, \ and\
  \bibinfo {author} {\bibfnamefont {P.~L.}\ \bibnamefont {McEuen}},\ }\bibfield
   {title} {\enquote {\bibinfo {title} {Coupling of spin and orbital motion of
  electrons in carbon nanotubes},}\ }\href@noop {} {\bibfield  {journal}
  {\bibinfo  {journal} {Nature}\ }\textbf {\bibinfo {volume} {452}},\ \bibinfo
  {pages} {448} (\bibinfo {year} {2008})}\BibitemShut {NoStop}%
\bibitem [{\citenamefont {Jhang}\ \emph {et~al.}(2010)\citenamefont {Jhang},
  \citenamefont {Marganska}, \citenamefont {Skourski}, \citenamefont
  {Preusche}, \citenamefont {Witkamp}, \citenamefont {Grifoni}, \citenamefont
  {van~der Zant}, \citenamefont {Wosnitza},\ and\ \citenamefont
  {Strunk}}]{Jhang-2010-07}%
  \BibitemOpen
  \bibfield  {author} {\bibinfo {author} {\bibfnamefont {S.~H.}\ \bibnamefont
  {Jhang}}, \bibinfo {author} {\bibfnamefont {M.}~\bibnamefont {Marganska}},
  \bibinfo {author} {\bibfnamefont {Y.}~\bibnamefont {Skourski}}, \bibinfo
  {author} {\bibfnamefont {D.}~\bibnamefont {Preusche}}, \bibinfo {author}
  {\bibfnamefont {B.}~\bibnamefont {Witkamp}}, \bibinfo {author} {\bibfnamefont
  {M.}~\bibnamefont {Grifoni}}, \bibinfo {author} {\bibfnamefont
  {H.}~\bibnamefont {van~der Zant}}, \bibinfo {author} {\bibfnamefont
  {J.}~\bibnamefont {Wosnitza}}, \ and\ \bibinfo {author} {\bibfnamefont
  {C.}~\bibnamefont {Strunk}},\ }\bibfield  {title} {\enquote {\bibinfo {title}
  {Spin-orbit interaction in chiral carbon nanotubes probed in pulsed magnetic
  fields},}\ }\href@noop {} {\bibfield  {journal} {\bibinfo  {journal} {Phys.
  Rev. B}\ }\textbf {\bibinfo {volume} {82}},\ \bibinfo {pages} {041404(R)}
  (\bibinfo {year} {2010})}\BibitemShut {NoStop}%
\bibitem [{\citenamefont {Jespersen}\ \emph {et~al.}(2011)\citenamefont
  {Jespersen}, \citenamefont {Grove-Rasmussen}, \citenamefont {Paaske},
  \citenamefont {Muraki}, \citenamefont {Fujisawa}, \citenamefont
  {Nyg{\aa}rd},\ and\ \citenamefont {Flensberg}}]{Jespersen-2011-04}%
  \BibitemOpen
  \bibfield  {author} {\bibinfo {author} {\bibfnamefont {T.~S.}\ \bibnamefont
  {Jespersen}}, \bibinfo {author} {\bibfnamefont {K.}~\bibnamefont
  {Grove-Rasmussen}}, \bibinfo {author} {\bibfnamefont {J.}~\bibnamefont
  {Paaske}}, \bibinfo {author} {\bibfnamefont {K.}~\bibnamefont {Muraki}},
  \bibinfo {author} {\bibfnamefont {T.}~\bibnamefont {Fujisawa}}, \bibinfo
  {author} {\bibfnamefont {J.}~\bibnamefont {Nyg{\aa}rd}}, \ and\ \bibinfo
  {author} {\bibfnamefont {K.}~\bibnamefont {Flensberg}},\ }\bibfield  {title}
  {\enquote {\bibinfo {title} {Gate-dependent spin--orbit coupling in
  multielectron carbon nanotubes},}\ }\href@noop {} {\bibfield  {journal}
  {\bibinfo  {journal} {Nat. Phys.}\ }\textbf {\bibinfo {volume} {7}},\
  \bibinfo {pages} {348} (\bibinfo {year} {2011})}\BibitemShut {NoStop}%
\bibitem [{\citenamefont {Steele}\ \emph {et~al.}(2013)\citenamefont {Steele},
  \citenamefont {Pei}, \citenamefont {Laird}, \citenamefont {Jol},
  \citenamefont {Meerwaldt},\ and\ \citenamefont
  {Kouwenhoven}}]{Steele:2013fk}%
  \BibitemOpen
  \bibfield  {author} {\bibinfo {author} {\bibfnamefont {G.~A.}\ \bibnamefont
  {Steele}}, \bibinfo {author} {\bibfnamefont {F.}~\bibnamefont {Pei}},
  \bibinfo {author} {\bibfnamefont {E.~A.}\ \bibnamefont {Laird}}, \bibinfo
  {author} {\bibfnamefont {J.~M.}\ \bibnamefont {Jol}}, \bibinfo {author}
  {\bibfnamefont {H.~B.}\ \bibnamefont {Meerwaldt}}, \ and\ \bibinfo {author}
  {\bibfnamefont {L.~P.}\ \bibnamefont {Kouwenhoven}},\ }\bibfield  {title}
  {\enquote {\bibinfo {title} {Large spin-orbit coupling in carbon
  nanotubes},}\ }\href@noop {} {\bibfield  {journal} {\bibinfo  {journal} {Nat.
  Commun.}\ }\textbf {\bibinfo {volume} {4}},\ \bibinfo {pages} {1573}
  (\bibinfo {year} {2013})}\BibitemShut {NoStop}%
\bibitem [{\citenamefont {Ando}(2000)}]{Ando-2000-06}%
  \BibitemOpen
  \bibfield  {author} {\bibinfo {author} {\bibfnamefont {T.}~\bibnamefont
  {Ando}},\ }\bibfield  {title} {\enquote {\bibinfo {title} {Spin-orbit
  interaction in carbon nanotubes},}\ }\href@noop {} {\bibfield  {journal}
  {\bibinfo  {journal} {J. Phys. Soc. Jpn.}\ }\textbf {\bibinfo {volume}
  {69}},\ \bibinfo {pages} {1757} (\bibinfo {year} {2000})}\BibitemShut
  {NoStop}%
\bibitem [{\citenamefont {Chico}\ \emph {et~al.}(2004)\citenamefont {Chico},
  \citenamefont {Lopez-Sancho},\ and\ \citenamefont {Munoz}}]{Chico-2004-10}%
  \BibitemOpen
  \bibfield  {author} {\bibinfo {author} {\bibfnamefont {L.}~\bibnamefont
  {Chico}}, \bibinfo {author} {\bibfnamefont {M.~P.}\ \bibnamefont
  {Lopez-Sancho}}, \ and\ \bibinfo {author} {\bibfnamefont {M.~C.}\
  \bibnamefont {Munoz}},\ }\bibfield  {title} {\enquote {\bibinfo {title} {Spin
  splitting induced by spin-orbit interaction in chiral nanotubes},}\
  }\href@noop {} {\bibfield  {journal} {\bibinfo  {journal} {Phys. Rev. Lett.}\
  }\textbf {\bibinfo {volume} {93}},\ \bibinfo {pages} {176402} (\bibinfo
  {year} {2004})}\BibitemShut {NoStop}%
\bibitem [{\citenamefont {Huertas-Hernando}\ \emph {et~al.}(2006)\citenamefont
  {Huertas-Hernando}, \citenamefont {Guinea},\ and\ \citenamefont
  {Brataas}}]{Huertas-Hernando-2006-10}%
  \BibitemOpen
  \bibfield  {author} {\bibinfo {author} {\bibfnamefont {D.}~\bibnamefont
  {Huertas-Hernando}}, \bibinfo {author} {\bibfnamefont {F.}~\bibnamefont
  {Guinea}}, \ and\ \bibinfo {author} {\bibfnamefont {A.}~\bibnamefont
  {Brataas}},\ }\bibfield  {title} {\enquote {\bibinfo {title} {Spin-orbit
  coupling in curved graphene, fullerenes, nanotubes, and nanotube caps},}\
  }\href@noop {} {\bibfield  {journal} {\bibinfo  {journal} {Phys. Rev. B}\
  }\textbf {\bibinfo {volume} {74}},\ \bibinfo {pages} {155426} (\bibinfo
  {year} {2006})}\BibitemShut {NoStop}%
\bibitem [{\citenamefont {Chico}\ \emph {et~al.}(2009)\citenamefont {Chico},
  \citenamefont {L{\'o}pez-Sancho},\ and\ \citenamefont
  {Mu{\~n}oz}}]{Chico-2009-06}%
  \BibitemOpen
  \bibfield  {author} {\bibinfo {author} {\bibfnamefont {L.}~\bibnamefont
  {Chico}}, \bibinfo {author} {\bibfnamefont {M.~P.}\ \bibnamefont
  {L{\'o}pez-Sancho}}, \ and\ \bibinfo {author} {\bibfnamefont {M.~C.}\
  \bibnamefont {Mu{\~n}oz}},\ }\bibfield  {title} {\enquote {\bibinfo {title}
  {Curvature-induced anisotropic spin-orbit splitting in carbon nanotubes},}\
  }\href@noop {} {\bibfield  {journal} {\bibinfo  {journal} {Phys. Rev. B}\
  }\textbf {\bibinfo {volume} {79}},\ \bibinfo {pages} {235423} (\bibinfo
  {year} {2009})}\BibitemShut {NoStop}%
\bibitem [{\citenamefont {Izumida}\ \emph {et~al.}(2009)\citenamefont
  {Izumida}, \citenamefont {Sato},\ and\ \citenamefont
  {Saito}}]{Izumida-2009-06}%
  \BibitemOpen
  \bibfield  {author} {\bibinfo {author} {\bibfnamefont {W.}~\bibnamefont
  {Izumida}}, \bibinfo {author} {\bibfnamefont {K.}~\bibnamefont {Sato}}, \
  and\ \bibinfo {author} {\bibfnamefont {R.}~\bibnamefont {Saito}},\ }\bibfield
   {title} {\enquote {\bibinfo {title} {Spin--orbit interaction in single wall
  carbon nanotubes: Symmetry adapted tight-binding calculation and effective
  model analysis},}\ }\href@noop {} {\bibfield  {journal} {\bibinfo  {journal}
  {J. Phys. Soc. Jpn.}\ }\textbf {\bibinfo {volume} {78}},\ \bibinfo {pages}
  {074707} (\bibinfo {year} {2009})}\BibitemShut {NoStop}%
\bibitem [{\citenamefont {Jeong}\ and\ \citenamefont
  {Lee}(2009)}]{Jeong-2009-08}%
  \BibitemOpen
  \bibfield  {author} {\bibinfo {author} {\bibfnamefont {J.-S.}\ \bibnamefont
  {Jeong}}\ and\ \bibinfo {author} {\bibfnamefont {H.-W.}\ \bibnamefont
  {Lee}},\ }\bibfield  {title} {\enquote {\bibinfo {title} {Curvature-enhanced
  spin-orbit coupling in a carbon nanotube},}\ }\href@noop {} {\bibfield
  {journal} {\bibinfo  {journal} {Phys. Rev. B}\ }\textbf {\bibinfo {volume}
  {80}},\ \bibinfo {pages} {075409} (\bibinfo {year} {2009})}\BibitemShut
  {NoStop}%
\bibitem [{\citenamefont {Izumida}\ \emph {et~al.}(2015)\citenamefont
  {Izumida}, \citenamefont {Okuyama},\ and\ \citenamefont
  {Saito}}]{Izumida-2015-06}%
  \BibitemOpen
  \bibfield  {author} {\bibinfo {author} {\bibfnamefont {W.}~\bibnamefont
  {Izumida}}, \bibinfo {author} {\bibfnamefont {R.}~\bibnamefont {Okuyama}}, \
  and\ \bibinfo {author} {\bibfnamefont {R.}~\bibnamefont {Saito}},\ }\bibfield
   {title} {\enquote {\bibinfo {title} {Valley coupling in finite-length
  metallic single-wall carbon nanotubes},}\ }\href {\doibase
  10.1103/PhysRevB.91.235442} {\bibfield  {journal} {\bibinfo  {journal} {Phys.
  Rev. B}\ }\textbf {\bibinfo {volume} {91}},\ \bibinfo {pages} {235442}
  (\bibinfo {year} {2015})}\BibitemShut {NoStop}%
\bibitem [{\citenamefont {Marganska}\ \emph {et~al.}(2015)\citenamefont
  {Marganska}, \citenamefont {Chudzinski},\ and\ \citenamefont
  {Grifoni}}]{PhysRevB.92.075433}%
  \BibitemOpen
  \bibfield  {author} {\bibinfo {author} {\bibfnamefont {M.}~\bibnamefont
  {Marganska}}, \bibinfo {author} {\bibfnamefont {P.}~\bibnamefont
  {Chudzinski}}, \ and\ \bibinfo {author} {\bibfnamefont {M.}~\bibnamefont
  {Grifoni}},\ }\bibfield  {title} {\enquote {\bibinfo {title} {The two classes
  of low-energy spectra in finite carbon nanotubes},}\ }\href {\doibase
  10.1103/PhysRevB.92.075433} {\bibfield  {journal} {\bibinfo  {journal} {Phys.
  Rev. B}\ }\textbf {\bibinfo {volume} {92}},\ \bibinfo {pages} {075433}
  (\bibinfo {year} {2015})}\BibitemShut {NoStop}%
\bibitem [{\citenamefont {Lunde}\ \emph {et~al.}(2005)\citenamefont {Lunde},
  \citenamefont {Flensberg},\ and\ \citenamefont {Jauho}}]{PhysRevB.71.125408}%
  \BibitemOpen
  \bibfield  {author} {\bibinfo {author} {\bibfnamefont {A.~M.}\ \bibnamefont
  {Lunde}}, \bibinfo {author} {\bibfnamefont {K.}~\bibnamefont {Flensberg}}, \
  and\ \bibinfo {author} {\bibfnamefont {A.-P.}\ \bibnamefont {Jauho}},\
  }\bibfield  {title} {\enquote {\bibinfo {title} {Intershell resistance in
  multiwall carbon nanotubes: A {C}oulomb drag study},}\ }\href@noop {}
  {\bibfield  {journal} {\bibinfo  {journal} {Phys. Rev. B}\ }\textbf {\bibinfo
  {volume} {71}},\ \bibinfo {pages} {125408} (\bibinfo {year}
  {2005})}\BibitemShut {NoStop}%
\bibitem [{\citenamefont {Schmid}\ \emph {et~al.}(2015)\citenamefont {Schmid},
  \citenamefont {Smirnov}, \citenamefont {Marga\ifmmode~\acute{n}\else
  \'{n}\fi{}ska}, \citenamefont {Dirnaichner}, \citenamefont {Stiller},
  \citenamefont {Grifoni}, \citenamefont {H\"uttel},\ and\ \citenamefont
  {Strunk}}]{PhysRevB.91.155435}%
  \BibitemOpen
  \bibfield  {author} {\bibinfo {author} {\bibfnamefont {D.~R.}\ \bibnamefont
  {Schmid}}, \bibinfo {author} {\bibfnamefont {S.}~\bibnamefont {Smirnov}},
  \bibinfo {author} {\bibfnamefont {M.}~\bibnamefont
  {Marga\ifmmode~\acute{n}\else \'{n}\fi{}ska}}, \bibinfo {author}
  {\bibfnamefont {A.}~\bibnamefont {Dirnaichner}}, \bibinfo {author}
  {\bibfnamefont {P.~L.}\ \bibnamefont {Stiller}}, \bibinfo {author}
  {\bibfnamefont {M.}~\bibnamefont {Grifoni}}, \bibinfo {author} {\bibfnamefont
  {A.~K.}\ \bibnamefont {H\"uttel}}, \ and\ \bibinfo {author} {\bibfnamefont
  {C.}~\bibnamefont {Strunk}},\ }\bibfield  {title} {\enquote {\bibinfo {title}
  {Broken {SU}(4) symmetry in a {K}ondo-correlated carbon nanotube},}\
  }\href@noop {} {\bibfield  {journal} {\bibinfo  {journal} {Phys. Rev. B}\
  }\textbf {\bibinfo {volume} {91}},\ \bibinfo {pages} {155435} (\bibinfo
  {year} {2015})}\BibitemShut {NoStop}%
\bibitem [{\citenamefont {Ferrier}\ \emph {et~al.}(2016)\citenamefont
  {Ferrier}, \citenamefont {Arakawa}, \citenamefont {Hata}, \citenamefont
  {Fujiwara}, \citenamefont {Delagrange}, \citenamefont {Weil}, \citenamefont
  {Deblock}, \citenamefont {Sakano}, \citenamefont {Oguri},\ and\ \citenamefont
  {Kobayashi}}]{Ferrier:2016aa}%
  \BibitemOpen
  \bibfield  {author} {\bibinfo {author} {\bibfnamefont {M.}~\bibnamefont
  {Ferrier}}, \bibinfo {author} {\bibfnamefont {T.}~\bibnamefont {Arakawa}},
  \bibinfo {author} {\bibfnamefont {T.}~\bibnamefont {Hata}}, \bibinfo {author}
  {\bibfnamefont {R.}~\bibnamefont {Fujiwara}}, \bibinfo {author}
  {\bibfnamefont {R.}~\bibnamefont {Delagrange}}, \bibinfo {author}
  {\bibfnamefont {R.}~\bibnamefont {Weil}}, \bibinfo {author} {\bibfnamefont
  {R.}~\bibnamefont {Deblock}}, \bibinfo {author} {\bibfnamefont
  {R.}~\bibnamefont {Sakano}}, \bibinfo {author} {\bibfnamefont
  {A.}~\bibnamefont {Oguri}}, \ and\ \bibinfo {author} {\bibfnamefont
  {K.}~\bibnamefont {Kobayashi}},\ }\bibfield  {title} {\enquote {\bibinfo
  {title} {Universality of non-equilibrium fluctuations in strongly correlated
  quantum liquids},}\ }\href {http://dx.doi.org/10.1038/nphys3556} {\bibfield
  {journal} {\bibinfo  {journal} {Nat Phys}\ }\textbf {\bibinfo {volume}
  {12}},\ \bibinfo {pages} {230--235} (\bibinfo {year} {2016})}\BibitemShut
  {NoStop}%
\bibitem [{\citenamefont {Jarillo-Herrero}\ \emph {et~al.}(2004)\citenamefont
  {Jarillo-Herrero}, \citenamefont {Sapmaz}, \citenamefont {Dekker},
  \citenamefont {Kouwenhoven},\ and\ \citenamefont {van~der
  Zant}}]{Jarillo-Herrero-2004-05}%
  \BibitemOpen
  \bibfield  {author} {\bibinfo {author} {\bibfnamefont {P.}~\bibnamefont
  {Jarillo-Herrero}}, \bibinfo {author} {\bibfnamefont {S.}~\bibnamefont
  {Sapmaz}}, \bibinfo {author} {\bibfnamefont {C.}~\bibnamefont {Dekker}},
  \bibinfo {author} {\bibfnamefont {L.~P.}\ \bibnamefont {Kouwenhoven}}, \ and\
  \bibinfo {author} {\bibfnamefont {H.~S.~J.}\ \bibnamefont {van~der Zant}},\
  }\bibfield  {title} {\enquote {\bibinfo {title} {Electron-hole symmetry in a
  semiconducting carbon nanotube quantum dot},}\ }\href@noop {} {\bibfield
  {journal} {\bibinfo  {journal} {Nature}\ }\textbf {\bibinfo {volume} {429}},\
  \bibinfo {pages} {389} (\bibinfo {year} {2004})}\BibitemShut {NoStop}%
\bibitem [{\citenamefont {Deshpande}\ and\ \citenamefont
  {Bockrath}(2008)}]{Deshpande-2008-04}%
  \BibitemOpen
  \bibfield  {author} {\bibinfo {author} {\bibfnamefont {V.~V.}\ \bibnamefont
  {Deshpande}}\ and\ \bibinfo {author} {\bibfnamefont {M.}~\bibnamefont
  {Bockrath}},\ }\bibfield  {title} {\enquote {\bibinfo {title} {The
  one-dimensional {W}igner crystal in carbon nanotubes},}\ }\href@noop {}
  {\bibfield  {journal} {\bibinfo  {journal} {Nat. Phys.}\ }\textbf {\bibinfo
  {volume} {4}},\ \bibinfo {pages} {314} (\bibinfo {year} {2008})}\BibitemShut
  {NoStop}%
\bibitem [{\citenamefont {Saito}\ \emph {et~al.}(1998)\citenamefont {Saito},
  \citenamefont {Dresselhaus},\ and\ \citenamefont {Dresselhaus}}]{Saito-1998}%
  \BibitemOpen
  \bibfield  {author} {\bibinfo {author} {\bibfnamefont {R.}~\bibnamefont
  {Saito}}, \bibinfo {author} {\bibfnamefont {G.}~\bibnamefont {Dresselhaus}},
  \ and\ \bibinfo {author} {\bibfnamefont {M.~S.}\ \bibnamefont
  {Dresselhaus}},\ }\href@noop {} {\emph {\bibinfo {title} {Physical Properties
  of Carbon Nanotubes}}}\ (\bibinfo  {publisher} {Imperial College Press},\
  \bibinfo {address} {London},\ \bibinfo {year} {1998})\BibitemShut {NoStop}%
\bibitem [{\citenamefont {Samsonidze}\ \emph {et~al.}(2003)\citenamefont
  {Samsonidze}, \citenamefont {Saito}, \citenamefont {Jorio}, \citenamefont
  {Pimenta}, \citenamefont {Souza~Filho}, \citenamefont {Gr{\"u}neis},
  \citenamefont {Dresselhaus},\ and\ \citenamefont
  {Dresselhaus}}]{Samsonidze:2003-12-01T00:00:00:1533-4880:431}%
  \BibitemOpen
  \bibfield  {author} {\bibinfo {author} {\bibfnamefont {Ge.~G.}\ \bibnamefont
  {Samsonidze}}, \bibinfo {author} {\bibfnamefont {R.}~\bibnamefont {Saito}},
  \bibinfo {author} {\bibfnamefont {A.}~\bibnamefont {Jorio}}, \bibinfo
  {author} {\bibfnamefont {M.~A.}\ \bibnamefont {Pimenta}}, \bibinfo {author}
  {\bibfnamefont {A.~G.}\ \bibnamefont {Souza~Filho}}, \bibinfo {author}
  {\bibfnamefont {A.}~\bibnamefont {Gr{\"u}neis}}, \bibinfo {author}
  {\bibfnamefont {G.}~\bibnamefont {Dresselhaus}}, \ and\ \bibinfo {author}
  {\bibfnamefont {M.~S.}\ \bibnamefont {Dresselhaus}},\ }\bibfield  {title}
  {\enquote {\bibinfo {title} {The concept of cutting lines in carbon nanotube
  science},}\ }\href@noop {} {\bibfield  {journal} {\bibinfo  {journal} {J.
  Nanosci. Nanotechnol.}\ }\textbf {\bibinfo {volume} {3}},\ \bibinfo {pages}
  {431--458} (\bibinfo {year} {2003})}\BibitemShut {NoStop}%
\bibitem [{\citenamefont {Barros}\ \emph {et~al.}(2006)\citenamefont {Barros},
  \citenamefont {Jorio}, \citenamefont {Samsonidze}, \citenamefont {Capaz},
  \citenamefont {Filho}, \citenamefont {Filho}, \citenamefont {Dresselhaus},\
  and\ \citenamefont {Dresselhaus}}]{Barros-2006-09}%
  \BibitemOpen
  \bibfield  {author} {\bibinfo {author} {\bibfnamefont {E.~B.}\ \bibnamefont
  {Barros}}, \bibinfo {author} {\bibfnamefont {A.}~\bibnamefont {Jorio}},
  \bibinfo {author} {\bibfnamefont {G.~G.}\ \bibnamefont {Samsonidze}},
  \bibinfo {author} {\bibfnamefont {R.~B.}\ \bibnamefont {Capaz}}, \bibinfo
  {author} {\bibfnamefont {A.~G.~Souza}\ \bibnamefont {Filho}}, \bibinfo
  {author} {\bibfnamefont {J.~M.}\ \bibnamefont {Filho}}, \bibinfo {author}
  {\bibfnamefont {G.}~\bibnamefont {Dresselhaus}}, \ and\ \bibinfo {author}
  {\bibfnamefont {M.~S.}\ \bibnamefont {Dresselhaus}},\ }\bibfield  {title}
  {\enquote {\bibinfo {title} {Review on the symmetry-related properties of
  carbon nanotubes},}\ }\href@noop {} {\bibfield  {journal} {\bibinfo
  {journal} {Phys. Rep.}\ }\textbf {\bibinfo {volume} {431}},\ \bibinfo {pages}
  {261} (\bibinfo {year} {2006})}\BibitemShut {NoStop}%
\bibitem [{\citenamefont {Samsonidze}\ \emph {et~al.}(2004)\citenamefont
  {Samsonidze}, \citenamefont {Gr\"uneis}, \citenamefont {Saito}, \citenamefont
  {Jorio}, \citenamefont {Souza~Filho}, \citenamefont {Dresselhaus},\ and\
  \citenamefont {Dresselhaus}}]{PhysRevB.69.205402}%
  \BibitemOpen
  \bibfield  {author} {\bibinfo {author} {\bibfnamefont {Ge.~G.}\ \bibnamefont
  {Samsonidze}}, \bibinfo {author} {\bibfnamefont {A.}~\bibnamefont
  {Gr\"uneis}}, \bibinfo {author} {\bibfnamefont {R.}~\bibnamefont {Saito}},
  \bibinfo {author} {\bibfnamefont {A.}~\bibnamefont {Jorio}}, \bibinfo
  {author} {\bibfnamefont {A.~G.}\ \bibnamefont {Souza~Filho}}, \bibinfo
  {author} {\bibfnamefont {G.}~\bibnamefont {Dresselhaus}}, \ and\ \bibinfo
  {author} {\bibfnamefont {M.~S.}\ \bibnamefont {Dresselhaus}},\ }\bibfield
  {title} {\enquote {\bibinfo {title} {Interband optical transitions in left-
  and right-handed single-wall carbon nanotubes},}\ }\href@noop {} {\bibfield
  {journal} {\bibinfo  {journal} {Phys. Rev. B}\ }\textbf {\bibinfo {volume}
  {69}},\ \bibinfo {pages} {205402} (\bibinfo {year} {2004})}\BibitemShut
  {NoStop}%
\bibitem [{\citenamefont {White}\ \emph {et~al.}(1993)\citenamefont {White},
  \citenamefont {Robertson},\ and\ \citenamefont {Mintmire}}]{White-1993-03}%
  \BibitemOpen
  \bibfield  {author} {\bibinfo {author} {\bibfnamefont {C.~T.}\ \bibnamefont
  {White}}, \bibinfo {author} {\bibfnamefont {D.~H.}\ \bibnamefont
  {Robertson}}, \ and\ \bibinfo {author} {\bibfnamefont {J.~W.}\ \bibnamefont
  {Mintmire}},\ }\bibfield  {title} {\enquote {\bibinfo {title} {Helical and
  rotational symmetries of nanoscale graphitic tubules},}\ }\href@noop {}
  {\bibfield  {journal} {\bibinfo  {journal} {Phys. Rev. B}\ }\textbf {\bibinfo
  {volume} {47}},\ \bibinfo {pages} {5485} (\bibinfo {year}
  {1993})}\BibitemShut {NoStop}%
\bibitem [{\citenamefont {Jishi}\ \emph {et~al.}(1993)\citenamefont {Jishi},
  \citenamefont {Dresselhaus},\ and\ \citenamefont
  {Dresselhaus}}]{PhysRevB.47.16671}%
  \BibitemOpen
  \bibfield  {author} {\bibinfo {author} {\bibfnamefont {R.~A.}\ \bibnamefont
  {Jishi}}, \bibinfo {author} {\bibfnamefont {M.~S.}\ \bibnamefont
  {Dresselhaus}}, \ and\ \bibinfo {author} {\bibfnamefont {G.}~\bibnamefont
  {Dresselhaus}},\ }\bibfield  {title} {\enquote {\bibinfo {title} {Symmetry
  properties of chiral carbon nanotubes},}\ }\href@noop {} {\bibfield
  {journal} {\bibinfo  {journal} {Phys. Rev. B}\ }\textbf {\bibinfo {volume}
  {47}},\ \bibinfo {pages} {16671} (\bibinfo {year} {1993})}\BibitemShut
  {NoStop}%
\bibitem [{\citenamefont {Jorio}\ \emph {et~al.}(2005)\citenamefont {Jorio},
  \citenamefont {Fantini}, \citenamefont {Pimenta}, \citenamefont {Capaz},
  \citenamefont {Samsonidze}, \citenamefont {Dresselhaus}, \citenamefont
  {Dresselhaus}, \citenamefont {Jiang}, \citenamefont {Kobayashi},
  \citenamefont {Gr{\"u}neis},\ and\ \citenamefont {Saito}}]{Jorio-2005-02}%
  \BibitemOpen
  \bibfield  {author} {\bibinfo {author} {\bibfnamefont {A.}~\bibnamefont
  {Jorio}}, \bibinfo {author} {\bibfnamefont {C.}~\bibnamefont {Fantini}},
  \bibinfo {author} {\bibfnamefont {M.~A.}\ \bibnamefont {Pimenta}}, \bibinfo
  {author} {\bibfnamefont {R.~B.}\ \bibnamefont {Capaz}}, \bibinfo {author}
  {\bibfnamefont {Ge.~G.}\ \bibnamefont {Samsonidze}}, \bibinfo {author}
  {\bibfnamefont {G.}~\bibnamefont {Dresselhaus}}, \bibinfo {author}
  {\bibfnamefont {M.~S.}\ \bibnamefont {Dresselhaus}}, \bibinfo {author}
  {\bibfnamefont {J.}~\bibnamefont {Jiang}}, \bibinfo {author} {\bibfnamefont
  {N.}~\bibnamefont {Kobayashi}}, \bibinfo {author} {\bibfnamefont
  {A.}~\bibnamefont {Gr{\"u}neis}}, \ and\ \bibinfo {author} {\bibfnamefont
  {R.}~\bibnamefont {Saito}},\ }\bibfield  {title} {\enquote {\bibinfo {title}
  {Resonance {R}aman spectroscopy $(n,m)$-dependent effects in small-diameter
  single-wall carbon nanotubes},}\ }\href@noop {} {\bibfield  {journal}
  {\bibinfo  {journal} {Phys. Rev. B}\ }\textbf {\bibinfo {volume} {71}},\
  \bibinfo {pages} {075401} (\bibinfo {year} {2005})}\BibitemShut {NoStop}%
\bibitem [{\citenamefont {Akhmerov}\ and\ \citenamefont
  {Beenakker}(2008)}]{Akhmerov-2008-02}%
  \BibitemOpen
  \bibfield  {author} {\bibinfo {author} {\bibfnamefont {A.~R.}\ \bibnamefont
  {Akhmerov}}\ and\ \bibinfo {author} {\bibfnamefont {C.~W.~J.}\ \bibnamefont
  {Beenakker}},\ }\bibfield  {title} {\enquote {\bibinfo {title} {Boundary
  conditions for {D}irac fermions on a terminated honeycomb lattice},}\
  }\href@noop {} {\bibfield  {journal} {\bibinfo  {journal} {Phys. Rev. B}\
  }\textbf {\bibinfo {volume} {77}},\ \bibinfo {pages} {085423} (\bibinfo
  {year} {2008})}\BibitemShut {NoStop}%
\bibitem [{\citenamefont {Porezag}\ \emph {et~al.}(1995)\citenamefont
  {Porezag}, \citenamefont {Frauenheim}, \citenamefont {K{\"o}hler},
  \citenamefont {Seifert},\ and\ \citenamefont {Kaschner}}]{Porezag-1995-05}%
  \BibitemOpen
  \bibfield  {author} {\bibinfo {author} {\bibfnamefont {D.}~\bibnamefont
  {Porezag}}, \bibinfo {author} {\bibfnamefont {Th.}\ \bibnamefont
  {Frauenheim}}, \bibinfo {author} {\bibfnamefont {Th.}\ \bibnamefont
  {K{\"o}hler}}, \bibinfo {author} {\bibfnamefont {G.}~\bibnamefont {Seifert}},
  \ and\ \bibinfo {author} {\bibfnamefont {R.}~\bibnamefont {Kaschner}},\
  }\bibfield  {title} {\enquote {\bibinfo {title} {Construction of
  tight-binding-like potentials on the basis of density-functional theory:
  Application to carbon},}\ }\href@noop {} {\bibfield  {journal} {\bibinfo
  {journal} {Phys. Rev. B}\ }\textbf {\bibinfo {volume} {51}},\ \bibinfo
  {pages} {12947} (\bibinfo {year} {1995})}\BibitemShut {NoStop}%
\bibitem [{\citenamefont {Fujita}\ \emph {et~al.}(1996)\citenamefont {Fujita},
  \citenamefont {Wakabayashi}, \citenamefont {Nakada},\ and\ \citenamefont
  {Kusakabe}}]{Fujita-1996}%
  \BibitemOpen
  \bibfield  {author} {\bibinfo {author} {\bibfnamefont {M.}~\bibnamefont
  {Fujita}}, \bibinfo {author} {\bibfnamefont {K.}~\bibnamefont {Wakabayashi}},
  \bibinfo {author} {\bibfnamefont {K.}~\bibnamefont {Nakada}}, \ and\ \bibinfo
  {author} {\bibfnamefont {K.}~\bibnamefont {Kusakabe}},\ }\bibfield  {title}
  {\enquote {\bibinfo {title} {Peculiar localized state at zigzag graphite
  edge},}\ }\href@noop {} {\bibfield  {journal} {\bibinfo  {journal} {J. Phys.
  Soc. Jpn.}\ }\textbf {\bibinfo {volume} {65}},\ \bibinfo {pages} {1920}
  (\bibinfo {year} {1996})}\BibitemShut {NoStop}%
\bibitem [{\citenamefont {Klusek}\ \emph {et~al.}(2000)\citenamefont {Klusek},
  \citenamefont {Waqar}, \citenamefont {Denisov}, \citenamefont {Kompaniets},
  \citenamefont {Makarenko}, \citenamefont {Titkov},\ and\ \citenamefont
  {Bhatti}}]{Klusek2000508}%
  \BibitemOpen
  \bibfield  {author} {\bibinfo {author} {\bibfnamefont {Z.}~\bibnamefont
  {Klusek}}, \bibinfo {author} {\bibfnamefont {Z.}~\bibnamefont {Waqar}},
  \bibinfo {author} {\bibfnamefont {E.~A.}\ \bibnamefont {Denisov}}, \bibinfo
  {author} {\bibfnamefont {T.~N.}\ \bibnamefont {Kompaniets}}, \bibinfo
  {author} {\bibfnamefont {I.~V.}\ \bibnamefont {Makarenko}}, \bibinfo {author}
  {\bibfnamefont {A.~N.}\ \bibnamefont {Titkov}}, \ and\ \bibinfo {author}
  {\bibfnamefont {A.~S.}\ \bibnamefont {Bhatti}},\ }\bibfield  {title}
  {\enquote {\bibinfo {title} {Observations of local electron states on the
  edges of the circular pits on hydrogen-etched graphite surface by scanning
  tunneling spectroscopy},}\ }\href {\doibase
  http://dx.doi.org/10.1016/S0169-4332(00)00374-3} {\bibfield  {journal}
  {\bibinfo  {journal} {Appl. Surf. Sci.}\ }\textbf {\bibinfo {volume} {161}},\
  \bibinfo {pages} {508} (\bibinfo {year} {2000})}\BibitemShut {NoStop}%
\bibitem [{\citenamefont {Kobayashi}\ \emph {et~al.}(2005)\citenamefont
  {Kobayashi}, \citenamefont {Fukui}, \citenamefont {Enoki}, \citenamefont
  {Kusakabe},\ and\ \citenamefont {Kaburagi}}]{PhysRevB.71.193406}%
  \BibitemOpen
  \bibfield  {author} {\bibinfo {author} {\bibfnamefont {Y.}~\bibnamefont
  {Kobayashi}}, \bibinfo {author} {\bibfnamefont {K.-I.}\ \bibnamefont
  {Fukui}}, \bibinfo {author} {\bibfnamefont {T.}~\bibnamefont {Enoki}},
  \bibinfo {author} {\bibfnamefont {K.}~\bibnamefont {Kusakabe}}, \ and\
  \bibinfo {author} {\bibfnamefont {Y.}~\bibnamefont {Kaburagi}},\ }\bibfield
  {title} {\enquote {\bibinfo {title} {Observation of zigzag and armchair edges
  of graphite using scanning tunneling microscopy and spectroscopy},}\ }\href
  {\doibase 10.1103/PhysRevB.71.193406} {\bibfield  {journal} {\bibinfo
  {journal} {Phys. Rev. B}\ }\textbf {\bibinfo {volume} {71}},\ \bibinfo
  {pages} {193406} (\bibinfo {year} {2005})}\BibitemShut {NoStop}%
\bibitem [{\citenamefont {Niimi}\ \emph {et~al.}(2006)\citenamefont {Niimi},
  \citenamefont {Matsui}, \citenamefont {Kambara}, \citenamefont {Tagami},
  \citenamefont {Tsukada},\ and\ \citenamefont
  {Fukuyama}}]{PhysRevB.73.085421}%
  \BibitemOpen
  \bibfield  {author} {\bibinfo {author} {\bibfnamefont {Y.}~\bibnamefont
  {Niimi}}, \bibinfo {author} {\bibfnamefont {T.}~\bibnamefont {Matsui}},
  \bibinfo {author} {\bibfnamefont {H.}~\bibnamefont {Kambara}}, \bibinfo
  {author} {\bibfnamefont {K.}~\bibnamefont {Tagami}}, \bibinfo {author}
  {\bibfnamefont {M.}~\bibnamefont {Tsukada}}, \ and\ \bibinfo {author}
  {\bibfnamefont {Hiroshi}\ \bibnamefont {Fukuyama}},\ }\bibfield  {title}
  {\enquote {\bibinfo {title} {Scanning tunneling microscopy and spectroscopy
  of the electronic local density of states of graphite surfaces near
  monoatomic step edges},}\ }\href {\doibase 10.1103/PhysRevB.73.085421}
  {\bibfield  {journal} {\bibinfo  {journal} {Phys. Rev. B}\ }\textbf {\bibinfo
  {volume} {73}},\ \bibinfo {pages} {085421} (\bibinfo {year}
  {2006})}\BibitemShut {NoStop}%
\bibitem [{\citenamefont {Sasaki}\ \emph {et~al.}(2006)\citenamefont {Sasaki},
  \citenamefont {Murakami},\ and\ \citenamefont {Saito}}]{Sasaki-2006-11}%
  \BibitemOpen
  \bibfield  {author} {\bibinfo {author} {\bibfnamefont {K.}~\bibnamefont
  {Sasaki}}, \bibinfo {author} {\bibfnamefont {S.}~\bibnamefont {Murakami}}, \
  and\ \bibinfo {author} {\bibfnamefont {R.}~\bibnamefont {Saito}},\ }\bibfield
   {title} {\enquote {\bibinfo {title} {Stabilization mechanism of edge states
  in graphene},}\ }\href {\doibase http://dx.doi.org/10.1063/1.2181274}
  {\bibfield  {journal} {\bibinfo  {journal} {Appl. Phys. Lett.}\ }\textbf
  {\bibinfo {volume} {88}},\ \bibinfo {pages} {13110} (\bibinfo {year}
  {2006})}\BibitemShut {NoStop}%
\bibitem [{\citenamefont {Izumida}\ \emph {et~al.}(2012)\citenamefont
  {Izumida}, \citenamefont {Vikstr\"om},\ and\ \citenamefont
  {Saito}}]{Izumida-2012-04}%
  \BibitemOpen
  \bibfield  {author} {\bibinfo {author} {\bibfnamefont {W.}~\bibnamefont
  {Izumida}}, \bibinfo {author} {\bibfnamefont {A.}~\bibnamefont {Vikstr\"om}},
  \ and\ \bibinfo {author} {\bibfnamefont {R.}~\bibnamefont {Saito}},\
  }\bibfield  {title} {\enquote {\bibinfo {title} {Asymmetric velocities of
  {D}irac particles and vernier spectrum in metallic single-wall carbon
  nanotubes},}\ }\href {\doibase 10.1103/PhysRevB.85.165430} {\bibfield
  {journal} {\bibinfo  {journal} {Phys. Rev. B}\ }\textbf {\bibinfo {volume}
  {85}},\ \bibinfo {pages} {165430} (\bibinfo {year} {2012})}\BibitemShut
  {NoStop}%
\bibitem [{\citenamefont {Wen}\ and\ \citenamefont {Zee}(1989)}]{Wen1989641}%
  \BibitemOpen
  \bibfield  {author} {\bibinfo {author} {\bibfnamefont {X.G.}\ \bibnamefont
  {Wen}}\ and\ \bibinfo {author} {\bibfnamefont {A.}~\bibnamefont {Zee}},\
  }\bibfield  {title} {\enquote {\bibinfo {title} {Winding number, family index
  theorem, and electron hopping in a magnetic field},}\ }\href {\doibase
  http://dx.doi.org/10.1016/0550-3213(89)90062-X} {\bibfield  {journal}
  {\bibinfo  {journal} {Nucl. Phys. B}\ }\textbf {\bibinfo {volume} {316}},\
  \bibinfo {pages} {641} (\bibinfo {year} {1989})}\BibitemShut {NoStop}%
\bibitem [{\citenamefont {Ryu}\ and\ \citenamefont
  {Hatsugai}(2002)}]{PhysRevLett.89.077002}%
  \BibitemOpen
  \bibfield  {author} {\bibinfo {author} {\bibfnamefont {S.}~\bibnamefont
  {Ryu}}\ and\ \bibinfo {author} {\bibfnamefont {Y.}~\bibnamefont {Hatsugai}},\
  }\bibfield  {title} {\enquote {\bibinfo {title} {Topological origin of
  zero-energy edge states in particle-hole symmetric systems},}\ }\href
  {\doibase 10.1103/PhysRevLett.89.077002} {\bibfield  {journal} {\bibinfo
  {journal} {Phys. Rev. Lett.}\ }\textbf {\bibinfo {volume} {89}},\ \bibinfo
  {pages} {077002} (\bibinfo {year} {2002})}\BibitemShut {NoStop}%
\bibitem [{\citenamefont {Sasaki}\ \emph {et~al.}(2005)\citenamefont {Sasaki},
  \citenamefont {Murakami}, \citenamefont {Saito},\ and\ \citenamefont
  {Kawazoe}}]{PhysRevB.71.195401}%
  \BibitemOpen
  \bibfield  {author} {\bibinfo {author} {\bibfnamefont {K.}~\bibnamefont
  {Sasaki}}, \bibinfo {author} {\bibfnamefont {S.}~\bibnamefont {Murakami}},
  \bibinfo {author} {\bibfnamefont {R.}~\bibnamefont {Saito}}, \ and\ \bibinfo
  {author} {\bibfnamefont {Y.}~\bibnamefont {Kawazoe}},\ }\bibfield  {title}
  {\enquote {\bibinfo {title} {Controlling edge states of zigzag carbon
  nanotubes by the {A}haronov-{B}ohm flux},}\ }\href@noop {} {\bibfield
  {journal} {\bibinfo  {journal} {Phys. Rev. B}\ }\textbf {\bibinfo {volume}
  {71}},\ \bibinfo {pages} {195401} (\bibinfo {year} {2005})}\BibitemShut
  {NoStop}%
\bibitem [{\citenamefont {Schnyder}\ \emph {et~al.}(2008)\citenamefont
  {Schnyder}, \citenamefont {Ryu}, \citenamefont {Furusaki},\ and\
  \citenamefont {Ludwig}}]{PhysRevB.78.195125}%
  \BibitemOpen
  \bibfield  {author} {\bibinfo {author} {\bibfnamefont {A.~P.}\ \bibnamefont
  {Schnyder}}, \bibinfo {author} {\bibfnamefont {S.}~\bibnamefont {Ryu}},
  \bibinfo {author} {\bibfnamefont {A.}~\bibnamefont {Furusaki}}, \ and\
  \bibinfo {author} {\bibfnamefont {A.~W.~W.}\ \bibnamefont {Ludwig}},\
  }\bibfield  {title} {\enquote {\bibinfo {title} {Classification of
  topological insulators and superconductors in three spatial dimensions},}\
  }\href {\doibase 10.1103/PhysRevB.78.195125} {\bibfield  {journal} {\bibinfo
  {journal} {Phys. Rev. B}\ }\textbf {\bibinfo {volume} {78}},\ \bibinfo
  {pages} {195125} (\bibinfo {year} {2008})}\BibitemShut {NoStop}%
\bibitem [{\citenamefont {Kitaev}(2009)}]{Kitaev-AIPConfProc2009}%
  \BibitemOpen
  \bibfield  {author} {\bibinfo {author} {\bibfnamefont {A.}~\bibnamefont
  {Kitaev}},\ }\bibfield  {title} {\enquote {\bibinfo {title} {Periodic table
  for topological insulators and superconductors},}\ }\href {\doibase
  http://dx.doi.org/10.1063/1.3149495} {\bibfield  {journal} {\bibinfo
  {journal} {AIP Conf. Proc.}\ }\textbf {\bibinfo {volume} {1134}},\ \bibinfo
  {pages} {22} (\bibinfo {year} {2009})}\BibitemShut {NoStop}%
\bibitem [{\citenamefont {Ryu}\ \emph {et~al.}(2010)\citenamefont {Ryu},
  \citenamefont {Schnyder}, \citenamefont {Furusaki},\ and\ \citenamefont
  {Ludwig}}]{Ryu-NJP-2010}%
  \BibitemOpen
  \bibfield  {author} {\bibinfo {author} {\bibfnamefont {S.}~\bibnamefont
  {Ryu}}, \bibinfo {author} {\bibfnamefont {A.~P.}\ \bibnamefont {Schnyder}},
  \bibinfo {author} {\bibfnamefont {A.}~\bibnamefont {Furusaki}}, \ and\
  \bibinfo {author} {\bibfnamefont {A.~W.~W.}\ \bibnamefont {Ludwig}},\
  }\bibfield  {title} {\enquote {\bibinfo {title} {Topological insulators and
  superconductors: tenfold way and dimensional hierarchy},}\ }\href
  {http://stacks.iop.org/1367-2630/12/i=6/a=065010} {\bibfield  {journal}
  {\bibinfo  {journal} {New J. Phys.}\ }\textbf {\bibinfo {volume} {12}},\
  \bibinfo {pages} {065010} (\bibinfo {year} {2010})}\BibitemShut {NoStop}%
\bibitem [{\citenamefont {Thouless}\ \emph {et~al.}(1982)\citenamefont
  {Thouless}, \citenamefont {Kohmoto}, \citenamefont {Nightingale},\ and\
  \citenamefont {den Nijs}}]{PhysRevLett.49.405}%
  \BibitemOpen
  \bibfield  {author} {\bibinfo {author} {\bibfnamefont {D.~J.}\ \bibnamefont
  {Thouless}}, \bibinfo {author} {\bibfnamefont {M.}~\bibnamefont {Kohmoto}},
  \bibinfo {author} {\bibfnamefont {M.~P.}\ \bibnamefont {Nightingale}}, \ and\
  \bibinfo {author} {\bibfnamefont {M.}~\bibnamefont {den Nijs}},\ }\bibfield
  {title} {\enquote {\bibinfo {title} {Quantized {H}all conductance in a
  two-dimensional periodic potential},}\ }\href {\doibase
  10.1103/PhysRevLett.49.405} {\bibfield  {journal} {\bibinfo  {journal} {Phys.
  Rev. Lett.}\ }\textbf {\bibinfo {volume} {49}},\ \bibinfo {pages} {405}
  (\bibinfo {year} {1982})}\BibitemShut {NoStop}%
\bibitem [{\citenamefont {Kohmoto}(1985)}]{KOHMOTO1985343}%
  \BibitemOpen
  \bibfield  {author} {\bibinfo {author} {\bibfnamefont {M.}~\bibnamefont
  {Kohmoto}},\ }\bibfield  {title} {\enquote {\bibinfo {title} {Topological
  invariant and the quantization of the {H}all conductance},}\ }\href {\doibase
  http://dx.doi.org/10.1016/0003-4916(85)90148-4} {\bibfield  {journal}
  {\bibinfo  {journal} {Ann. Phys.}\ }\textbf {\bibinfo {volume} {160}},\
  \bibinfo {pages} {343} (\bibinfo {year} {1985})}\BibitemShut {NoStop}%
\bibitem [{\citenamefont {Laughlin}(1981)}]{PhysRevB.23.5632}%
  \BibitemOpen
  \bibfield  {author} {\bibinfo {author} {\bibfnamefont {R.~B.}\ \bibnamefont
  {Laughlin}},\ }\bibfield  {title} {\enquote {\bibinfo {title} {Quantized
  {H}all conductivity in two dimensions},}\ }\href {\doibase
  10.1103/PhysRevB.23.5632} {\bibfield  {journal} {\bibinfo  {journal} {Phys.
  Rev. B}\ }\textbf {\bibinfo {volume} {23}},\ \bibinfo {pages} {5632--5633}
  (\bibinfo {year} {1981})}\BibitemShut {NoStop}%
\bibitem [{\citenamefont {Halperin}(1982)}]{PhysRevB.25.2185}%
  \BibitemOpen
  \bibfield  {author} {\bibinfo {author} {\bibfnamefont {B.~I.}\ \bibnamefont
  {Halperin}},\ }\bibfield  {title} {\enquote {\bibinfo {title} {Quantized
  {H}all conductance, current-carrying edge states, and the existence of
  extended states in a two-dimensional disordered potential},}\ }\href
  {\doibase 10.1103/PhysRevB.25.2185} {\bibfield  {journal} {\bibinfo
  {journal} {Phys. Rev. B}\ }\textbf {\bibinfo {volume} {25}},\ \bibinfo
  {pages} {2185} (\bibinfo {year} {1982})}\BibitemShut {NoStop}%
\bibitem [{\citenamefont {Hatsugai}(1993{\natexlab{a}})}]{PhysRevB.48.11851}%
  \BibitemOpen
  \bibfield  {author} {\bibinfo {author} {\bibfnamefont {Y.}~\bibnamefont
  {Hatsugai}},\ }\bibfield  {title} {\enquote {\bibinfo {title} {Edge states in
  the integer quantum {H}all effect and the {R}iemann surface of the {B}loch
  function},}\ }\href {\doibase 10.1103/PhysRevB.48.11851} {\bibfield
  {journal} {\bibinfo  {journal} {Phys. Rev. B}\ }\textbf {\bibinfo {volume}
  {48}},\ \bibinfo {pages} {11851} (\bibinfo {year}
  {1993}{\natexlab{a}})}\BibitemShut {NoStop}%
\bibitem [{\citenamefont {Hatsugai}(1993{\natexlab{b}})}]{PhysRevLett.71.3697}%
  \BibitemOpen
  \bibfield  {author} {\bibinfo {author} {\bibfnamefont {Y.}~\bibnamefont
  {Hatsugai}},\ }\bibfield  {title} {\enquote {\bibinfo {title} {{C}hern number
  and edge states in the integer quantum {H}all effect},}\ }\href {\doibase
  10.1103/PhysRevLett.71.3697} {\bibfield  {journal} {\bibinfo  {journal}
  {Phys. Rev. Lett.}\ }\textbf {\bibinfo {volume} {71}},\ \bibinfo {pages}
  {3697} (\bibinfo {year} {1993}{\natexlab{b}})}\BibitemShut {NoStop}%
\bibitem [{\citenamefont {Gurarie}(2011)}]{PhysRevB.83.085426}%
  \BibitemOpen
  \bibfield  {author} {\bibinfo {author} {\bibfnamefont {V.}~\bibnamefont
  {Gurarie}},\ }\bibfield  {title} {\enquote {\bibinfo {title} {Single-particle
  {G}reen's functions and interacting topological insulators},}\ }\href
  {\doibase 10.1103/PhysRevB.83.085426} {\bibfield  {journal} {\bibinfo
  {journal} {Phys. Rev. B}\ }\textbf {\bibinfo {volume} {83}},\ \bibinfo
  {pages} {085426} (\bibinfo {year} {2011})}\BibitemShut {NoStop}%
\bibitem [{\citenamefont {Sato}\ \emph {et~al.}(2011)\citenamefont {Sato},
  \citenamefont {Tanaka}, \citenamefont {Yada},\ and\ \citenamefont
  {Yokoyama}}]{PhysRevB.83.224511}%
  \BibitemOpen
  \bibfield  {author} {\bibinfo {author} {\bibfnamefont {M.}~\bibnamefont
  {Sato}}, \bibinfo {author} {\bibfnamefont {Y.}~\bibnamefont {Tanaka}},
  \bibinfo {author} {\bibfnamefont {K.}~\bibnamefont {Yada}}, \ and\ \bibinfo
  {author} {\bibfnamefont {T.}~\bibnamefont {Yokoyama}},\ }\bibfield  {title}
  {\enquote {\bibinfo {title} {Topology of {A}ndreev bound states with flat
  dispersion},}\ }\href {\doibase 10.1103/PhysRevB.83.224511} {\bibfield
  {journal} {\bibinfo  {journal} {Phys. Rev. B}\ }\textbf {\bibinfo {volume}
  {83}},\ \bibinfo {pages} {224511} (\bibinfo {year} {2011})}\BibitemShut
  {NoStop}%
\bibitem [{\citenamefont {Essin}\ and\ \citenamefont
  {Gurarie}(2011)}]{PhysRevB.84.125132}%
  \BibitemOpen
  \bibfield  {author} {\bibinfo {author} {\bibfnamefont {A.~M.}\ \bibnamefont
  {Essin}}\ and\ \bibinfo {author} {\bibfnamefont {V.}~\bibnamefont
  {Gurarie}},\ }\bibfield  {title} {\enquote {\bibinfo {title} {Bulk-boundary
  correspondence of topological insulators from their respective {G}reen's
  functions},}\ }\href {\doibase 10.1103/PhysRevB.84.125132} {\bibfield
  {journal} {\bibinfo  {journal} {Phys. Rev. B}\ }\textbf {\bibinfo {volume}
  {84}},\ \bibinfo {pages} {125132} (\bibinfo {year} {2011})}\BibitemShut
  {NoStop}%
\bibitem [{\citenamefont {Liu}\ \emph {et~al.}(2011)\citenamefont {Liu},
  \citenamefont {Nishide}, \citenamefont {Tanaka},\ and\ \citenamefont
  {Kataura}}]{Liu:2011aa}%
  \BibitemOpen
  \bibfield  {author} {\bibinfo {author} {\bibfnamefont {H.}~\bibnamefont
  {Liu}}, \bibinfo {author} {\bibfnamefont {D.}~\bibnamefont {Nishide}},
  \bibinfo {author} {\bibfnamefont {T.}~\bibnamefont {Tanaka}}, \ and\ \bibinfo
  {author} {\bibfnamefont {H.}~\bibnamefont {Kataura}},\ }\bibfield  {title}
  {\enquote {\bibinfo {title} {Large-scale single-chirality separation of
  single-wall carbon nanotubes by simple gel chromatography},}\ }\href
  {http://dx.doi.org/10.1038/ncomms1313} {\bibfield  {journal} {\bibinfo
  {journal} {Nat Commun}\ }\textbf {\bibinfo {volume} {2}},\ \bibinfo {pages}
  {309} (\bibinfo {year} {2011})}\BibitemShut {NoStop}%
\bibitem [{\citenamefont {Sanchez-Valencia}\ \emph {et~al.}(2014)\citenamefont
  {Sanchez-Valencia}, \citenamefont {Dienel}, \citenamefont {Groning},
  \citenamefont {Shorubalko}, \citenamefont {Mueller}, \citenamefont {Jansen},
  \citenamefont {Amsharov}, \citenamefont {Ruffieux},\ and\ \citenamefont
  {Fasel}}]{Sanchez-Valencia:2014aa}%
  \BibitemOpen
  \bibfield  {author} {\bibinfo {author} {\bibfnamefont {J.~R.}\ \bibnamefont
  {Sanchez-Valencia}}, \bibinfo {author} {\bibfnamefont {T.}~\bibnamefont
  {Dienel}}, \bibinfo {author} {\bibfnamefont {O.}~\bibnamefont {Groning}},
  \bibinfo {author} {\bibfnamefont {I.}~\bibnamefont {Shorubalko}}, \bibinfo
  {author} {\bibfnamefont {A.}~\bibnamefont {Mueller}}, \bibinfo {author}
  {\bibfnamefont {M.}~\bibnamefont {Jansen}}, \bibinfo {author} {\bibfnamefont
  {K.}~\bibnamefont {Amsharov}}, \bibinfo {author} {\bibfnamefont
  {P.}~\bibnamefont {Ruffieux}}, \ and\ \bibinfo {author} {\bibfnamefont
  {R.}~\bibnamefont {Fasel}},\ }\bibfield  {title} {\enquote {\bibinfo {title}
  {Controlled synthesis of single-chirality carbon nanotubes},}\ }\href
  {http://dx.doi.org/10.1038/nature13607} {\bibfield  {journal} {\bibinfo
  {journal} {Nature}\ }\textbf {\bibinfo {volume} {512}},\ \bibinfo {pages}
  {61} (\bibinfo {year} {2014})}\BibitemShut {NoStop}%
\bibitem [{\citenamefont {Wild{\"o}er}\ \emph {et~al.}(1998)\citenamefont
  {Wild{\"o}er}, \citenamefont {Venema}, \citenamefont {Rinzler}, \citenamefont
  {Smalley},\ and\ \citenamefont {Dekker}}]{Wilder-1998-01}%
  \BibitemOpen
  \bibfield  {author} {\bibinfo {author} {\bibfnamefont {J.~W.~G.}\
  \bibnamefont {Wild{\"o}er}}, \bibinfo {author} {\bibfnamefont {L.~C.}\
  \bibnamefont {Venema}}, \bibinfo {author} {\bibfnamefont {A.~G.}\
  \bibnamefont {Rinzler}}, \bibinfo {author} {\bibfnamefont {R.~E.}\
  \bibnamefont {Smalley}}, \ and\ \bibinfo {author} {\bibfnamefont
  {C.}~\bibnamefont {Dekker}},\ }\bibfield  {title} {\enquote {\bibinfo {title}
  {Electronic structure of atomically resolved carbon nanotubes},}\ }\href@noop
  {} {\bibfield  {journal} {\bibinfo  {journal} {Nature}\ }\textbf {\bibinfo
  {volume} {391}},\ \bibinfo {pages} {59} (\bibinfo {year} {1998})}\BibitemShut
  {NoStop}%
\bibitem [{\citenamefont {Odom}\ \emph {et~al.}(1998)\citenamefont {Odom},
  \citenamefont {Huang}, \citenamefont {Kim},\ and\ \citenamefont
  {Lieber}}]{odom-1998-01}%
  \BibitemOpen
  \bibfield  {author} {\bibinfo {author} {\bibfnamefont {T.~W.}\ \bibnamefont
  {Odom}}, \bibinfo {author} {\bibfnamefont {J.-L.}\ \bibnamefont {Huang}},
  \bibinfo {author} {\bibfnamefont {P.}~\bibnamefont {Kim}}, \ and\ \bibinfo
  {author} {\bibfnamefont {C.~M.}\ \bibnamefont {Lieber}},\ }\bibfield  {title}
  {\enquote {\bibinfo {title} {Atomic structure and electronic properties of
  single-walled carbon nanotubes},}\ }\href@noop {} {\bibfield  {journal}
  {\bibinfo  {journal} {Nature}\ }\textbf {\bibinfo {volume} {391}},\ \bibinfo
  {pages} {62} (\bibinfo {year} {1998})}\BibitemShut {NoStop}%
\bibitem [{\citenamefont {Hartschuh}\ \emph {et~al.}(2003)\citenamefont
  {Hartschuh}, \citenamefont {S\'anchez}, \citenamefont {Xie},\ and\
  \citenamefont {Novotny}}]{PhysRevLett.90.095503}%
  \BibitemOpen
  \bibfield  {author} {\bibinfo {author} {\bibfnamefont {A.}~\bibnamefont
  {Hartschuh}}, \bibinfo {author} {\bibfnamefont {E.~J.}\ \bibnamefont
  {S\'anchez}}, \bibinfo {author} {\bibfnamefont {X.~S.}\ \bibnamefont {Xie}},
  \ and\ \bibinfo {author} {\bibfnamefont {L.}~\bibnamefont {Novotny}},\
  }\bibfield  {title} {\enquote {\bibinfo {title} {High-resolution near-field
  raman microscopy of single-walled carbon nanotubes},}\ }\href {\doibase
  10.1103/PhysRevLett.90.095503} {\bibfield  {journal} {\bibinfo  {journal}
  {Phys. Rev. Lett.}\ }\textbf {\bibinfo {volume} {90}},\ \bibinfo {pages}
  {095503} (\bibinfo {year} {2003})}\BibitemShut {NoStop}%
\bibitem [{\citenamefont {Koshino}\ \emph {et~al.}(2014)\citenamefont
  {Koshino}, \citenamefont {Morimoto},\ and\ \citenamefont
  {Sato}}]{PhysRevB.90.115207}%
  \BibitemOpen
  \bibfield  {author} {\bibinfo {author} {\bibfnamefont {M.}~\bibnamefont
  {Koshino}}, \bibinfo {author} {\bibfnamefont {T.}~\bibnamefont {Morimoto}}, \
  and\ \bibinfo {author} {\bibfnamefont {M.}~\bibnamefont {Sato}},\ }\bibfield
  {title} {\enquote {\bibinfo {title} {Topological zero modes and {D}irac
  points protected by spatial symmetry and chiral symmetry},}\ }\href {\doibase
  10.1103/PhysRevB.90.115207} {\bibfield  {journal} {\bibinfo  {journal} {Phys.
  Rev. B}\ }\textbf {\bibinfo {volume} {90}},\ \bibinfo {pages} {115207}
  (\bibinfo {year} {2014})}\BibitemShut {NoStop}%
\bibitem [{\citenamefont {Sakurai}\ and\ \citenamefont
  {Napolitano}(2010)}]{JJSakurai-MQP-2010}%
  \BibitemOpen
  \bibfield  {author} {\bibinfo {author} {\bibfnamefont {J.~J.}\ \bibnamefont
  {Sakurai}}\ and\ \bibinfo {author} {\bibfnamefont {J.~J.}\ \bibnamefont
  {Napolitano}},\ }\href@noop {} {\emph {\bibinfo {title} {Modern Quantum
  Mechanics (Second Edition)}}}\ (\bibinfo  {publisher} {Addison-Wesley},\
  \bibinfo {address} {San Francisco},\ \bibinfo {year} {2010})\BibitemShut
  {NoStop}%
\bibitem [{\citenamefont {{\it et. al.,}~in preparation}()}]{Yamakage-inPrep}%
  \BibitemOpen
  \bibfield  {author} {\bibinfo {author} {\bibfnamefont {A.~Yamakage}\
  \bibnamefont {{\it et. al.,}~in preparation}},\ }\href@noop {} {}\BibitemShut
  {NoStop}%
\bibitem [{\citenamefont {Su}\ \emph {et~al.}(1979)\citenamefont {Su},
  \citenamefont {Schrieffer},\ and\ \citenamefont
  {Heeger}}]{PhysRevLett.42.1698}%
  \BibitemOpen
  \bibfield  {author} {\bibinfo {author} {\bibfnamefont {W.~P.}\ \bibnamefont
  {Su}}, \bibinfo {author} {\bibfnamefont {J.~R.}\ \bibnamefont {Schrieffer}},
  \ and\ \bibinfo {author} {\bibfnamefont {A.~J.}\ \bibnamefont {Heeger}},\
  }\bibfield  {title} {\enquote {\bibinfo {title} {Solitons in
  polyacetylene},}\ }\href {\doibase 10.1103/PhysRevLett.42.1698} {\bibfield
  {journal} {\bibinfo  {journal} {Phys. Rev. Lett.}\ }\textbf {\bibinfo
  {volume} {42}},\ \bibinfo {pages} {1698} (\bibinfo {year}
  {1979})}\BibitemShut {NoStop}%
\bibitem [{\citenamefont {Hasan}\ and\ \citenamefont
  {Kane}(2010)}]{RevModPhys.82.3045}%
  \BibitemOpen
  \bibfield  {author} {\bibinfo {author} {\bibfnamefont {M.~Z.}\ \bibnamefont
  {Hasan}}\ and\ \bibinfo {author} {\bibfnamefont {C.~L.}\ \bibnamefont
  {Kane}},\ }\bibfield  {title} {\enquote {\bibinfo {title}
  {\textit{Colloquium} : Topological insulators},}\ }\href {\doibase
  10.1103/RevModPhys.82.3045} {\bibfield  {journal} {\bibinfo  {journal} {Rev.
  Mod. Phys.}\ }\textbf {\bibinfo {volume} {82}},\ \bibinfo {pages} {3045}
  (\bibinfo {year} {2010})}\BibitemShut {NoStop}%
\bibitem [{\citenamefont {Qi}\ and\ \citenamefont
  {Zhang}(2011)}]{RevModPhys.83.1057}%
  \BibitemOpen
  \bibfield  {author} {\bibinfo {author} {\bibfnamefont {X.-L.}\ \bibnamefont
  {Qi}}\ and\ \bibinfo {author} {\bibfnamefont {S.-C.}\ \bibnamefont {Zhang}},\
  }\bibfield  {title} {\enquote {\bibinfo {title} {Topological insulators and
  superconductors},}\ }\href {\doibase 10.1103/RevModPhys.83.1057} {\bibfield
  {journal} {\bibinfo  {journal} {Rev. Mod. Phys.}\ }\textbf {\bibinfo {volume}
  {83}},\ \bibinfo {pages} {1057} (\bibinfo {year} {2011})}\BibitemShut
  {NoStop}%
\bibitem [{\citenamefont {Kane}\ and\ \citenamefont
  {Mele}(2005)}]{Kane-2005-11}%
  \BibitemOpen
  \bibfield  {author} {\bibinfo {author} {\bibfnamefont {C.~L.}\ \bibnamefont
  {Kane}}\ and\ \bibinfo {author} {\bibfnamefont {E.~J.}\ \bibnamefont
  {Mele}},\ }\bibfield  {title} {\enquote {\bibinfo {title} {Quantum spin
  {H}all effect in graphene},}\ }\href@noop {} {\bibfield  {journal} {\bibinfo
  {journal} {Phys. Rev. Lett.}\ }\textbf {\bibinfo {volume} {95}},\ \bibinfo
  {pages} {226801} (\bibinfo {year} {2005})}\BibitemShut {NoStop}%
\bibitem [{\citenamefont {Kim}\ \emph {et~al.}(2015)\citenamefont {Kim},
  \citenamefont {Wieder}, \citenamefont {Kane},\ and\ \citenamefont
  {Rappe}}]{PhysRevLett.115.036806}%
  \BibitemOpen
  \bibfield  {author} {\bibinfo {author} {\bibfnamefont {Y.}~\bibnamefont
  {Kim}}, \bibinfo {author} {\bibfnamefont {B.~J.}\ \bibnamefont {Wieder}},
  \bibinfo {author} {\bibfnamefont {C.~L.}\ \bibnamefont {Kane}}, \ and\
  \bibinfo {author} {\bibfnamefont {A.~M.}\ \bibnamefont {Rappe}},\ }\bibfield
  {title} {\enquote {\bibinfo {title} {Dirac line nodes in inversion-symmetric
  crystals},}\ }\href {\doibase 10.1103/PhysRevLett.115.036806} {\bibfield
  {journal} {\bibinfo  {journal} {Phys. Rev. Lett.}\ }\textbf {\bibinfo
  {volume} {115}},\ \bibinfo {pages} {036806} (\bibinfo {year}
  {2015})}\BibitemShut {NoStop}%
\bibitem [{\citenamefont {Yamakage}\ \emph {et~al.}(2016)\citenamefont
  {Yamakage}, \citenamefont {Yamakawa}, \citenamefont {Tanaka},\ and\
  \citenamefont {Okamoto}}]{Yamakage-JPSJ-2016-01}%
  \BibitemOpen
  \bibfield  {author} {\bibinfo {author} {\bibfnamefont {A.}~\bibnamefont
  {Yamakage}}, \bibinfo {author} {\bibfnamefont {Y.}~\bibnamefont {Yamakawa}},
  \bibinfo {author} {\bibfnamefont {Y.}~\bibnamefont {Tanaka}}, \ and\ \bibinfo
  {author} {\bibfnamefont {Y.}~\bibnamefont {Okamoto}},\ }\bibfield  {title}
  {\enquote {\bibinfo {title} {Line-node {D}irac semimetal and topological
  insulating phase in noncentrosymmetric pnictides {C}a{A}g{$X$} ({$X$} = {P},
  {A}s)},}\ }\href {\doibase 10.7566/JPSJ.85.013708} {\bibfield  {journal}
  {\bibinfo  {journal} {J. Phys. Soc. Jpn.}\ }\textbf {\bibinfo {volume}
  {85}},\ \bibinfo {pages} {013708} (\bibinfo {year} {2016})}\BibitemShut
  {NoStop}%
\end{thebibliography}
%

\end{document}